\documentclass[3p, 12pt, short&compress]{elsarticle}
\pdfoutput = 1
\usepackage{graphics}
\usepackage{morefloats}
\usepackage{natbib}
\usepackage{amssymb,amsmath, slashed}
\usepackage{epsfig}
\usepackage{color}
\usepackage{fontenc}
\allowdisplaybreaks[2]
\newcommand{\bea}{\begin{eqnarray}}
\newcommand{\eea}{\end{eqnarray}}

\newcommand{\rw}{\rightarrow}

 \makeatletter
\def\alt{\mathrel{\mathpalette\gl@align<}}
\def\agt{\mathrel{\mathpalette\gl@align>}}
\def\gl@align#1#2{\lower.6ex\vbox{\baselineskip\z@skip\lineskip\z@
\ialign{$\m@th#1\hfil##\hfil$\crcr#2\crcr\sim\crcr}}} \makeatother

\begin{document}
\begin{center}
\baselineskip 20pt 
{\Large\bf 
Improved bounds on the heavy neutrino productions at the LHC
}
\vspace{1cm}

{\large 
Arindam Das\footnote{adas8@ua.edu} 
and 
Nobuchika Okada\footnote{okadan@ua.edu}
} \vspace{.5cm}

{\baselineskip 20pt \it
Department of Physics and Astronomy, 
University of Alabama, \\ 
Tuscaloosa,  Alabama 35487, USA 
}

\vspace{.5cm}

\vspace{1.5cm} 
{\bf Abstract}
\end{center}
The Majorana neutrino in type-I seesaw and the pseudo-Dirac neutrinos in the inverse seesaw 
can have sizable mixings with the light neutrinos in the standard model (SM), through which the heavy neutrinos can be produced at the Large Hadron Collider (LHC).
In producing the heavy neutrinos we study a variety of initial states such as quark-quark, quark-gluon and gluon-gluon as well as photon mediated processes.
For the Majorana heavy neutrino production we consider same-sign di-lepton plus di-jet as the signal events.
Using the recent ATLAS and CMS data at $\sqrt{s}=$ 8 TeV 
with 20.3 fb$^{-1}$ and 19.7 fb$^{-1}$ luminosities, respectively, we obtain direct upper bounds on the light-heavy neutrino mixing angles.
For the pseudo-Dirac heavy neutrino production we consider the final sate with tri-lepton plus missing energy as the signal events. 
Using the recent anomalous multi-lepton search by CMS at $\sqrt{s}=$ 8 TeV with 19.5 fb$^{-1}$ luminosity we obtain upper bounds on the mixing angles.
Taking the varieties of initial states into account, the previously obtained upper bounds on the mixing angles have been improved.
We scale our results at the 8 TeV LHC to obtain a prospective
search reach at the 14 TeV LHC  with high luminosities.
\thispagestyle{empty}

\newpage

\addtocounter{page}{-1}
\setcounter{footnote}{0}

\baselineskip 18pt

\section{Introduction}
The current neutrino oscillation data~\cite{Neut1}-\cite{Neut6} 
  have established the existence of the neutrino mass 
  and the flavor mixing. 
This requires us to extend the Standard Model $\left(\rm SM\right)$. 
The seesaw extension \cite{typeI-seesaw1}-\cite{typeI-seesaw5}
  of the SM is probably the simplest idea 
  which can naturally explain the tiny neutrino mass.
In case of type-I seesaw, the SM is extended by introducing 
  SM-singlet heavy Majorana neutrino 
  which induce a dimension five operator, leading to 
  a very small light Majorana neutrino masses. 
If such heavy neutrino mass lies in the electroweak scale,
  then the heavy neutrinos can be produced in the high
  energy colliders such as the Large Hadron Collider $\left( \rm LHC\right)$
  and International Linear Collider $\left(\rm ILC\right)$ \cite{ILC1, ILC2}. 
As the heavy neutrinos are singlet under the SM gauge group,
  they obtain the couplings with the SM gauge bosons only
  through the mixing with light SM neutrinos via the Dirac Yukawa coupling. 
In the general parameterization for the seesaw formula \cite{Ibarra1},
  the Dirac Yukawa coupling can be sizable for the electroweak scale
  heavy neutrinos, while reproducing the neutrino oscillation data.  
 
There is another kind of seesaw mechanism, so-called
  inverse seesaw \cite{inverse-seesaw1, inverse-seesaw2},
  where the small Majorana neutrino mass originates 
  from tiny lepton number violating parameters,
  rather than the suppression by the heavy neutrino mass
  in the conventional seesaw mechanism. 
In this case the heavy neutrinos are pseudo-Dirac particles and
  their Dirac Yukawa couplings can be even order one,
  while reproducing the tiny neutrino mass.
Thus, at the high energy colliders the heavy pseudo-Dirac neutrino
  can be produced through a sizable mixing with the SM neutrinos.

In this paper we study the heavy neutrino production
  at the LHC through a variety of initial states 
  such as quark-quark $\left(q\overline{q^{'}}\right)$, quark-gluon $\left(qg\right)$
  and gluon-gluon $\left(gg\right)$
  as well as the photon mediated processes. 
For fixed heavy neutrino masses $\left(m_{N}\right)$ 
  we calculate the individual cross section of the 
  respective processes normalized by the square of 
  the mixing angles between light and the heavy neutrinos.
We also study the kinematic distributions of the signal events produced by 
 different initial states and find the characteristic distributions corresponding
  to the individual initial states.
  
For the Majorana heavy neutrino case we consider the same 
  sign di-lepton plus two jet signal. 
Using the recent collider studies \cite{ATLAS7}-\cite{CMS8}
  we put an upper limit on the light-heavy mixing angles.
For the pseudo-Dirac heavy neutrino we consider
  tri-lepton plus missing energy signal.
We compare the upper bounds with those obtained 
  by other experiments \cite{CMS-trilep}-\cite{EWPD3}. 
Though some updated analysis could be found in \cite{Oliver1, Oliver2} 
Using all the initial states the upper bounds
  given in \cite{DDO1} are improved.

This paper is organized as follows. 
In Sec.~2, we introduce the models 
  for the type-I and the inverse seesaws. 
We calculate the production cross sections 
  of the heavy neutrinos at the LHC with a variety of
  initial states in Sec~3. 
In Sec.~4, we study the kinematic distributions 
  of the final state particles associated with the heavy
  neutrino productions for the individual initial states.
In Sec.~5 we simulate the signal events for the heavy
  neutrino productions. 
Comparing the generated events with the current LHC data,
  we obtain improved upper bounds on the mixing angles
  between the light-heavy neutrinos for the type-I and
  inverse seesaws, respectively.
Sec.~6 is devoted for conclusions.  

\section{The Models}
In type-I seesaw \cite{seesaw1, seesaw2}, we introduce
  SM gauge-singlet right handed Majorana neutrinos $N_R^{\beta}$,    
  where $\beta$ is the flavor index. $N_R^{\beta}$ couple with SM lepton doublets 
  $\ell_{L}^{\alpha}$ and the SM Higgs doublet $H$.
The relevant part of the Lagrangian is
\bea
\mathcal{L} \supset -Y_D^{\alpha\beta} \overline{\ell_L^{\alpha}}H N_R^{\beta} 
                   -\frac{1}{2} m_N^{\alpha \beta} \overline{N_R^{\alpha C}} N_R^{\beta}  + H. c. .
\label{typeI}
\eea
After the spontaneous electroweak symmetry breaking
   by the vacuum expectation value (VEV), 
   $ H =\begin{pmatrix} \frac{v}{\sqrt{2}} \\  0 \end{pmatrix}$, 
    we obtain the Dirac mass matrix as $M_{D}= \frac{Y_D v}{\sqrt{2}}$.
Using the Dirac and Majorana mass matrices 
  we can write the neutrino mass matrix as 
\bea
M_{\nu}=\begin{pmatrix}
0&&M_{D}\\
M_{D}^{T}&&m_{N}
\end{pmatrix}.
\label{typeInu}
\eea
Diagonalizing this matrix we obtain the seesaw formula
 for the light Majorana neutrinos as 
\bea
m_{\nu} \simeq - M_{D} m_{N}^{-1} M_{D}^{T}.
\label{seesawI}
\eea
For $m_{N}\sim 100$ GeV, we may find $Y_{D} \sim 10^{-6}$  with $m_{\nu}\sim 0.1$ eV.
However, in the general parameterization for the seesaw formula \cite{Ibarra1},  
  $Y_{D}$ can be as large as 1, and this is the case we consider in this paper.

There is another seesaw mechanism, so-called inverse seesaw \cite{inverse-seesaw1, inverse-seesaw2},
   where the light Majorana neutrino mass is generated through tiny 
   lepton number violation.
The relevant part of the Lagrangian is given by
\bea
\mathcal{L} \supset - Y_D^{\alpha\beta} \overline{\ell_L^{\alpha}} H N_R^{\beta}- m_N^{\alpha \beta} \overline{S_L^{\alpha}} N_R^{\beta} -\frac{1}{2} \mu_{\alpha \beta} \overline{S_L^{\alpha}}S_L^{\beta^{C}} + H. c. ,
\label{InvYuk}
\eea 
where  $N_R^{\alpha}$ and $S_L^{\beta}$ are two SM-singlet heavy neutrinos
   with the same lepton numbers, $m_N$ is the Dirac mass matrix, and
   $\mu$ is a small Majorana mass matrix violating the lepton numbers.
After the electroweak symmetry breaking we obtain the neutrino mass matrix as 
\bea
M_{\nu}=\begin{pmatrix}
0&&M_{D}&&0\\
M_{D}^{T}&&0&&m_{N}^{T}\\
0&&m_{N}&&\mu
\end{pmatrix}.
\label{InvMat}
\eea
Diagonalizing this mass matrix we obtain the light neutrino mass matrix 
\bea
M_{\nu} \simeq M_{D} m_{N}^{-1}\mu m_{N}^{-1^{T}} M_{D}^{T}.
\label{numass}
\eea
Note that the smallness of the light neutrino mass originates 
  from the small lepton number violating term $\mu$. 
The smallness of $\mu$ allows the $m_{D}m_{N}^{-1}$ parameter
  to be order one even for an electroweak scale heavy neutrino.
Since the scale of $\mu$ is much smaller than the scale of $m_{N}$,
  the heavy neutrinos become the pseudo-Dirac particles. 
This is the main difference between the type-I and the inverse seesaws. 
See, for example, \cite{ADO} for the 
  interaction Lagrangians and the partial decay widths 
  of the heavy neutrinos.

\section{Heavy neutrino production at the LHC}
We implement our model in the generator 
   {\tt MadGraph5-aMC@NLO} \cite{MG}-\cite{aMC} and
   calculate the production cross section of the heavy neutrino
   with a variety of initial states such as 
   $q\overline{q'}$, $qg$ and $gg$ 
   as well as photon mediated processes. 
For the final states we consider $N\ell+n{\rm -jet}$, 
   where the number of jets is $n=0, 1, 2$.
We separately calculate the production cross sections for
   individual initial states to understand which initial
   states dominantly contribute to the production process
   of the heavy neutrino. See \cite{Majorana1}-\cite{Brian1}for recent studies on 
   heavy neutrino production at high energy colliders.

\subsection{$N\ell$ \rm{\textbf{and}} $N\ell j$ {\rm \textbf{production processes from $q\overline{q'}$ and $qg$ initial states}}}
\begin{figure}
\begin{center}
\includegraphics[scale=0.47]{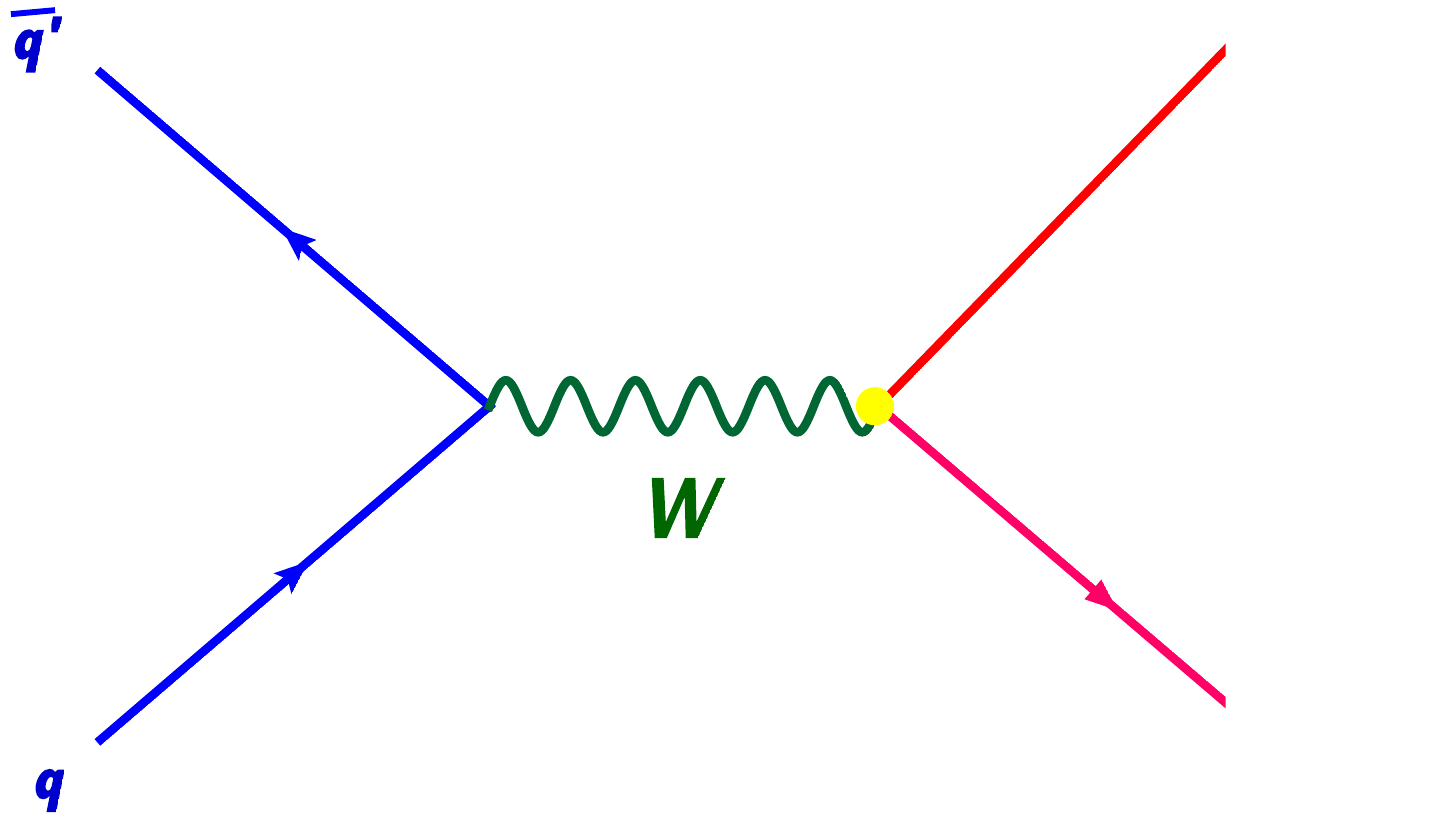}
\end{center}
\caption{Feynman diagram for $N\ell$ production from the $q\overline{q^{'}}$ annihilation.}
\label{fig1}
\end{figure}

\begin{figure}
\begin{center}
\includegraphics[scale=0.47]{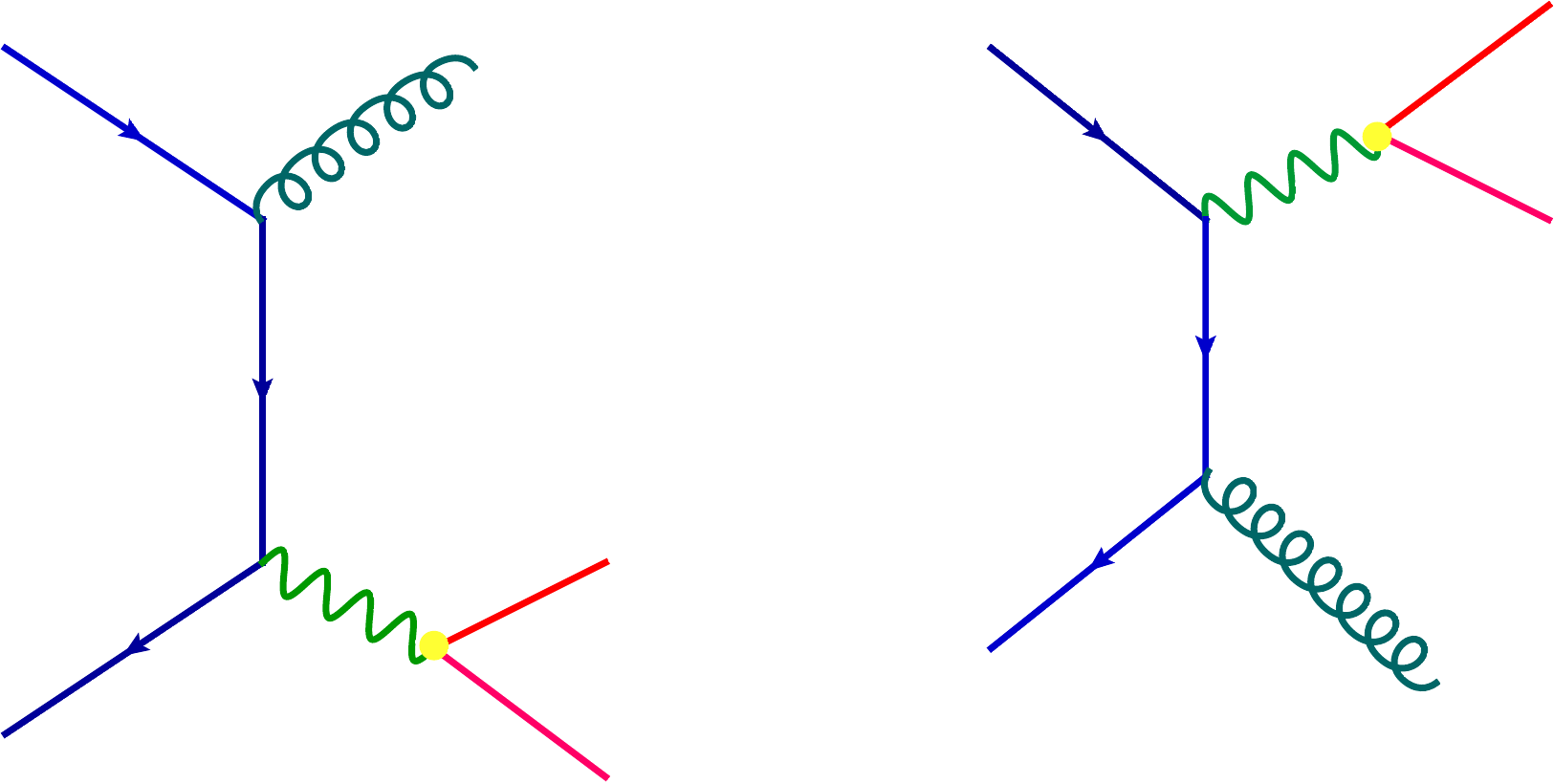}
\end{center}
\caption{Feynman diagrams for $N\ell j$ production from the $q\overline{q^{'}}$ annihilation.}
\label{fig2}
\end{figure}

\begin{figure}
\begin{center}
\includegraphics[scale=0.47]{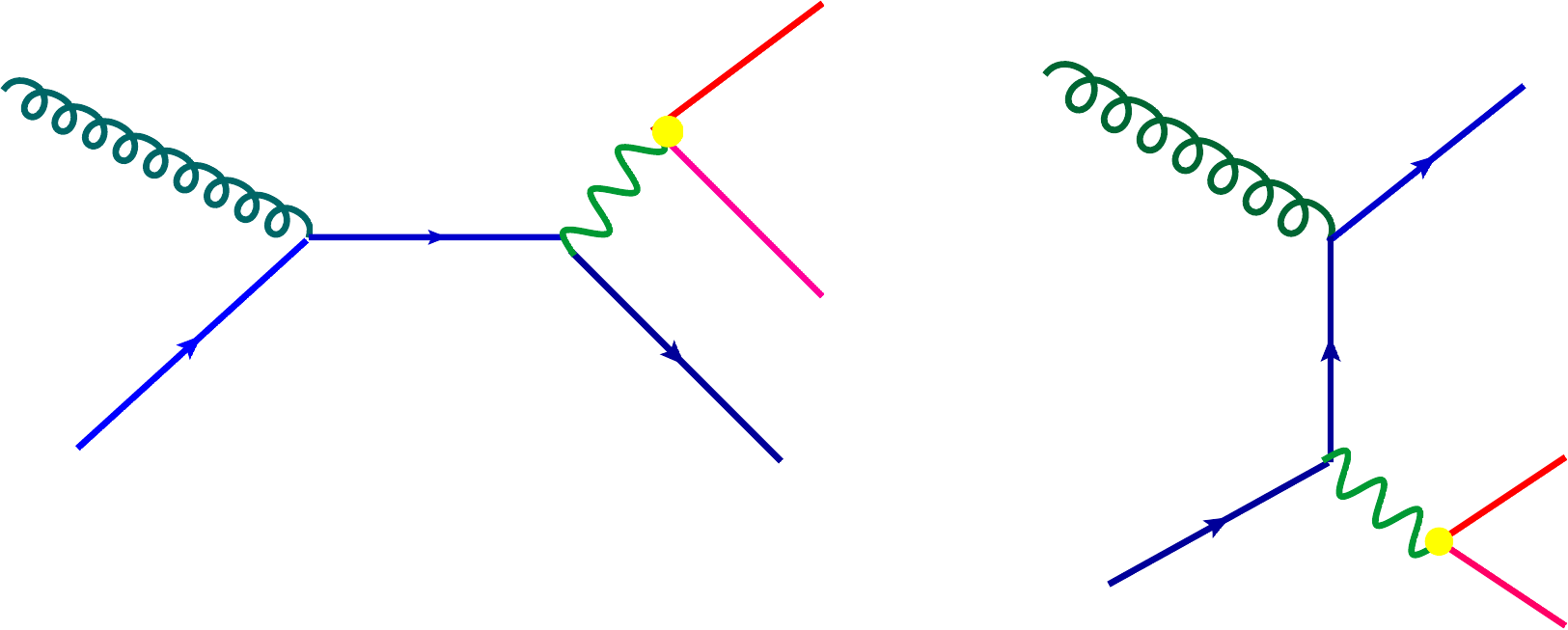}
\end{center}
\caption{Feynman diagrams for $N\ell j$ production from the $qg$ fusion.}
\label{fig3}
\end{figure}

We first consider the final state $N\ell$ from $q\overline{q'}$, 
  and  $N\ell j$ from both of $q\overline{q'}$
  and $qg$ initial states. 
The relevant Feynman diagrams with 
  the initial $q\overline{q'}$ state
  are shown in Figs.~\ref{fig1} and \ref{fig2}, 
  while those with the initial $qg$ state are 
  shown in Fig.~\ref{fig3}.
\begin{figure}
\begin{center}
\includegraphics[scale=0.47]{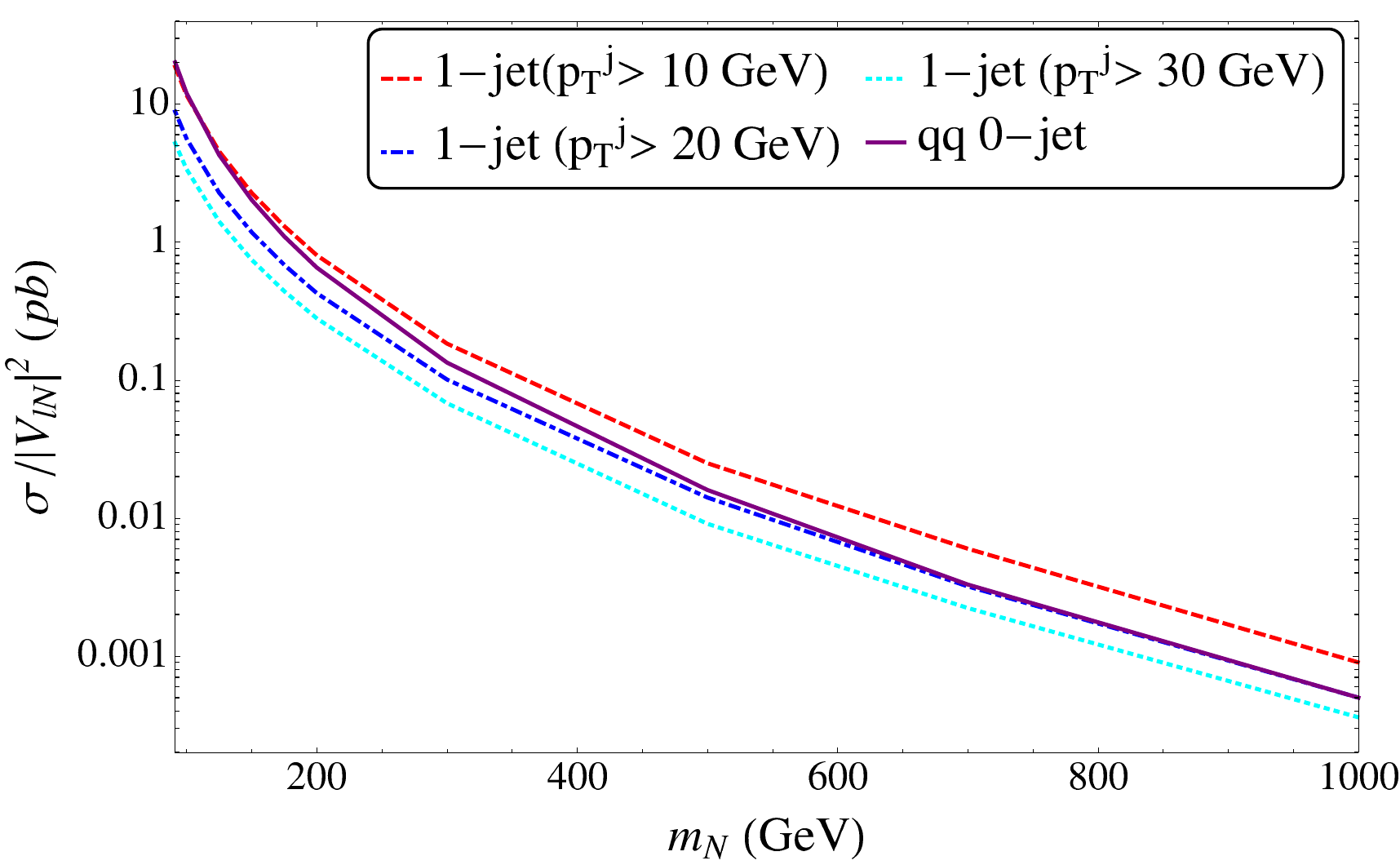}
\includegraphics[scale=0.47]{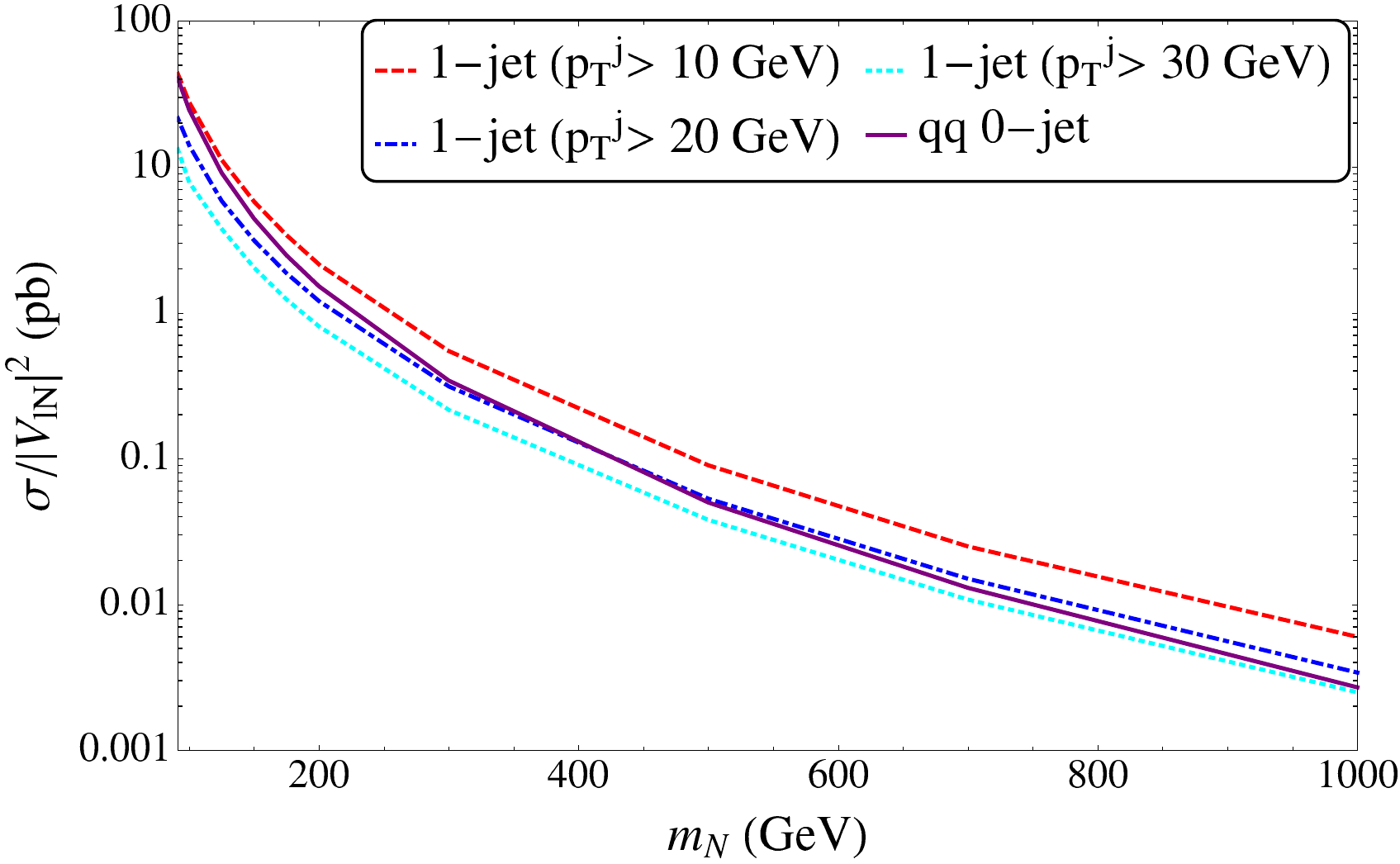}
\end{center}
\caption{The left panel shows the cross sections as a function of $m_{N}$ normalized by the square of the mixing angle for the $N\ell j$ final state at the $p_{T}^{j} >$ 10 GeV (dashed), $p_{T}^{j} >$20 GeV (dot-dashed) and $p_{T}^{j} >$ 30 GeV (dotted) from the $q\overline{q^{'}}$ and $qg$ initial states and the $N\ell$ final state from the $qq$ initial states (solid) at the 8 TeV LHC. The right panel is same as the left panel but at the 14 TeV LHC.}
\label{fig4}
\end{figure}
Fig.~\ref{fig4} shows the combined heavy neutrino production
   cross section normalized by the square of the mixing angle
   from the two initial states.
For the final state $N\ell j$ we impose 
  the minimum transverse momentum of the jet
  ($p_{T}^{j}$) as 10 GeV, 20 GeV and 30 GeV,
  respectively.
In generating the events we set the 
  {\tt Xqcut}$=p_{T}^{j}$ in MadGraph
  with the {\tt MLM} matching scheme ({\tt ickkw}$=1$)
  \cite{Matching}- \cite{Matching4}.
The left panel shows the results for $\sqrt{s}=$8 TeV, while 
  $\sqrt{s}=$14 TeV LHC results are shown in the right panel.
The $N\ell j$ cross section dominates over the production,
  while for a higher $p_{T}^{j}$ cut 
  the $N\ell$ cross section dominates.

\begin{figure}
\begin{center}
\includegraphics[scale=0.46]{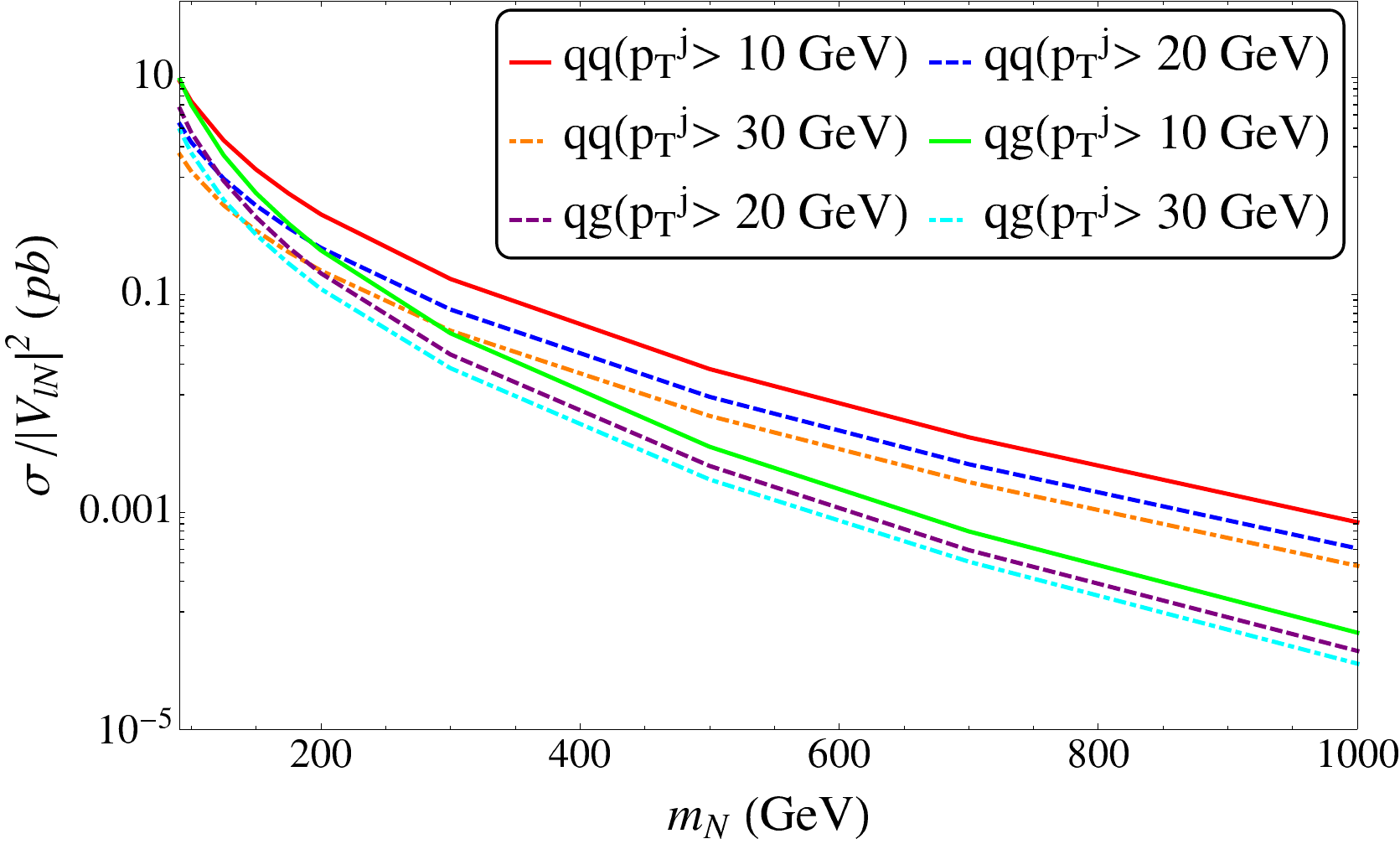}
\includegraphics[scale=0.45]{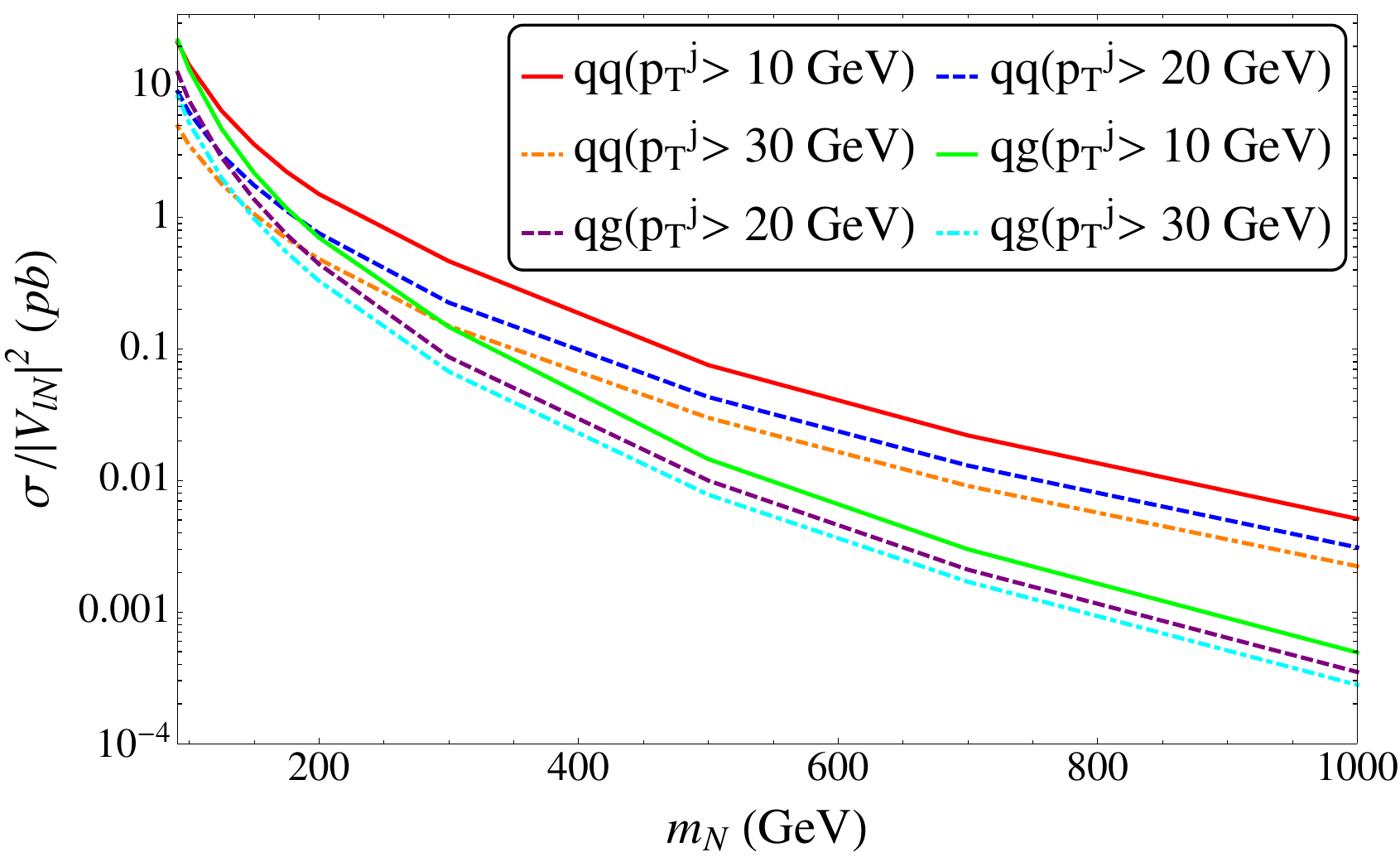}
\end{center}
\caption{The decomposition of the cross section for the individual initial states for different $p_{T}^{j}$ cuts such that 
$p_{T}^{j} >$ 10 GeV (solid), $p_{T}^{j} >$ 20 GeV (dashed) and $p_{T}^{j} >$ 30 GeV (dot-dashed). The cross sections at the 8 TeV LHC are shown 
in the left panel where as those at the 14 TeV LHC are shown in the right panel.}
\label{fig5}
\end{figure}

We show in Fig.~\ref{fig5} the decomposition of the cross sections shown
  in Fig.~\ref{fig4} for individual final states. 
The left panel shows the results for $\sqrt{s}=$8 TeV, while 
  $\sqrt{s}=$14 TeV results are shown in the right panel.
The cross section from the $q\overline{q'}$ initial state 
  dominates over the one from the $qg$ initial state for
  the cut of $p_{T}^{j} > 10$ GeV. 
With the increase of the $p_{T}^{j}$ cut the cross section from the 
  $qg$ initial state becomes dominant for the small mass region
  of the heavy neutrino, as pointed out in \cite{DDO1}. 
This is because the parton distribution function (PDF) of gluon is much
  larger than the PDFs of quarks at the low Bjorken scaling
  parameter. 
As the heavy neutrino mass becomes larger, the cross section from 
  $q\overline{q'}$ takes over the one from $qg$ because the 
  gluon PDF sharply drops toward the high Bjorken scaling parameter.

\subsection{$N\ell jj$ {\rm \textbf{ production processes from $q\overline{q'}$, $qg$ 
and $gg$ initial states }}}

\begin{figure}
\begin{center}
\includegraphics[scale=0.6]{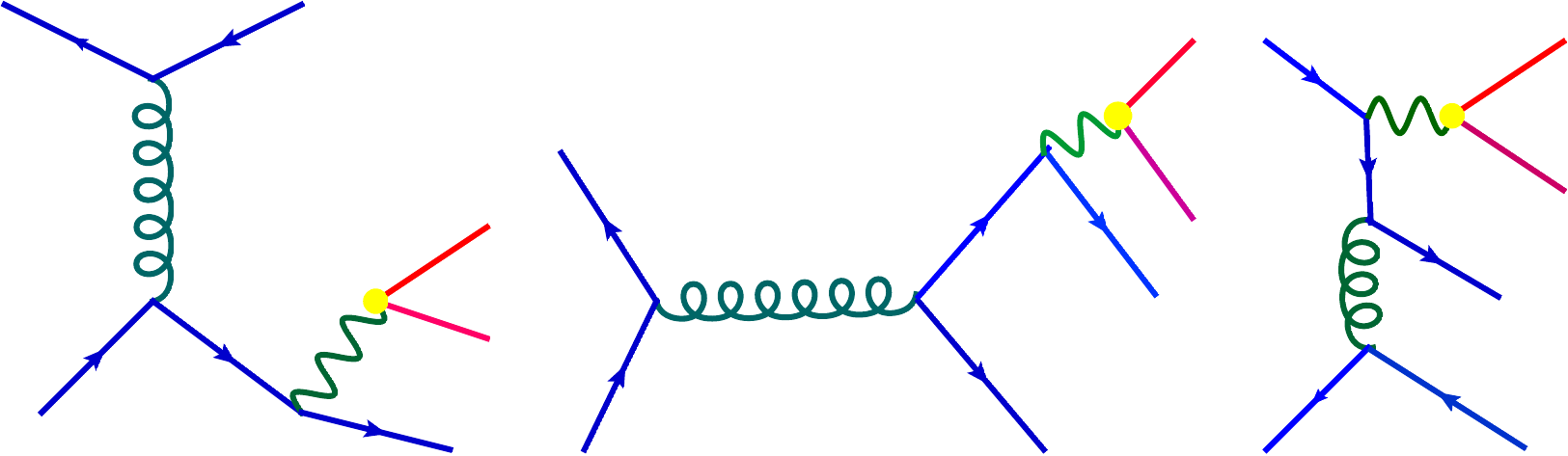}
\end{center}
\caption{Sample Feynman diagrams for $N\ell jj$ production processes from the $q\overline{q^{'}}$ initial state.}
\label{qq2j}
\end{figure}
\begin{figure}
\begin{center}
\includegraphics[scale=0.6]{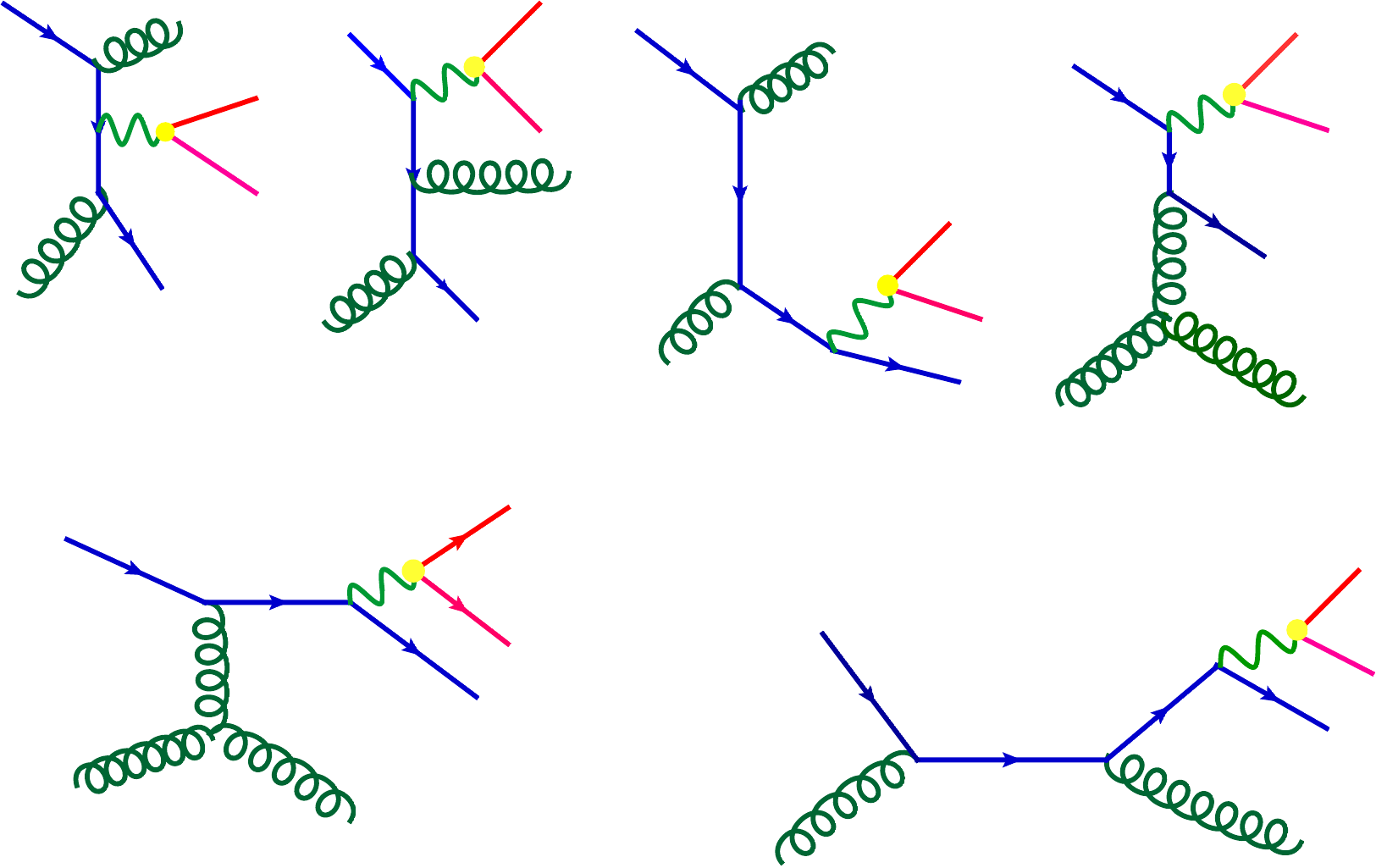}
\end{center}
\caption{Sample Feynman diagrams for $N\ell jj$ production processes from the $qg$ fusion.}
\label{qg2j}
\end{figure}
\begin{figure}
\begin{center}
\includegraphics[scale=0.6]{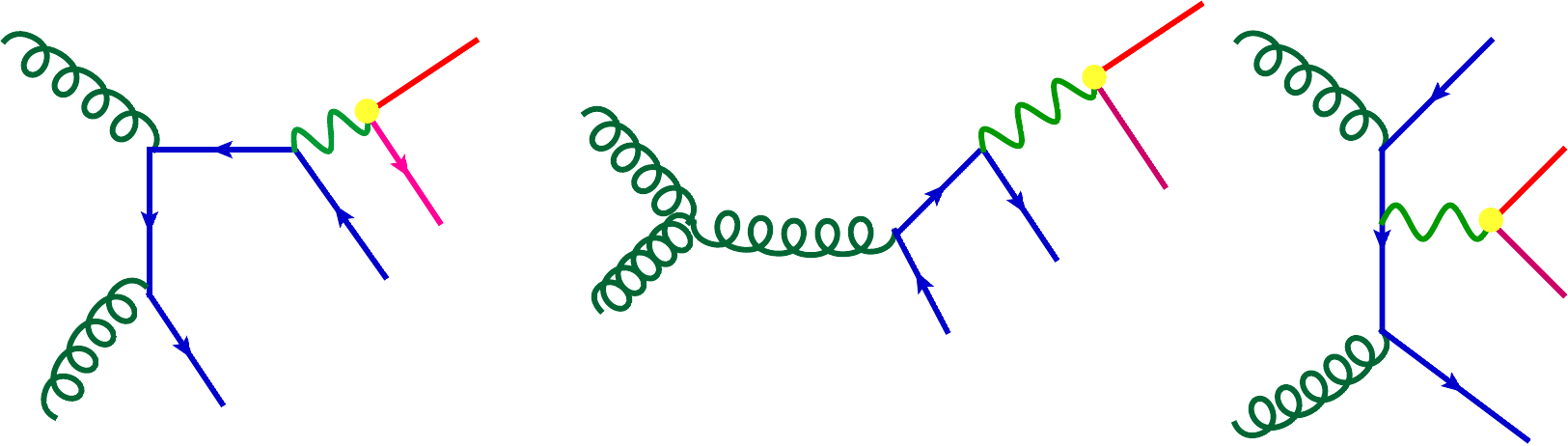}
\end{center}
\caption{Sample Feynman diagrams for $N\ell jj$ production processes from the $gg$ fusion.}
\label{gg2j}
\end{figure}
\begin{figure}
\begin{center}
\includegraphics[scale=0.47]{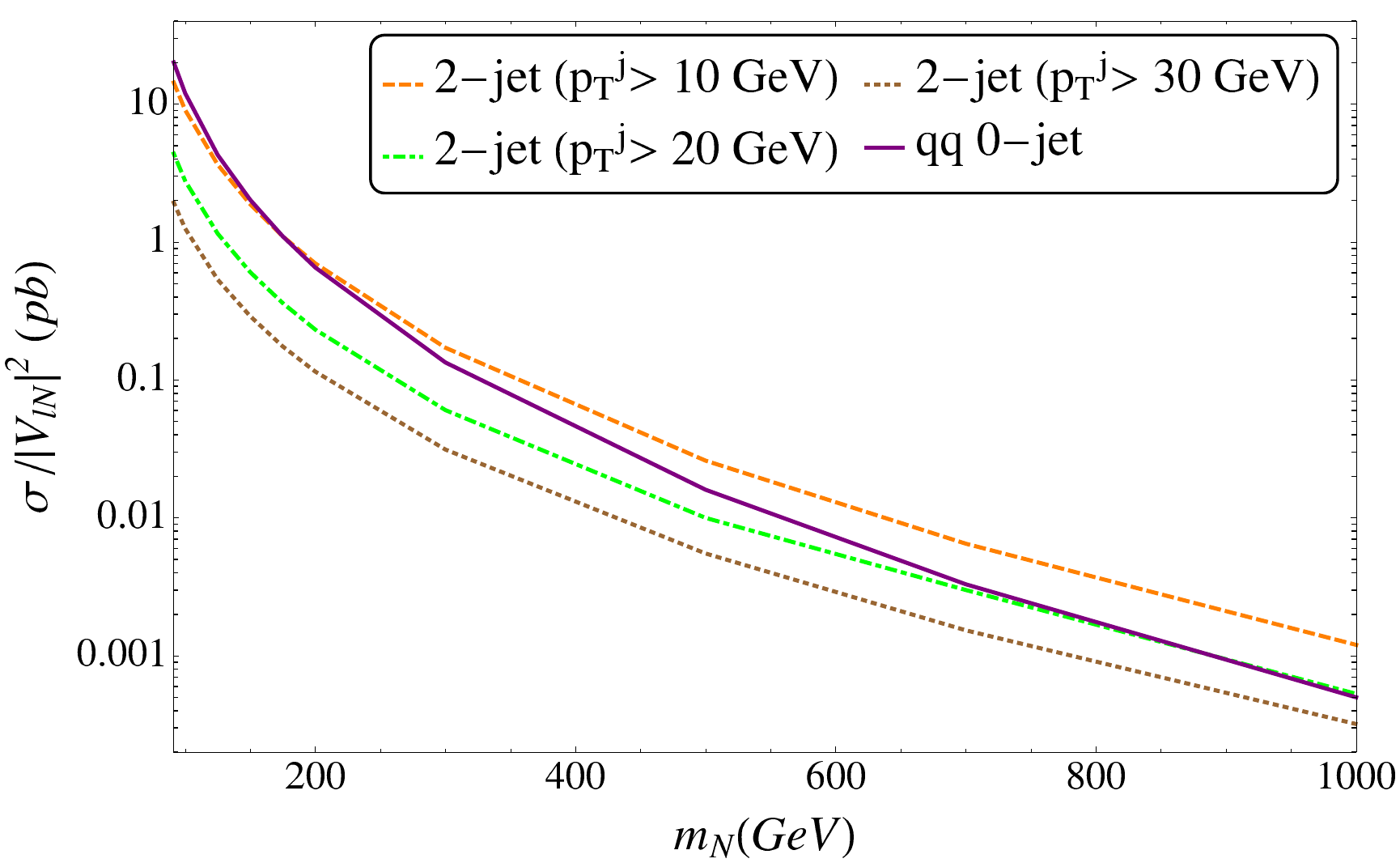}
\includegraphics[scale=0.47]{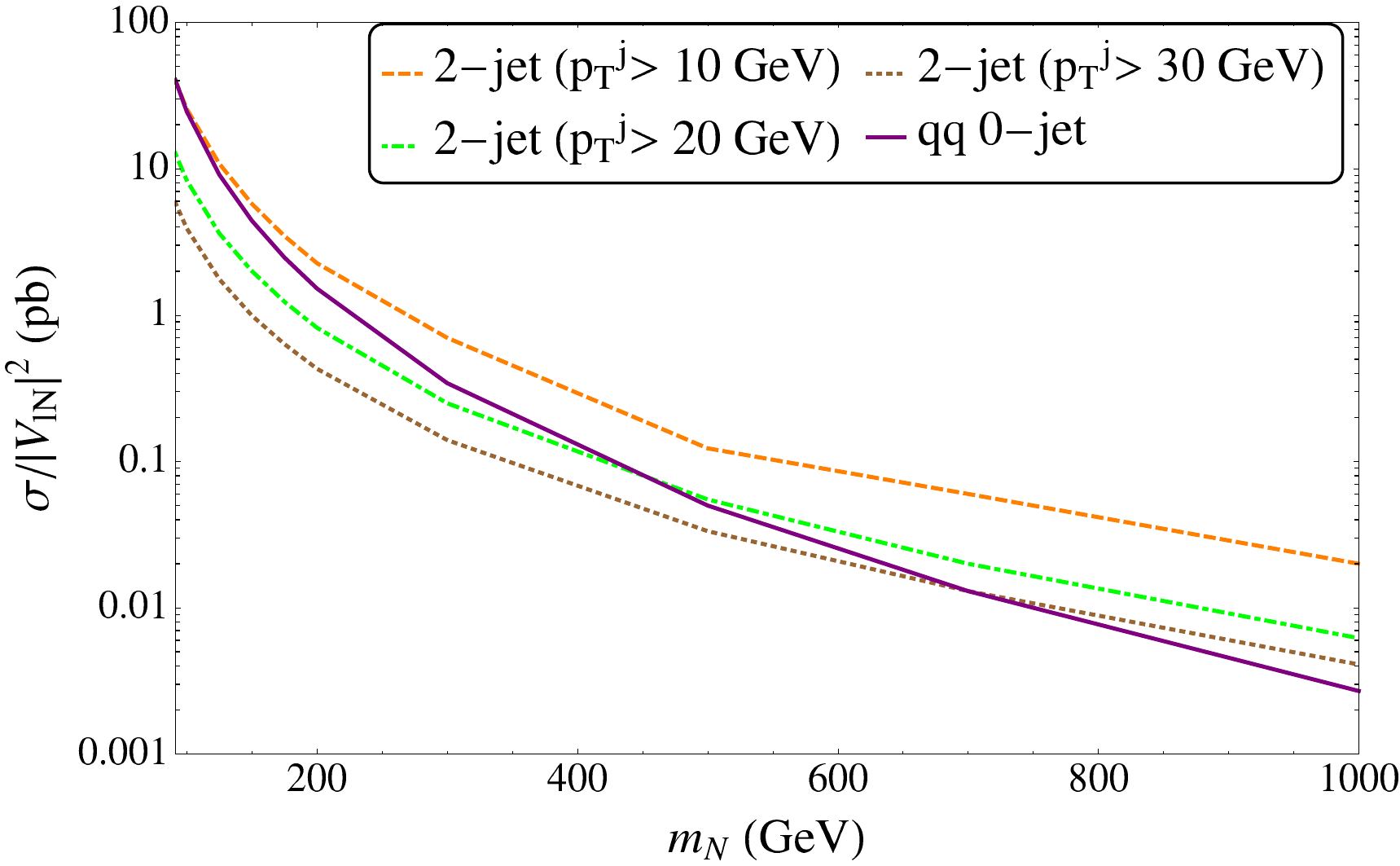}
\end{center}
\caption{The left panel shows the cross sections as a function of $m_{N}$ normalized by the square of the mixing angle for the $N\ell jj$ final state for $p_{T}^{j} >$ 10 GeV (dashed), $p_{T}^{j} >$20 GeV (dot-dashed) and $p_{T}^{j} >$ 30 GeV (dotted) from the $q\overline{q^{'}}$, $qg$ and $gg$ initial states and the $N\ell$ final state from the $q\overline{q^{'}}$ initial state (solid) at the 8 TeV LHC. The right panel shows the results for the 14 TeV LHC.}
\label{fig5a}
\end{figure}

Next we consider the $N\ell jj$ final state which comes from the 
  $q\overline{q'}$, $qg$ and $gg$ initial states. 
The relevant Feynman diagrams with the initial 
$q\overline{q'}$, $qg$ and $gg$ states
  are shown in Figs.~\ref{qq2j}-\ref{gg2j}.
Fig.~\ref{fig5a} shows the combined heavy neutrino production
   cross section normalized by the square of the mixing angle
   from the three initial states.
For the final state $N\ell jj$ we impose 
  the minimum transverse momentum for each jet
  ($p_{T}^{j}$) as 10 GeV, 20 GeV and 30 GeV,
  respectively.
As in the previous sub-section we set the 
  {\tt Xqcut}$=p_{T}^{j}$ in MadGraph
  with the {\tt MLM} matching scheme.
The left panel shows the results for $\sqrt{s}=$8 TeV, while those for 
  $\sqrt{s}=$14 TeV are shown in the right panel.
The $N\ell jj$ cross section dominates over the production 
  cross section for the final state with zero jet for the 10 GeV
  $p_{T}^{j}$ cut, while for a higher $p_{T}^{j}$ cut 
  the $N\ell$ cross section dominates.

\begin{figure}
\begin{center}
\includegraphics[scale=0.43]{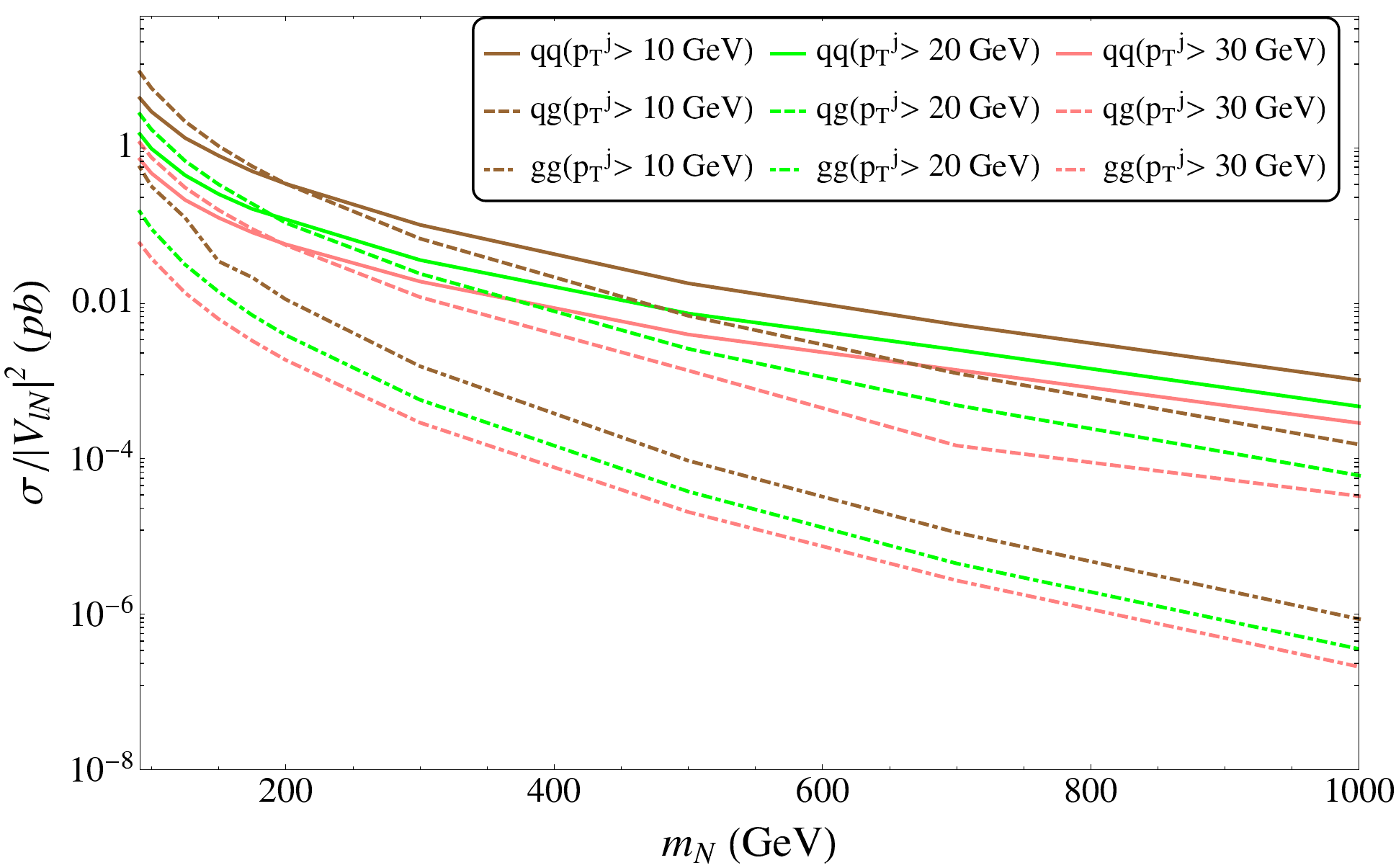}
\includegraphics[scale=0.44]{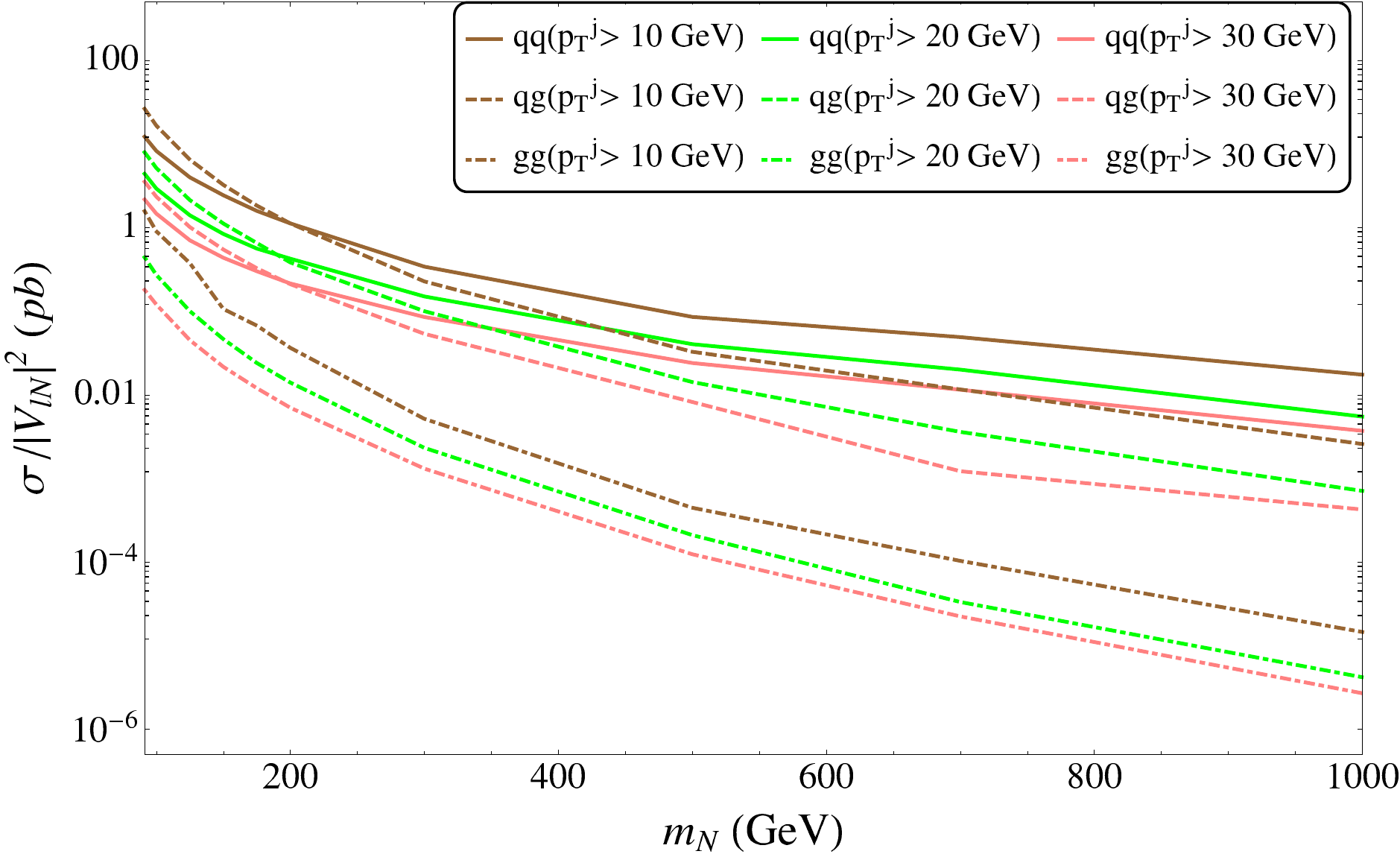}
\end{center}
\caption{The decomposition of the cross section for the individual initial states for different $p_{T}^{j}$ cuts such that 
$p_{T}^{j} >$ 10 GeV (solid), $p_{T}^{j} >$ 20 GeV (dashed) and $p_{T}^{j} >$ 30 GeV (dot-dashed). The cross sections at the 8 TeV are shown 
in the left panel where as those at the 14 TeV are shown in the right panel.}
\label{fig6}
\end{figure}

We show in Fig.~\ref{fig6} the decomposition of the cross sections shown
  in Fig.~\ref{fig5a} for individual initial states. 
The left panel shows the results for $\sqrt{s}=$8 TeV, while 
   results for $\sqrt{s}=$14 TeV are shown in the right panel.
The cross section from the $q\overline{q'}$ initial state 
  dominates over those from the $qg$ and $gg$ initial states for
  the cut of $p_{T}^{j} > 10$ GeV when $m_{N} \gtrsim 200$ GeV, 
  while for $m_{N} \lesssim 200$ GeV the $qg$ initial state 
  dominates.
This is in contrast with the results for the $N\ell j$, where $q\overline{q^{'}}$
  initial state always dominates the cross section.
Although we may think that the gluon fusion channel could dominate,
  it is found to be always smaller than the other channels.
This is because the Bjorken scaling parameters for two gluons
  can not be small simultaneously in order to produce 
  the heavy neutrino.
On the other hand, we expect the gluon fusion channel cross section
  becomes larger as we lower the heavy neutrino mass. 
We can see in Fig.~\ref{fig6} the gluon fusion channel rises sharper
  than the others towards the low mass region.
We see that as the heavy neutrino mass becomes larger, the cross section from 
  $q\overline{q'}$ takes over those from the $qg$ and $gg$ because the 
  gluon PDF sharply drops towards the high Bjorken scaling parameter.

\subsection{$N\ell j$ {\rm \textbf{and}} $N\ell jj$ {\rm \textbf{production processes from photon mediated processes}}}

\begin{figure}
\begin{center}
\includegraphics[scale=0.6]{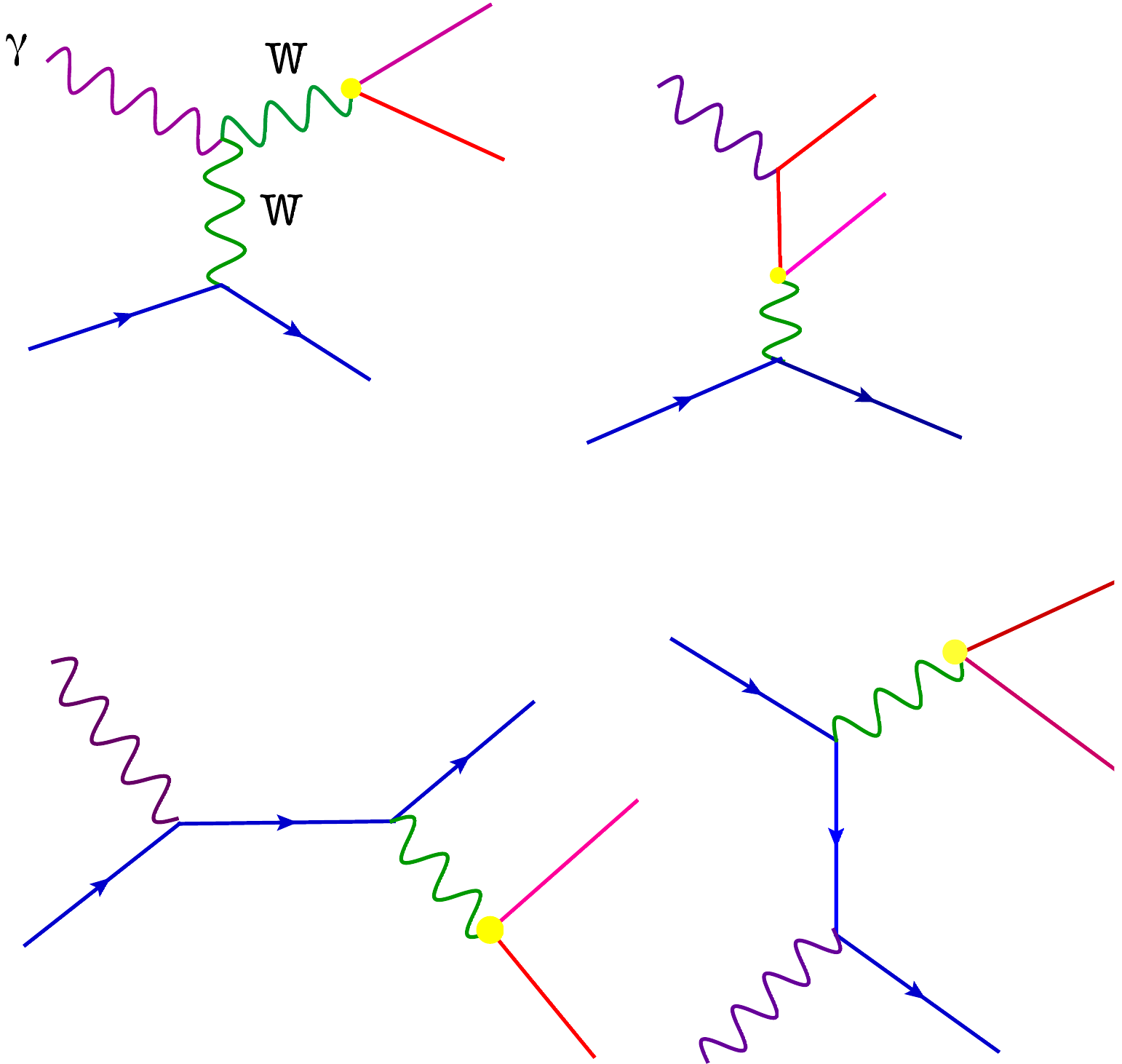}
\end{center}
\caption{The elastic or inelastic process for the $N\ell j$ final state.}
\label{Elasinelas}
\end{figure}
\begin{figure}
\begin{center}
\includegraphics[scale=1.0]{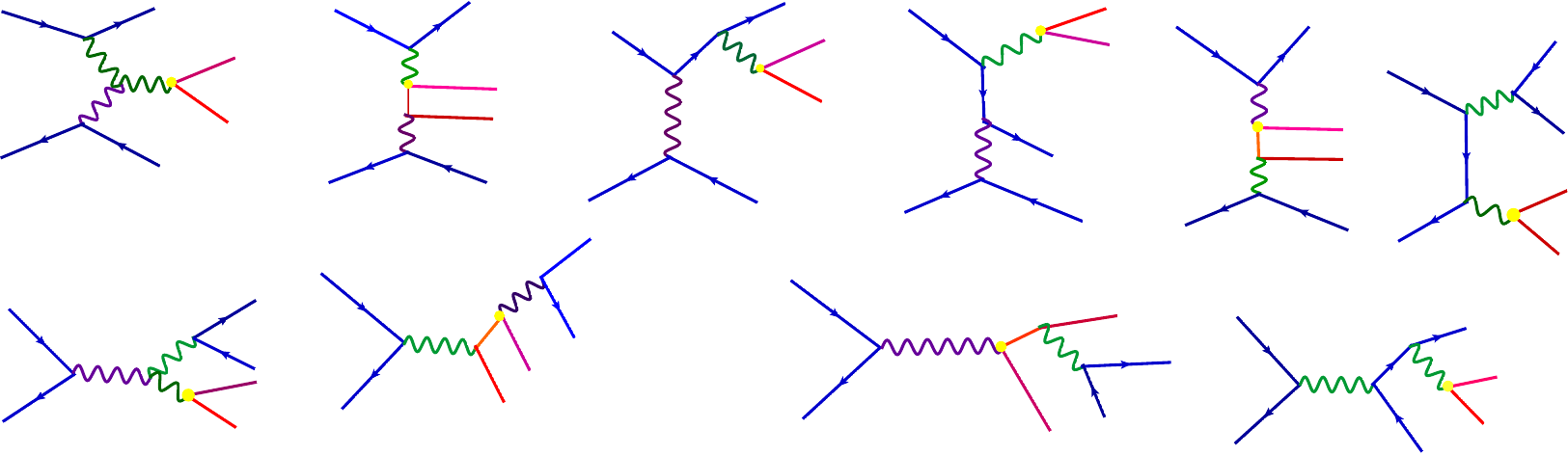}
\end{center}
\caption{The deep-inelastic processes for the $N\ell jj$ final state mediated by photon.}
\label{DIS}
\end{figure}
Apart from the QCD processes we also consider the heavy neutrino production at the LHC
   through the photon mediated processes. 
There are three types of photon mediated processes.
The first one is the elastic process where one photon is radiated from
  one proton and then scatters with a parton in the other proton.
The second one is the inelastic process where one photon is radiated from
  a parton inside one proton and scatters with a parton inside the other
  proton (for relevant Feynman diagrams, see Fig.~\ref{Elasinelas}).
The third one is the deep-inelastic scattering mediated by  photon with
  a large momentum transfer. (See Fig.~\ref{DIS} for relevant Feynman diagrams.)
Analysis of the deep inelastic process has been performed in \cite{p-photon}. 
All of these processes have been analyzed in \cite{ruiz}.

\begin{figure}
\begin{center}
\includegraphics[scale=0.6]{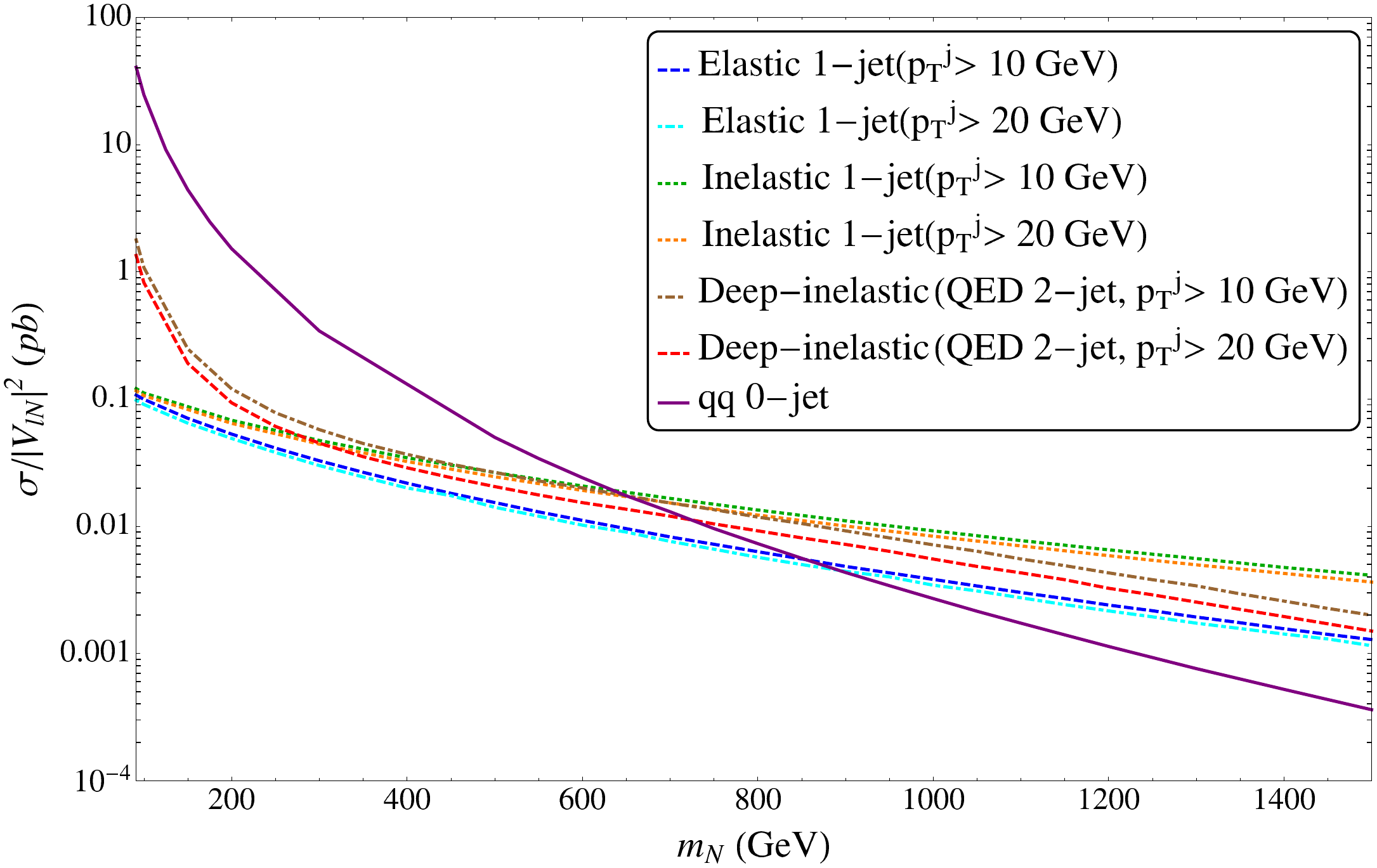}
\end{center}
\caption{Figure shows the production cross section of the heavy neutrino as a function of $m_{N}$ and normalized by the square of the mixing angle for the different photon mediated processes at the 14 TeV LHC. The solid line stands for the $N\ell$ final state for the $q\overline{q^{'}}$ initial state. The elastic process for different $p_T^{j}$ values are shown by the dashed lines. The inelastic processes are represented by the dotted lines. The deep inelastic process (QED 2-jet) is depicted by the dot-dashed lines.}
\label{fig7}
\end{figure}

We calculate the heavy neutrino production cross sections
 through the three photon mediated processes.
For the deep-inelastic processes we switch off the QCD vertices in {\tt MadGraph}.
Our results are shown in Fig.~\ref{fig7} for $\sqrt{s}= 14$ TeV LHC. 
Here we have imposed two different values of the $p_{T}^{j}$ cuts,
  $p_{T}^{j} > 10$ GeV and $20$ GeV.  
When $m_{N} \lesssim 400$ $(300)$ GeV, the deep-inelastic process dominates
  over the other photon mediated processes for $p_{T}^{j} > 10$ $(20)$ GeV and
otherwise, the inelastic process dominates.
The elastic process is always small compared to the others.
The analytic expressions of the radiated photon PDFs used in our calculations are given in \cite{ph11}-\cite{ph6}.

\section{Kinematic distributions of the heavy neutrino production }
Here we investigate the kinematic distributions of the leptons and
  jets associated with the heavy neutrino production. 
The different production processes show typical kinematic distributions.
To see this we generate the events for $m_{N}= 100$ GeV and $500$ GeV
  at the $\sqrt{s}=$ 14 TeV LHC.
We consider the transverse momenta $\left(p_{T}\right)$ and
  pseudo-rapidities $\left(\eta\right)$ for the leptons, jets
  and the heavy neutrino as kinematic variables.
We also consider the differential cross sections for the
  final states of $N\ell$, $N\ell j$ and $N\ell jj$ with 
  the cut $p_{T}^{j} >20$ GeV for the process/es with jet(s).

\begin{figure}
\begin{center}
\includegraphics[scale=0.58]{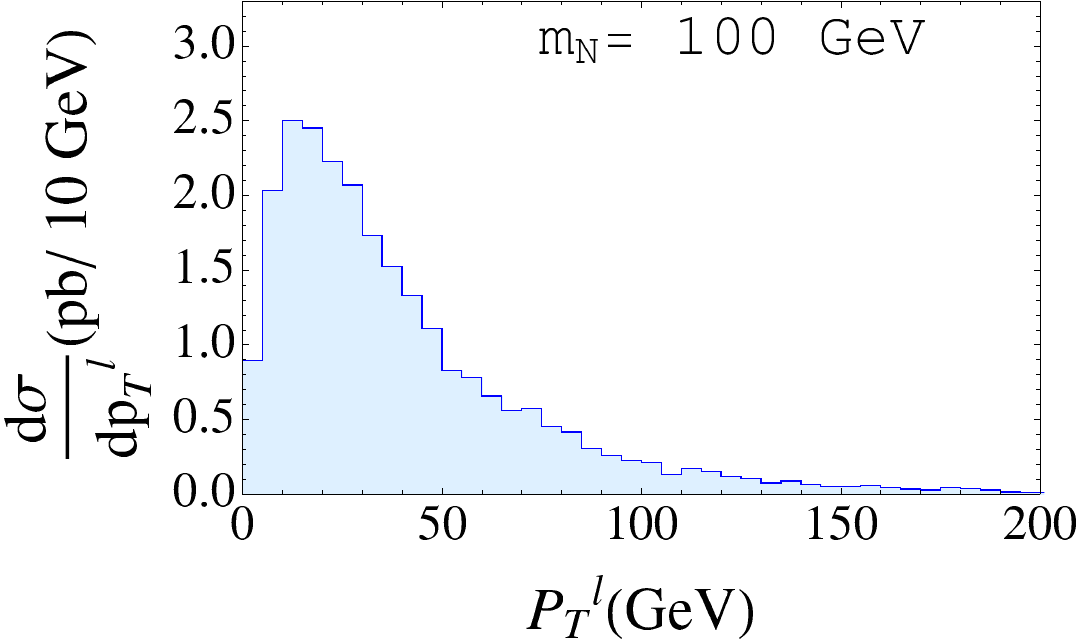}
\includegraphics[scale=0.55]{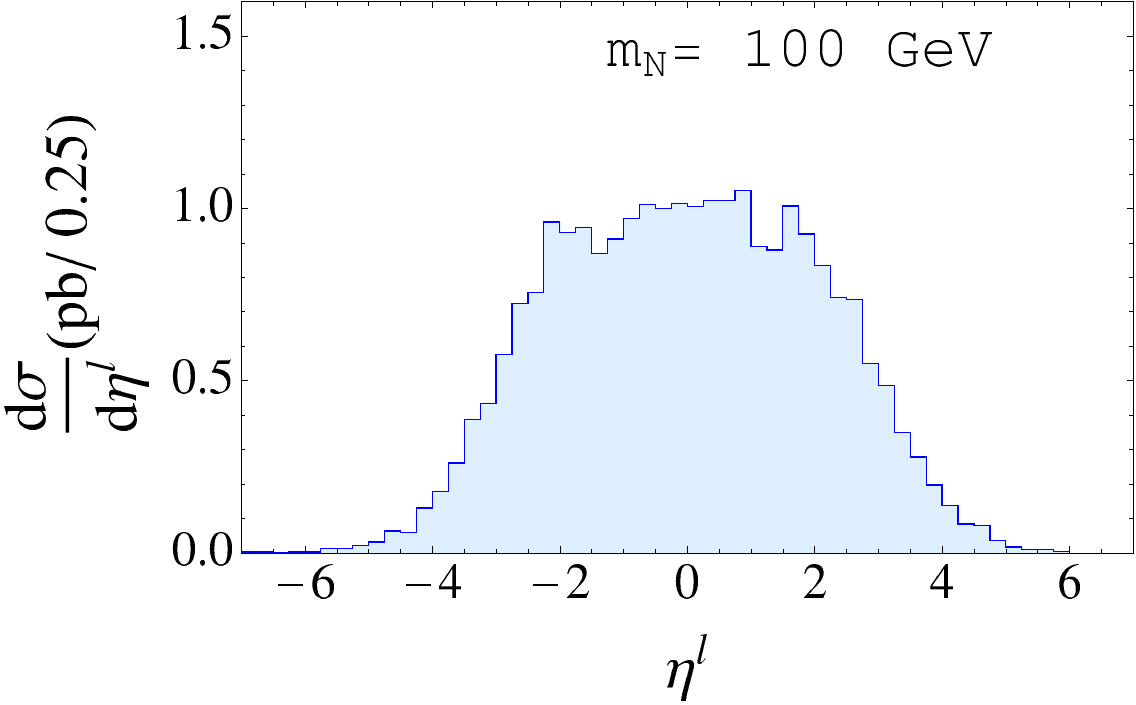}\\
\includegraphics[scale=0.5]{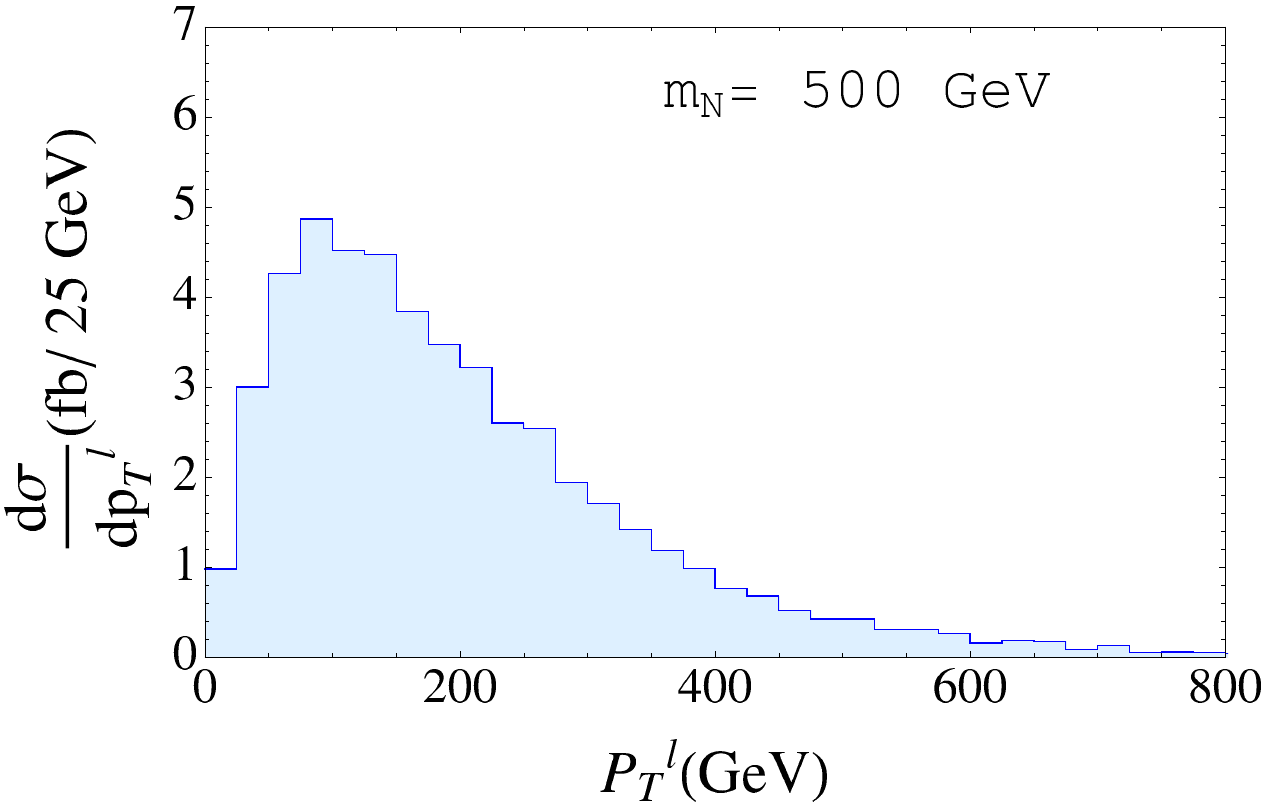}
\includegraphics[scale=0.57]{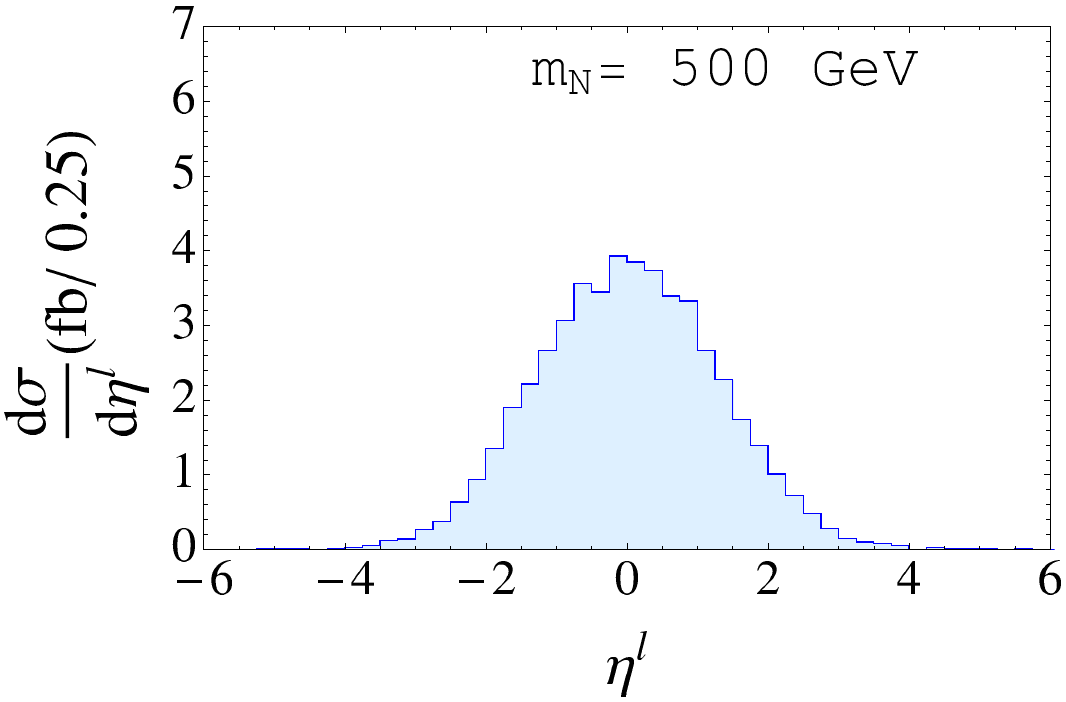}\\
\includegraphics[scale=0.51]{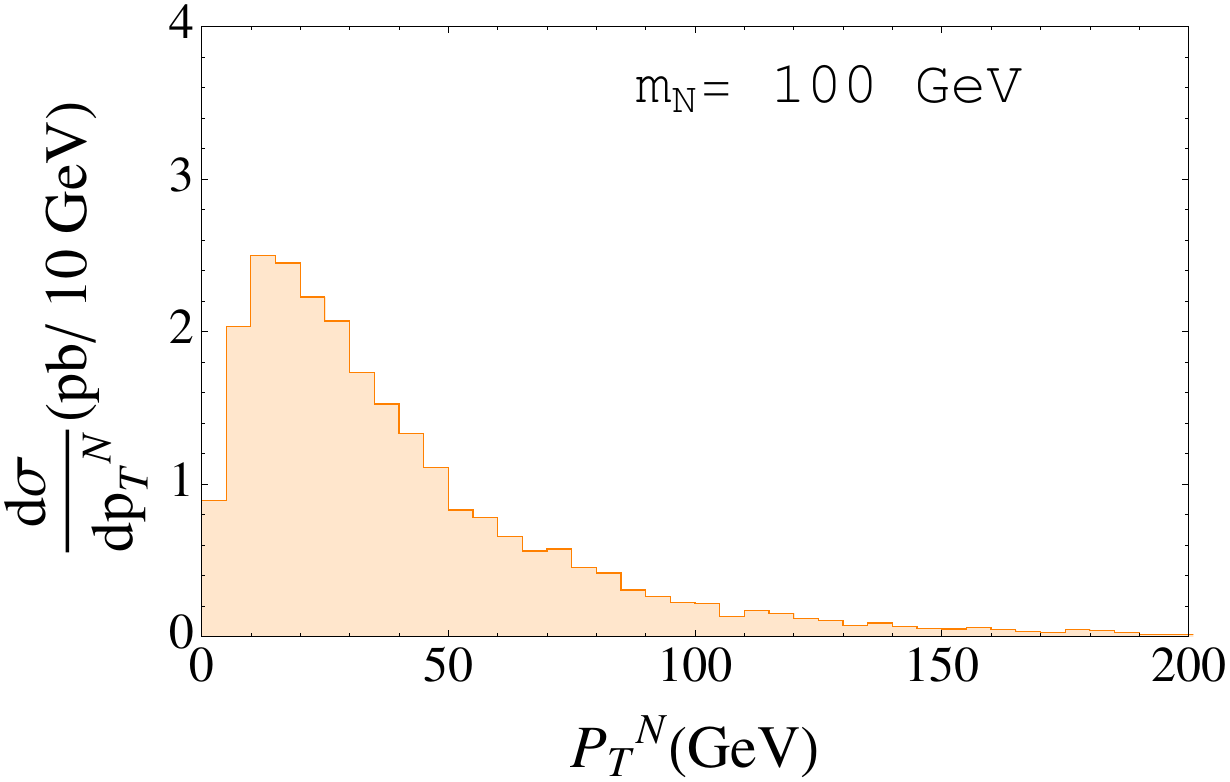}
\includegraphics[scale=0.46]{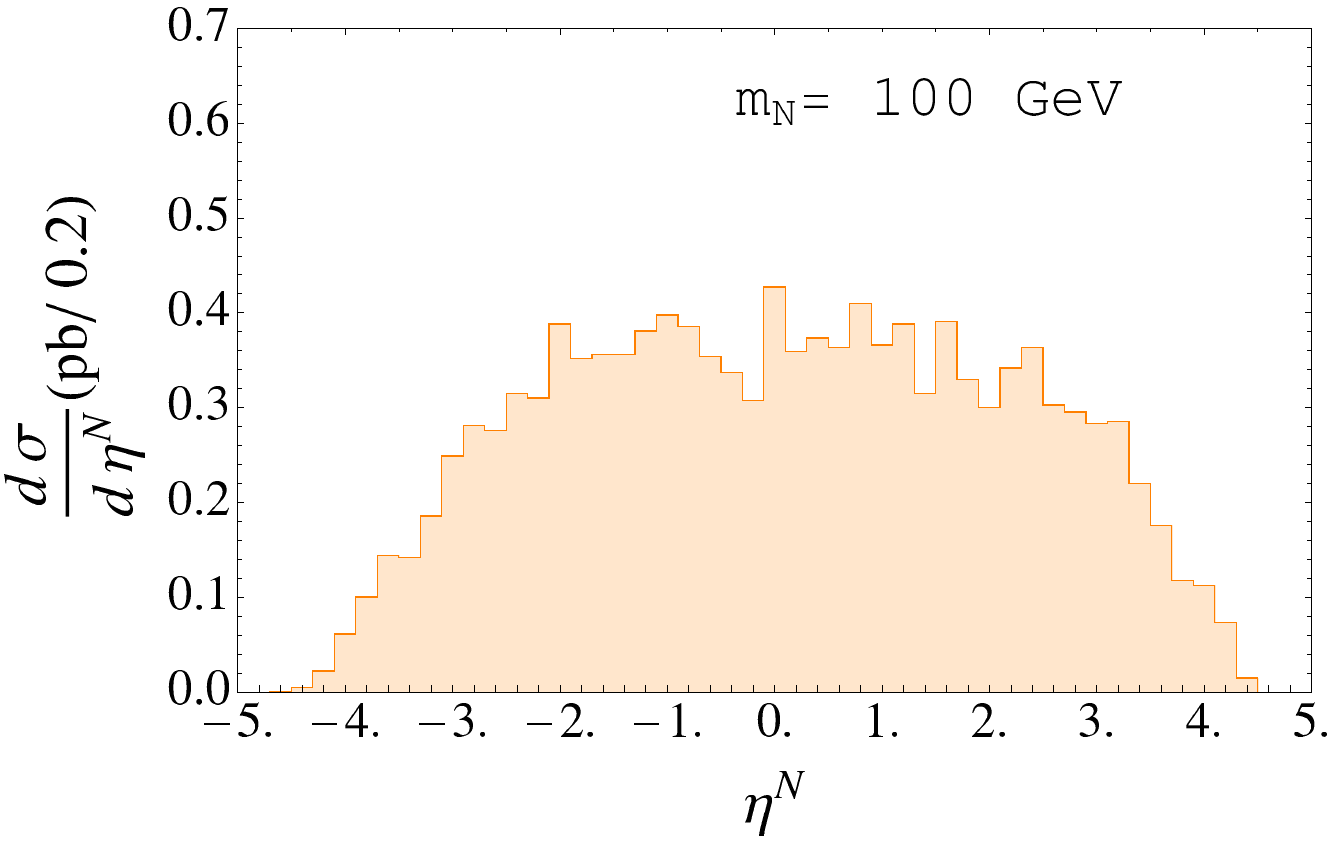}\\
\includegraphics[scale=0.55]{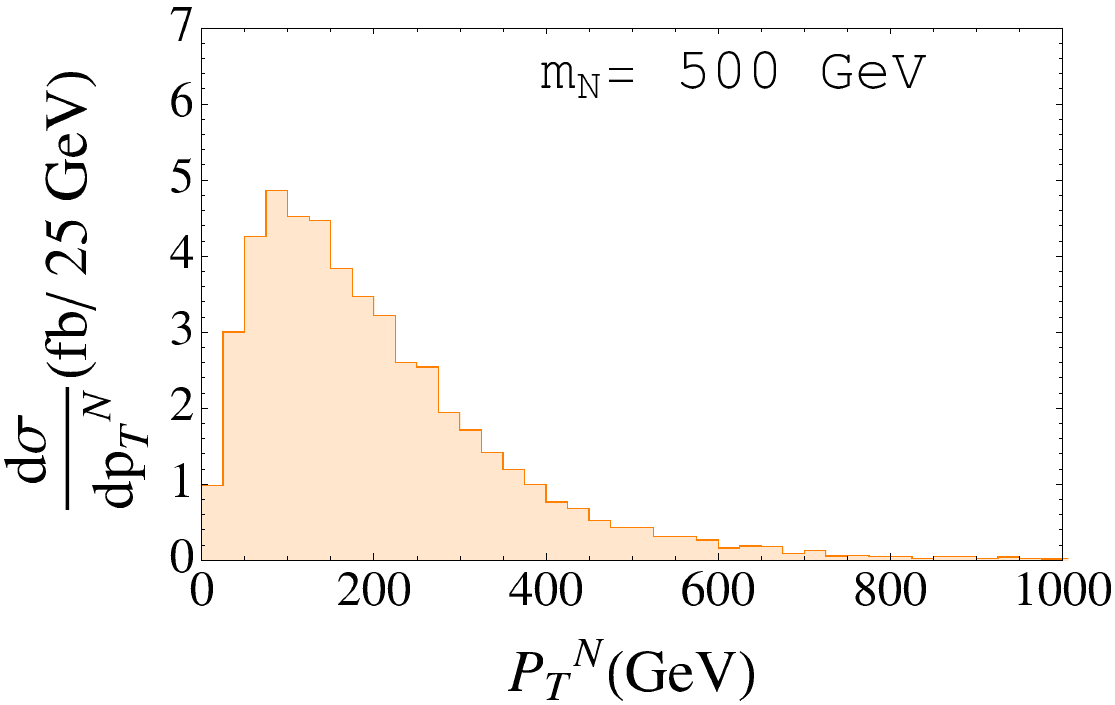}
\includegraphics[scale=0.35]{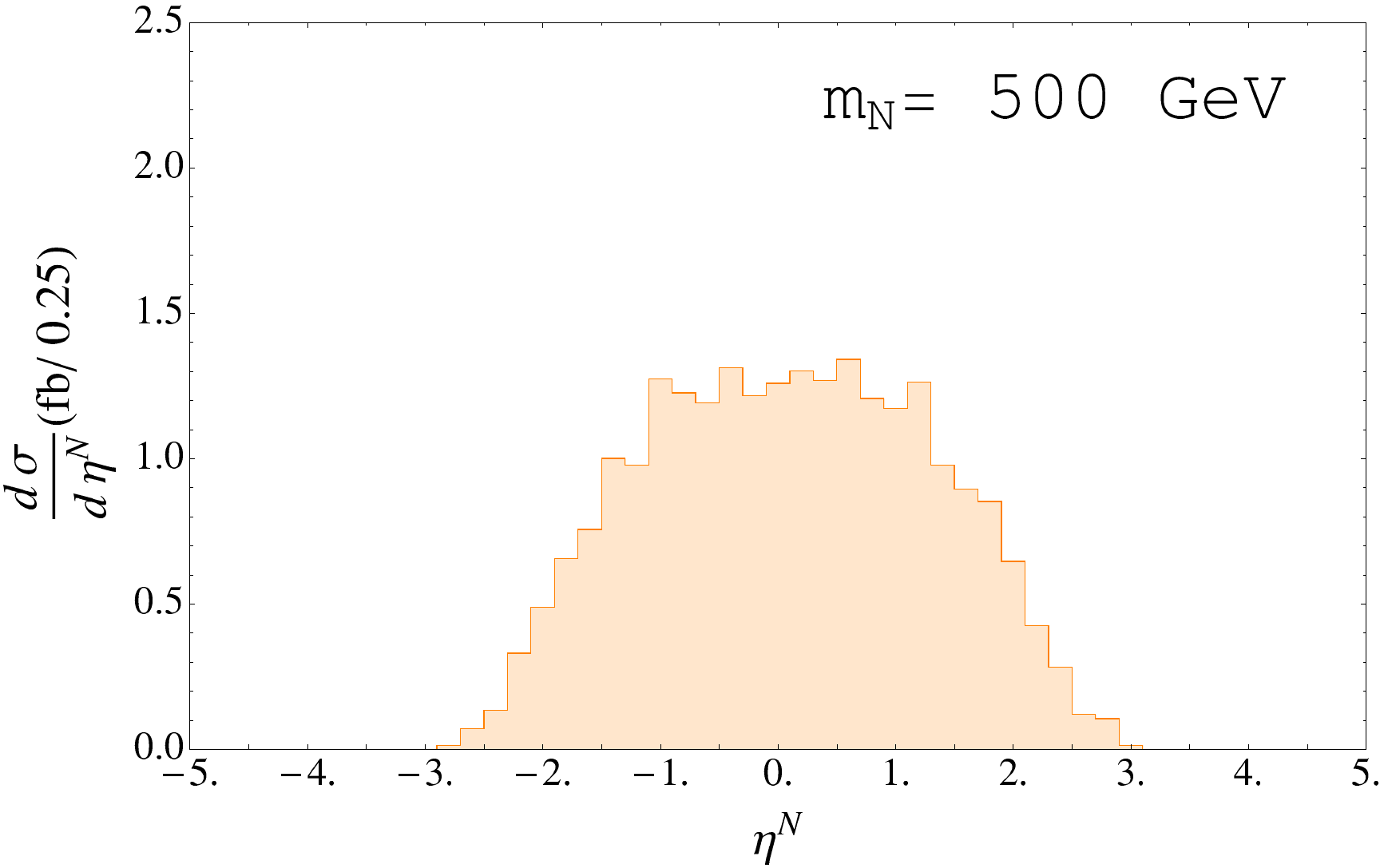}\\
\includegraphics[scale=0.5]{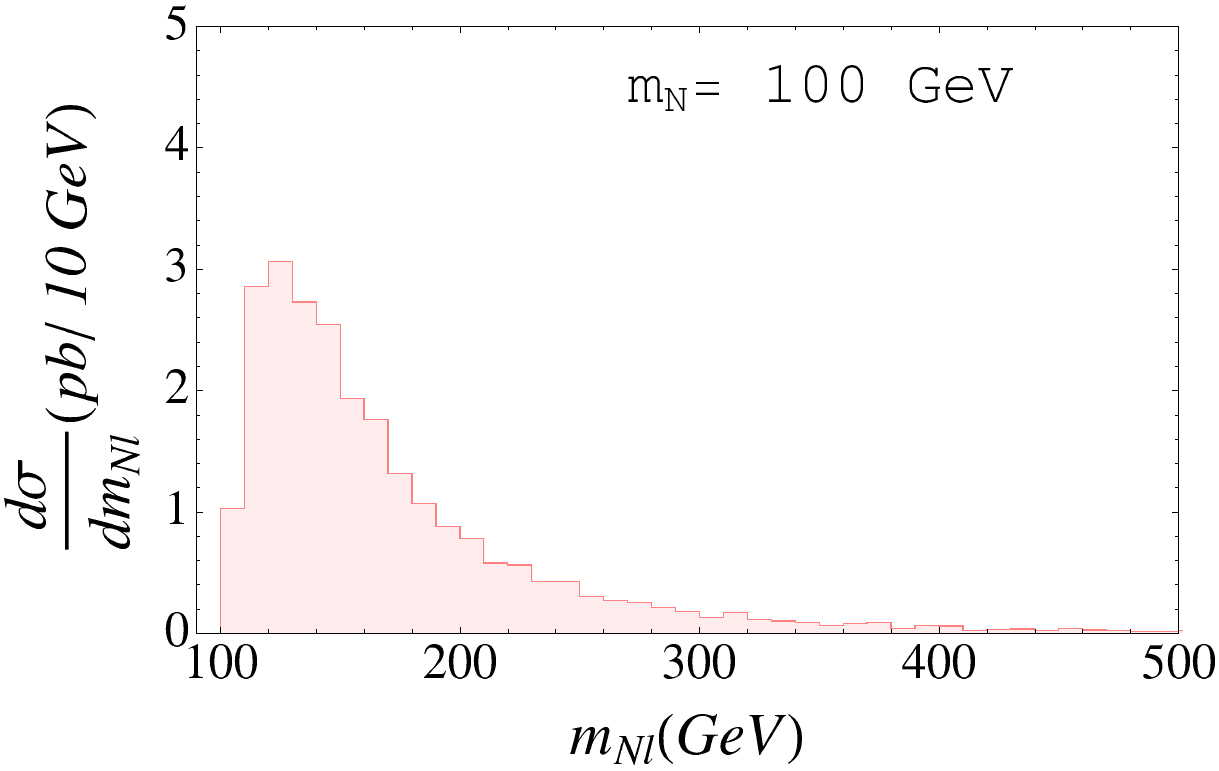}
\includegraphics[scale=0.53]{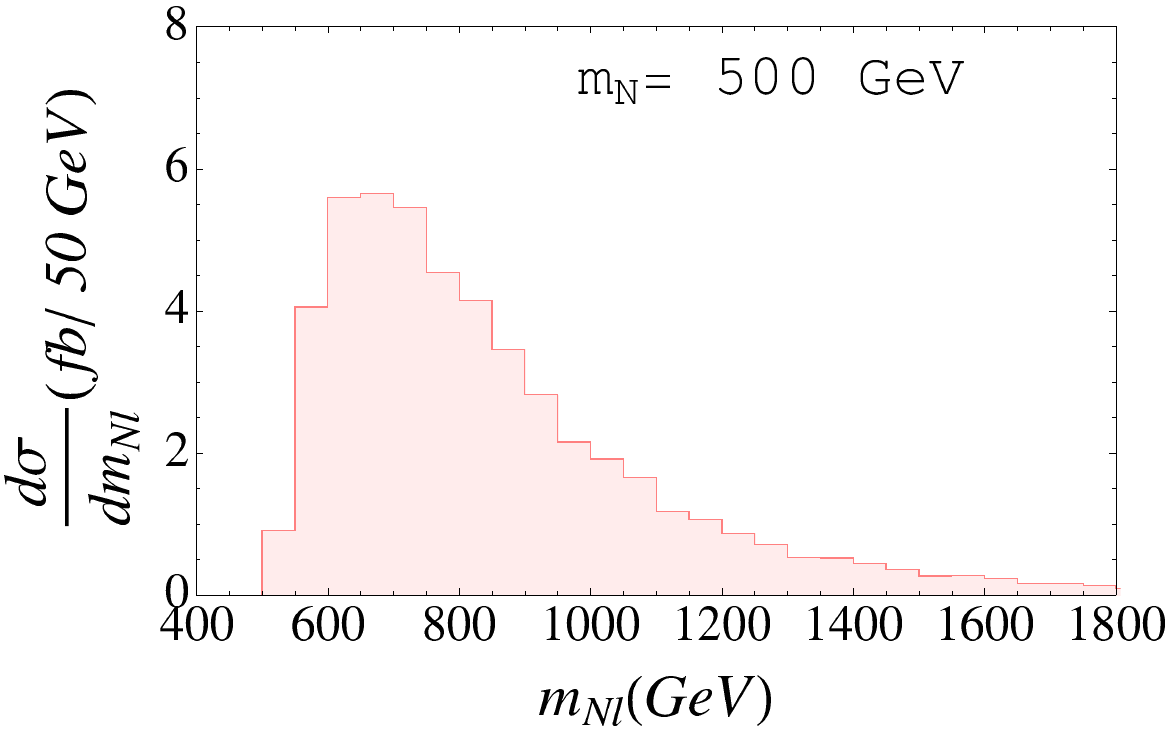}
\end{center}
\caption{The first and second rows show the $p_{T}$ and $\eta$ distributions of the lepton produced in the $N\ell$ final state from the $q\overline{q^{'}}$ 0-jet processes. The third and fourth rows show the same for the heavy neutrino. The fifth row shows the final state invariant mass $\left(m_{N\ell}\right)$ distributions. The left column stands for $m_{N}=100$ GeV whereas the right one is for $m_{N}=500$ GeV.}
\label{pp0j}
\end{figure}

\begin{figure}
\begin{center}
\includegraphics[scale=0.58]{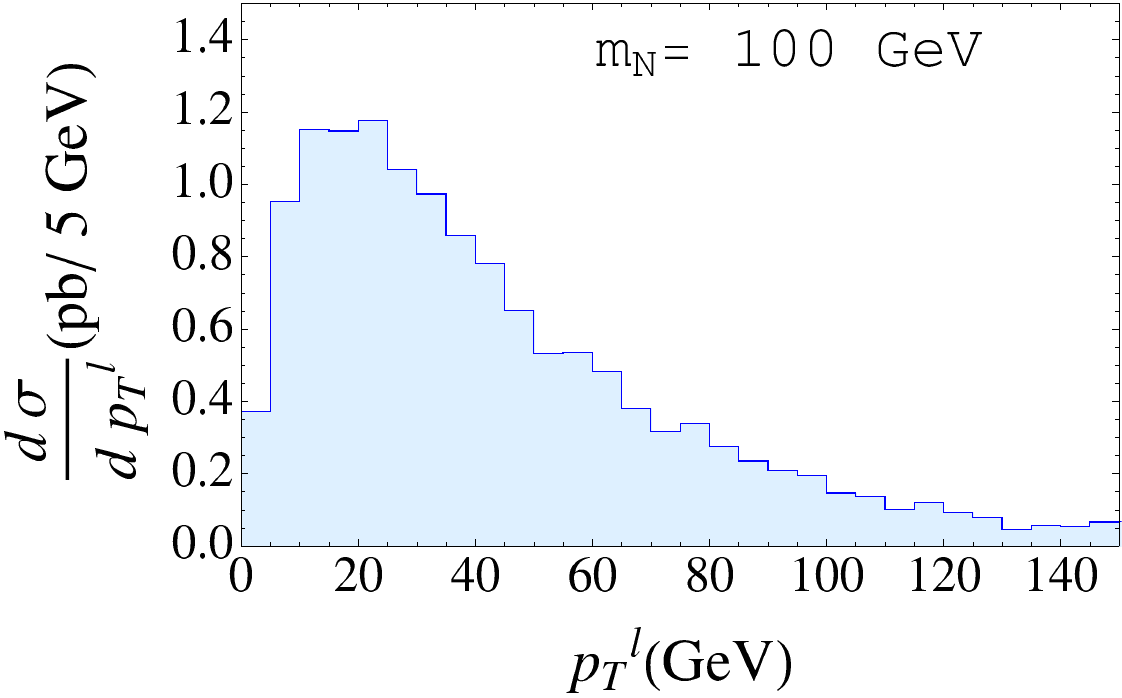}
\includegraphics[scale=0.58]{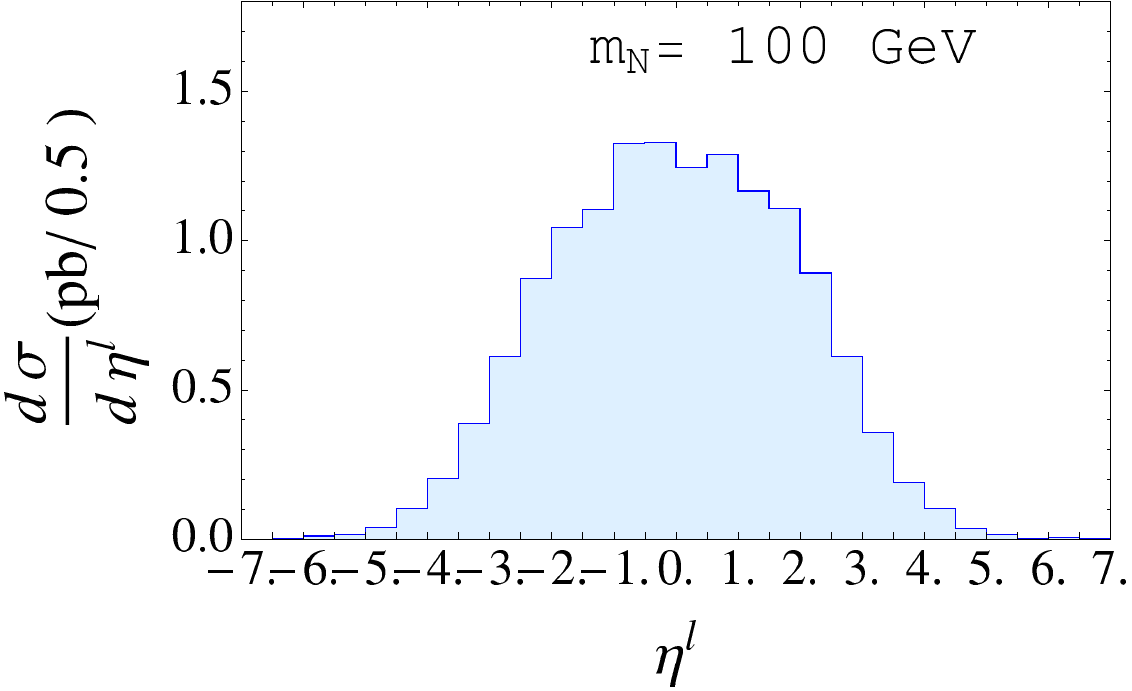}\\
\includegraphics[scale=0.58]{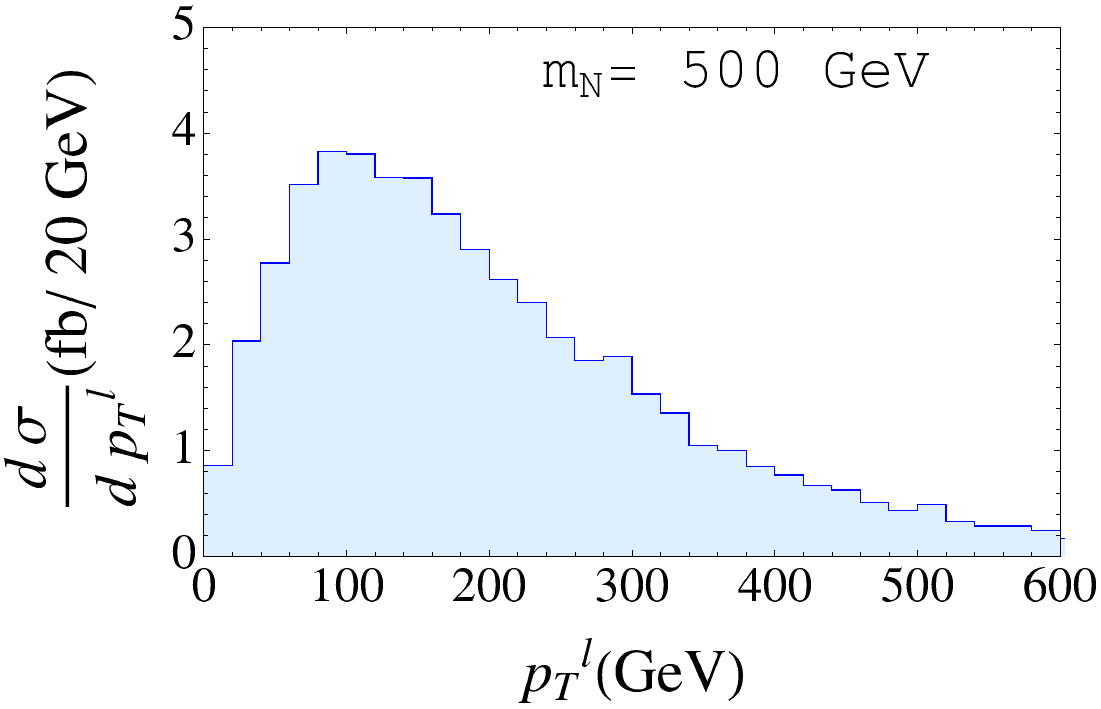}
\includegraphics[scale=0.58]{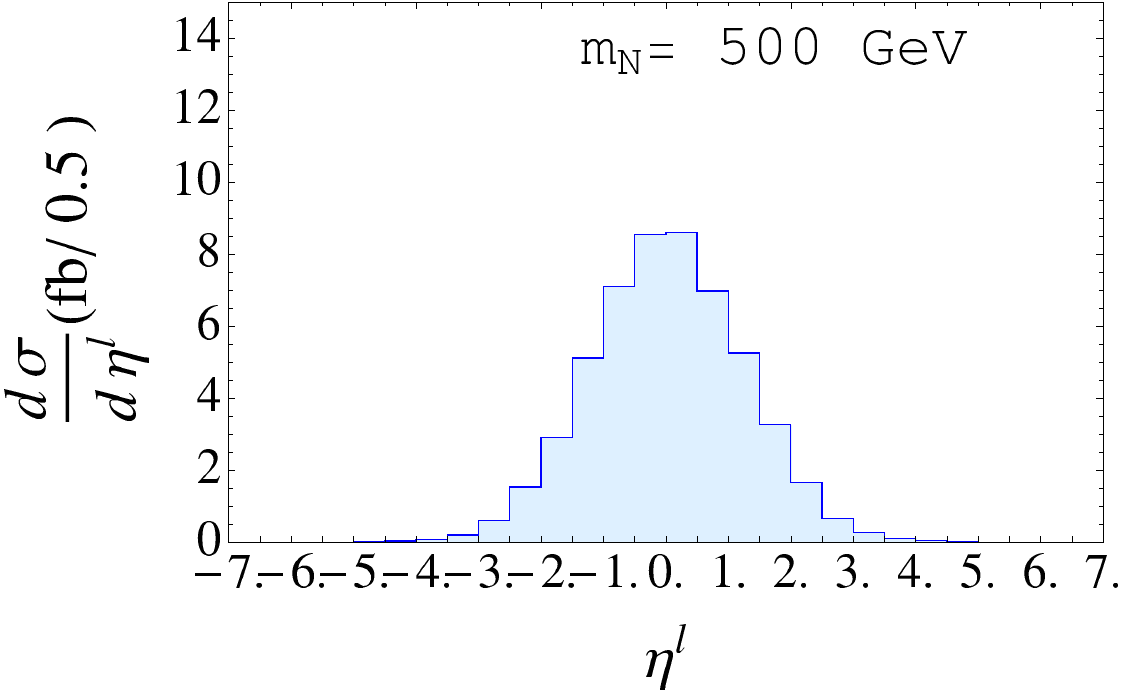}\\
\includegraphics[scale=0.5]{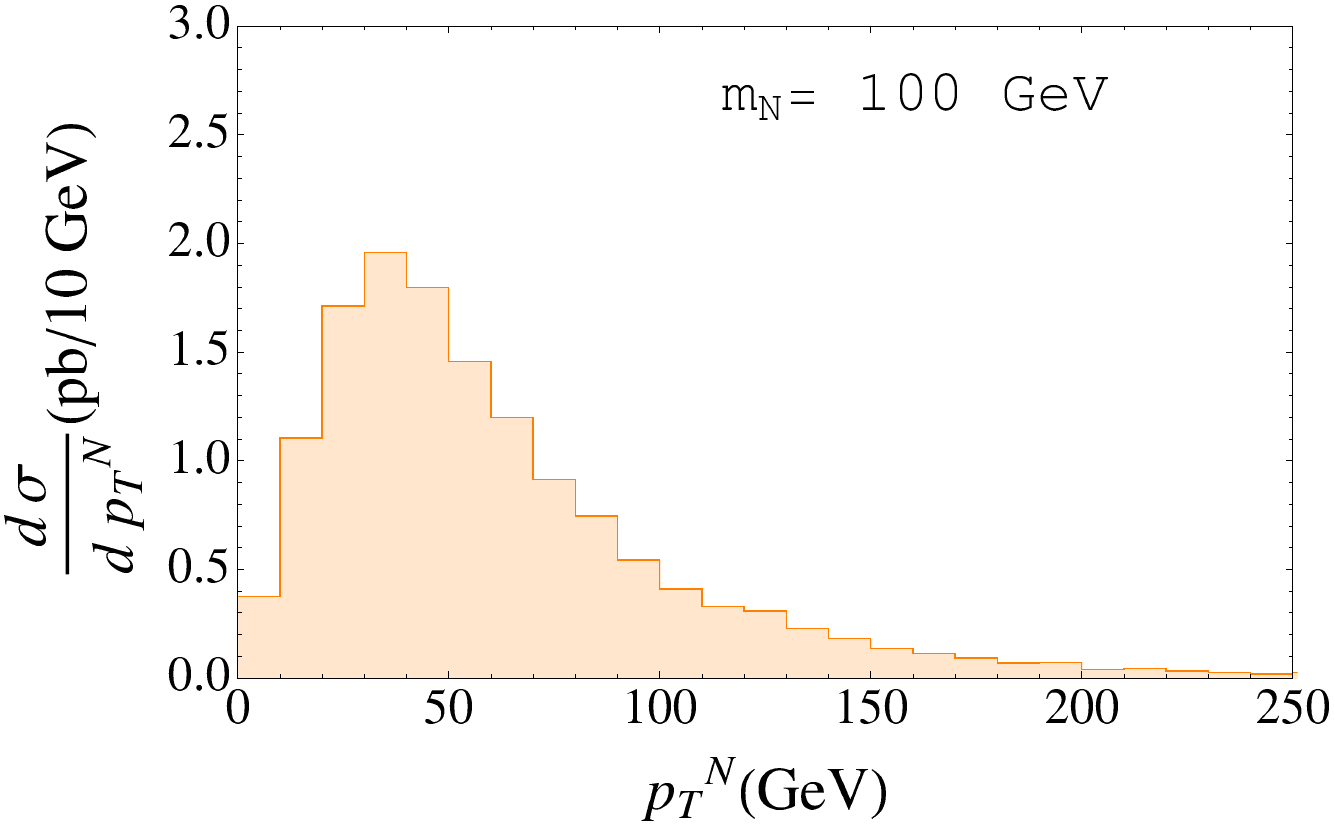}
\includegraphics[scale=0.55]{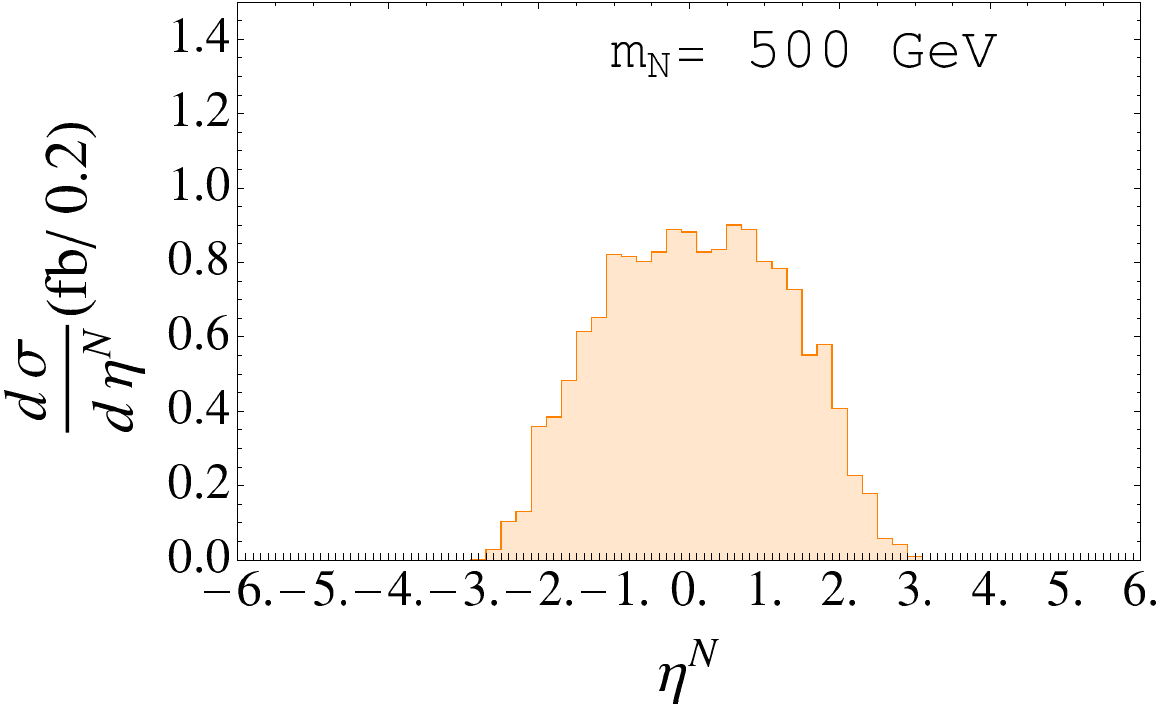}\\
\includegraphics[scale=0.55]{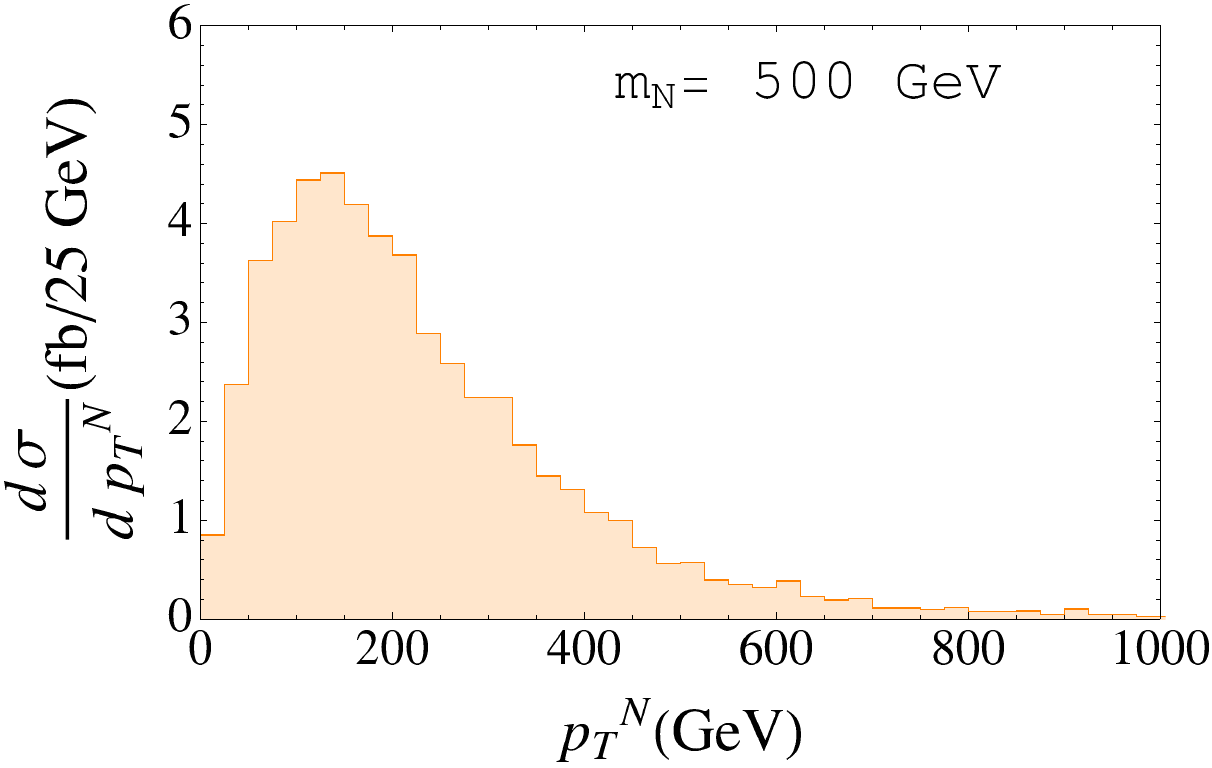}
\includegraphics[scale=0.55]{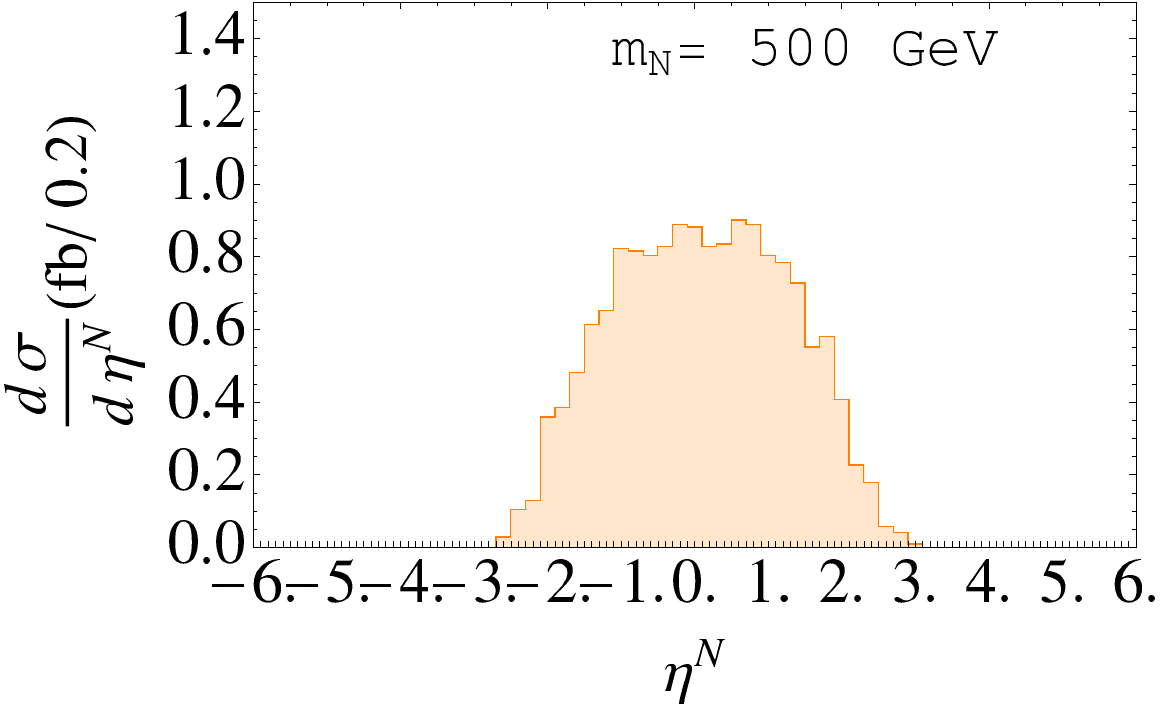}
\end{center}
\caption{The first and second rows show the $p_{T}$ and $\eta$ distributions of the lepton produced in the $N\ell j$ final state from the $q\overline{q^{'}}$ and $qg$ 1-jet processes. The third and fourth rows show the same for the heavy neutrino. The left column stands for $m_{N}=100$ GeV whereas the right one is for $m_{N}=500$ GeV.}
\label{pp1j1}
\end{figure}

\begin{figure}
\begin{center}
\includegraphics[scale=0.6]{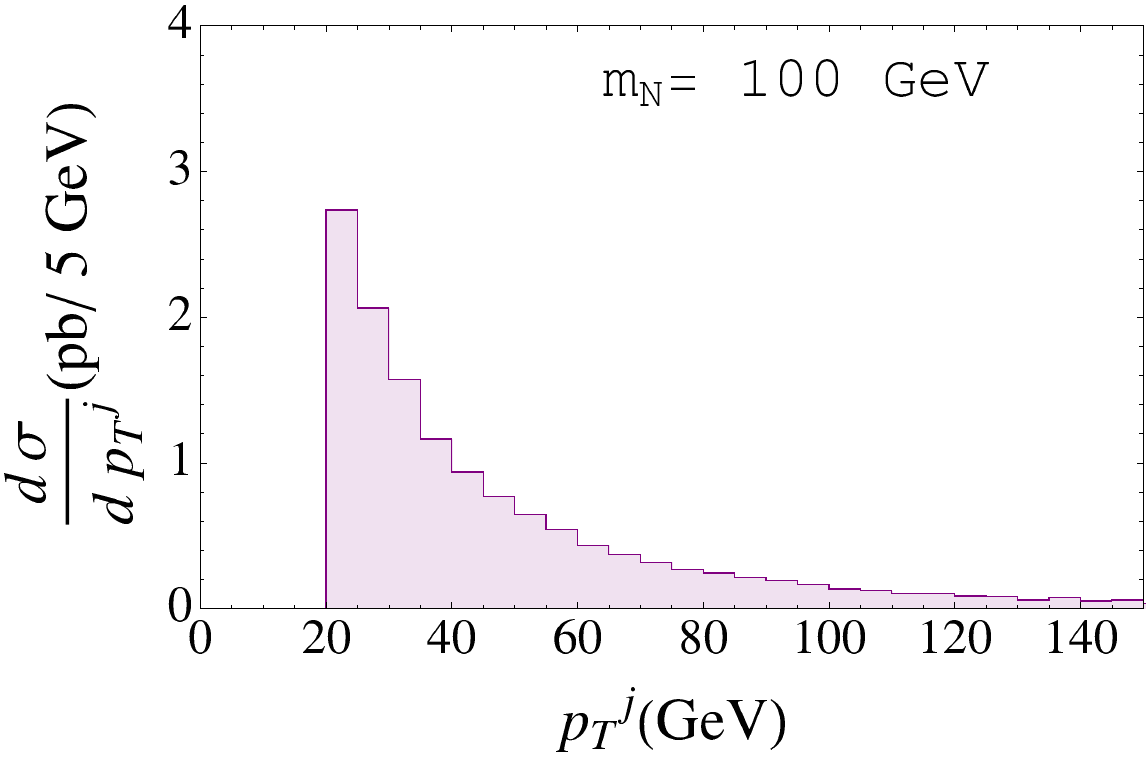}
\includegraphics[scale=0.56]{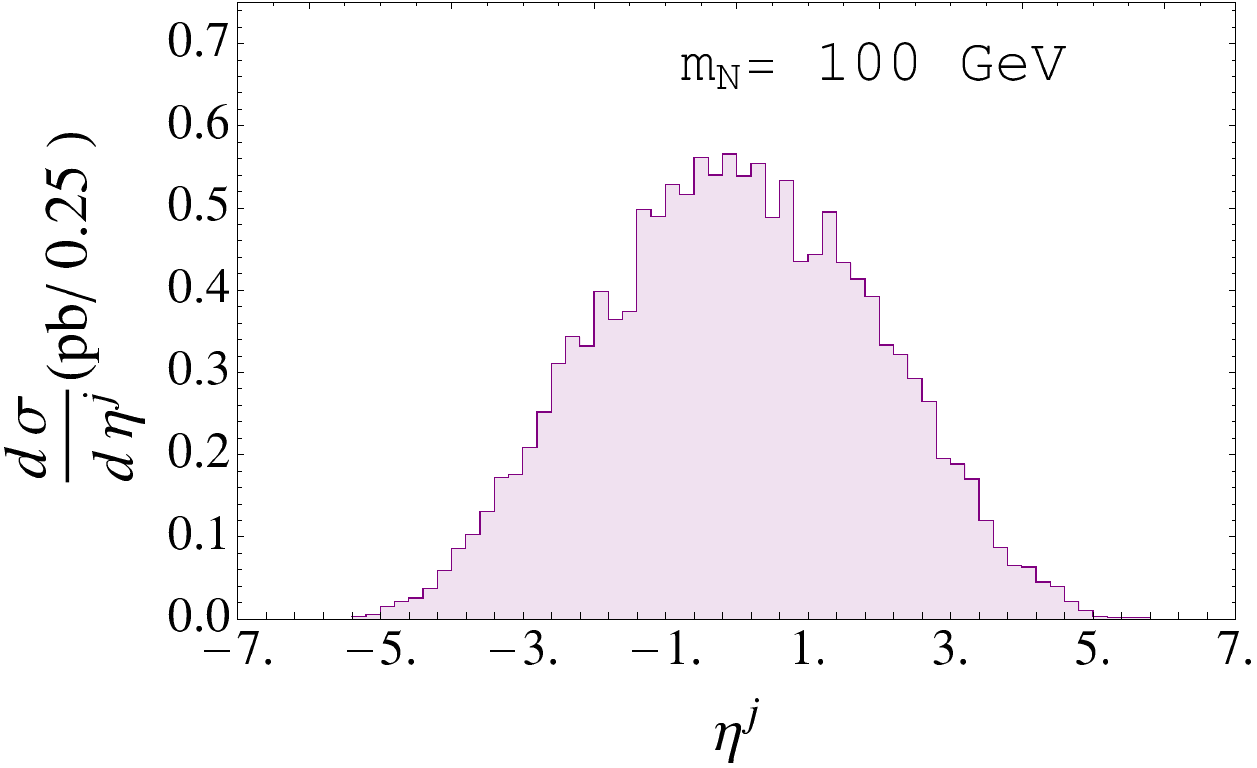}\\
\includegraphics[scale=0.62]{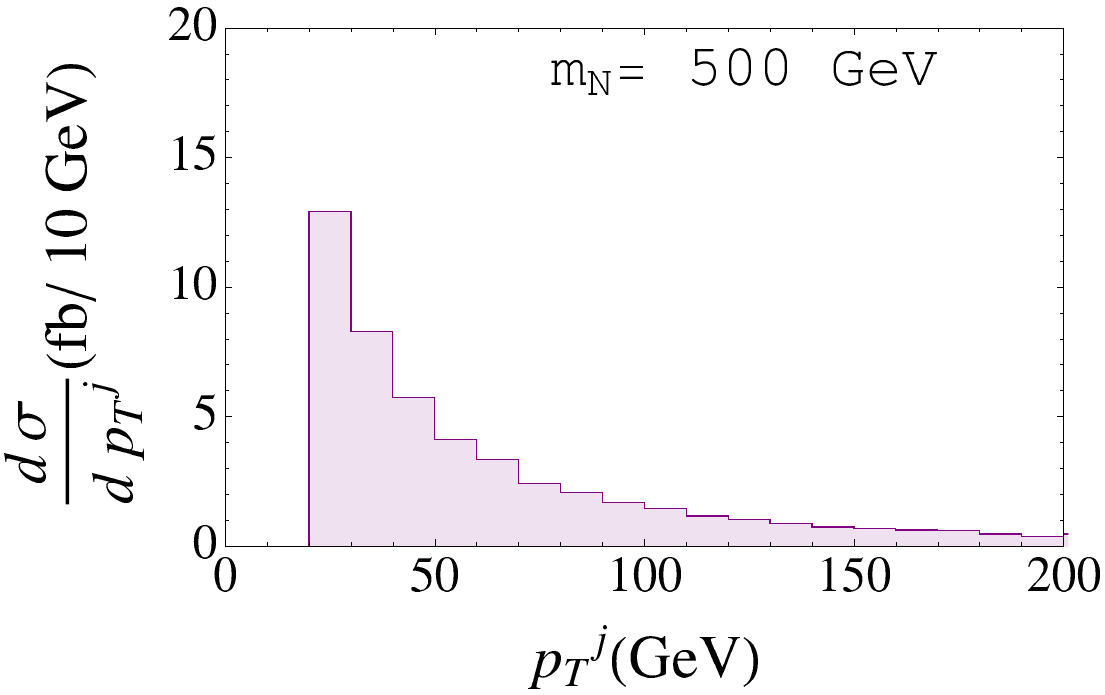}
\includegraphics[scale=0.6]{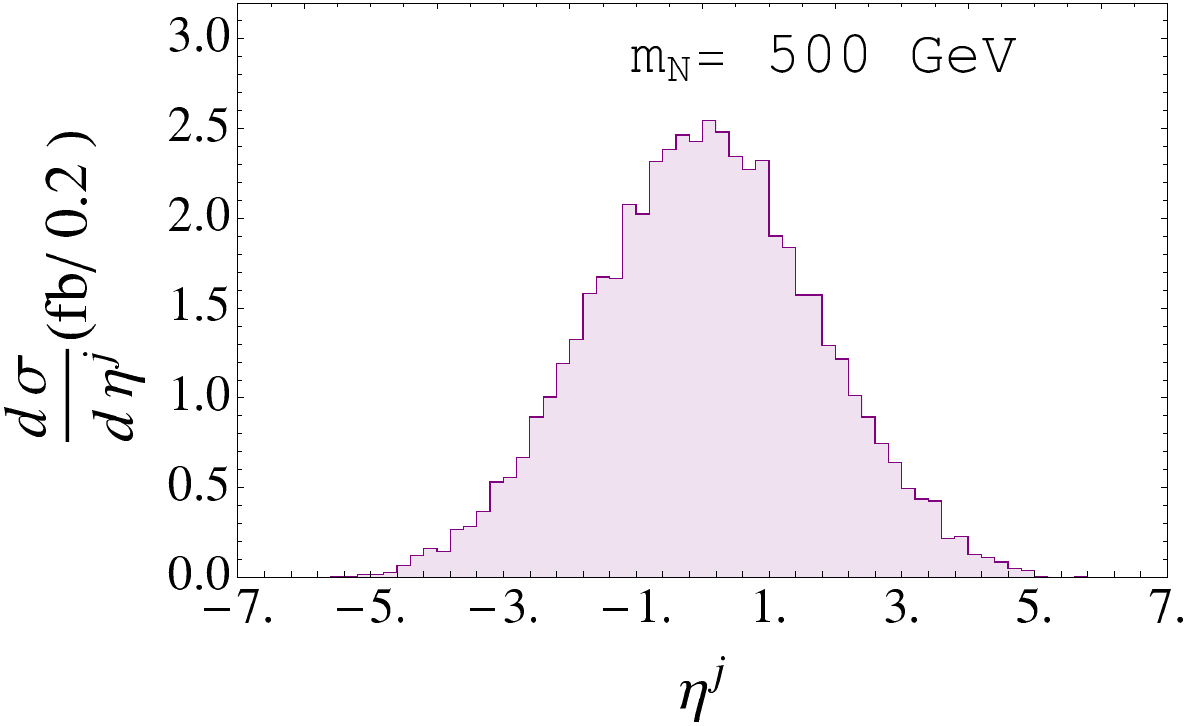}\\
\includegraphics[scale=0.6]{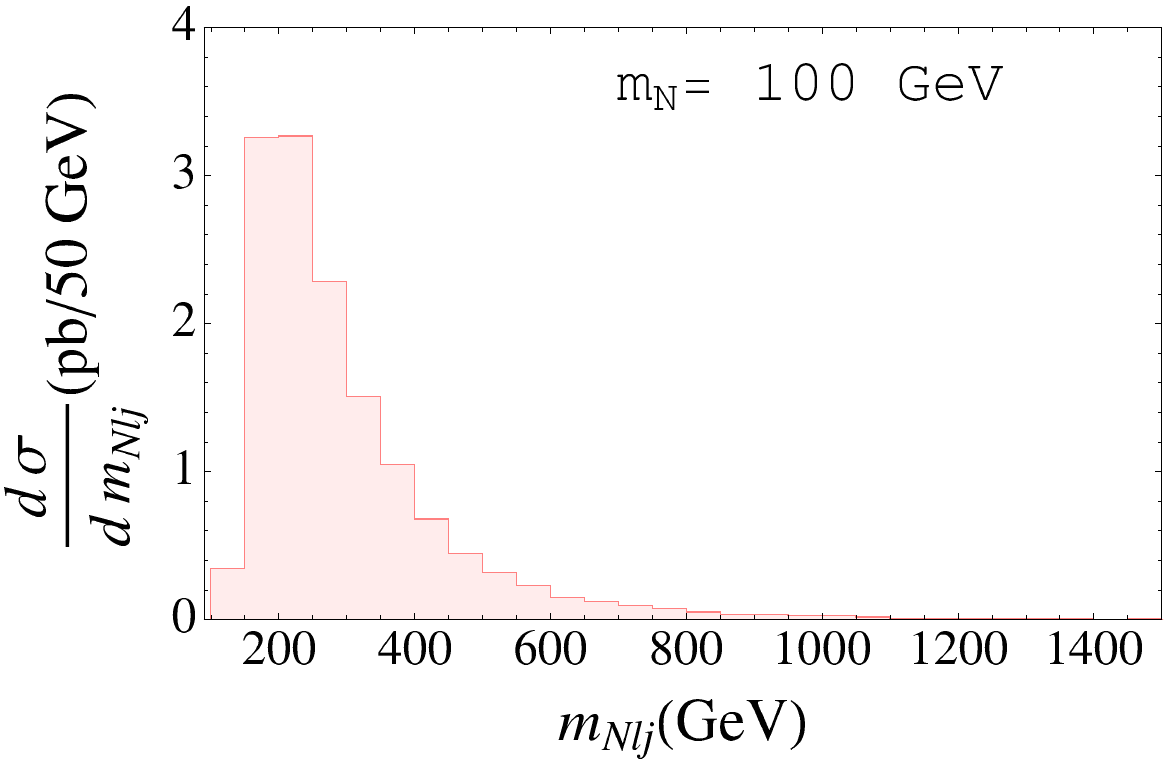}
\includegraphics[scale=0.6]{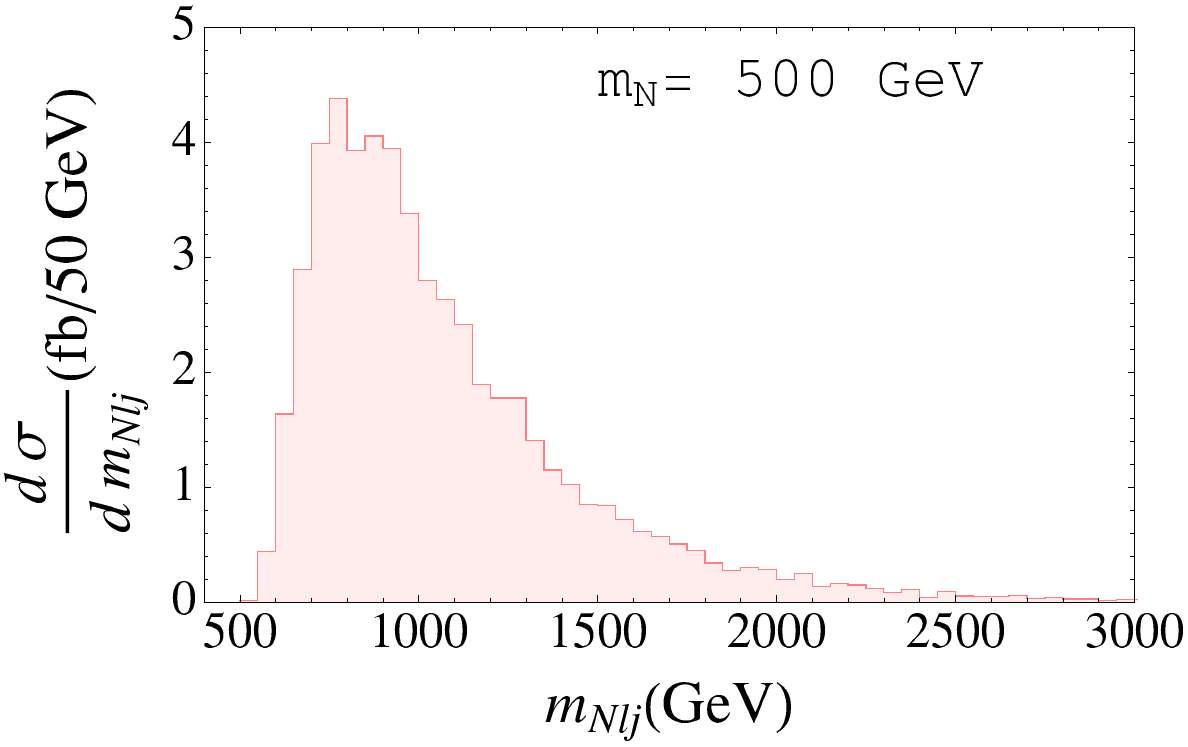}
\end{center}
\caption{The first and second rows show the $p_{T}$ and $\eta$ distributions of the jet produced in the $N\ell j$ final state from the $q\overline{q^{'}}$ and $qg$ 1-jet processes. The third row shows the final state invariant mass $\left(m_{N\ell j}\right)$ distributions. The left column stands for $m_{N}=100$ GeV whereas the right one is for $m_{N}=500$ GeV.}
\label{pp1j2}
\end{figure}

\begin{figure}
\begin{center}
\includegraphics[scale=0.57]{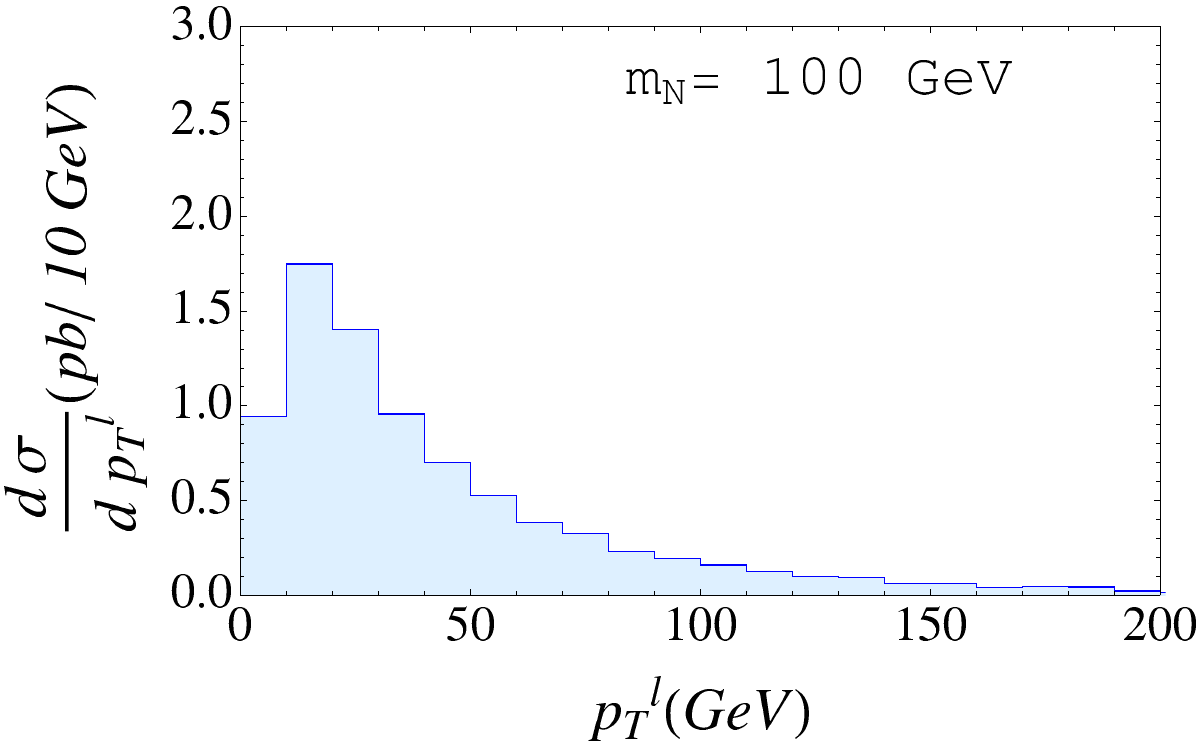}
\includegraphics[scale=0.57]{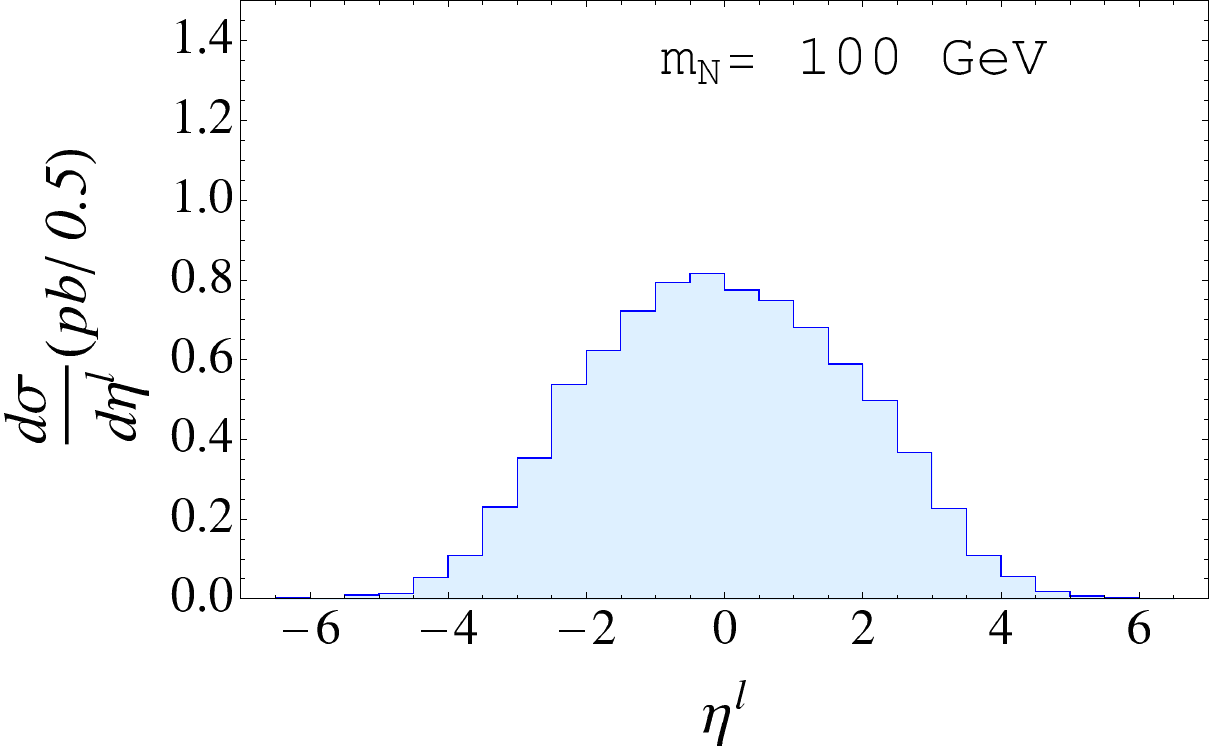}\\
\includegraphics[scale=0.67]{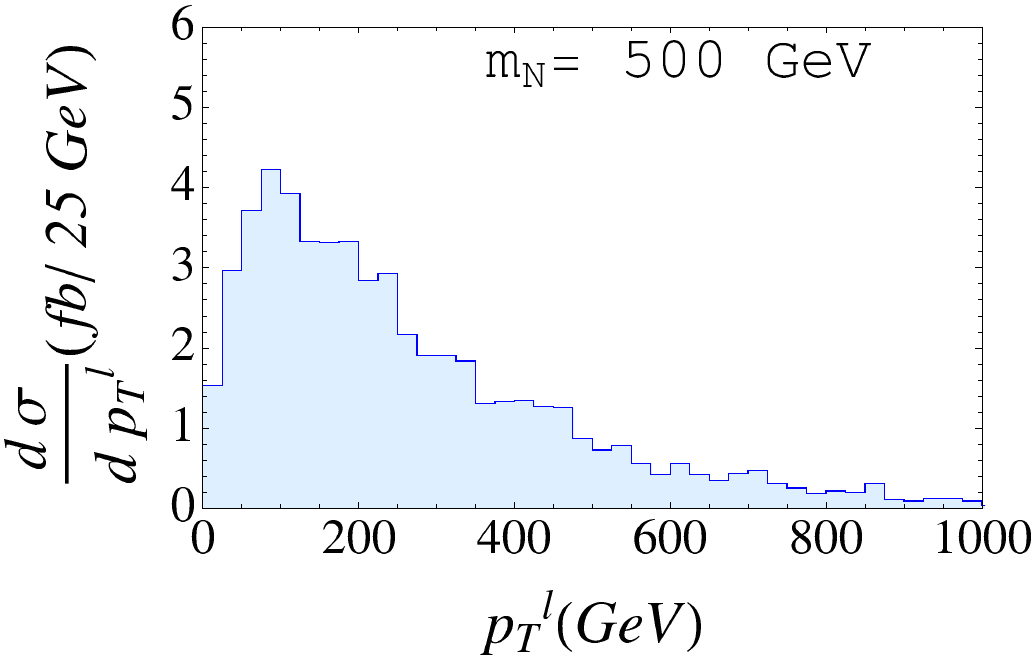}
\includegraphics[scale=0.49]{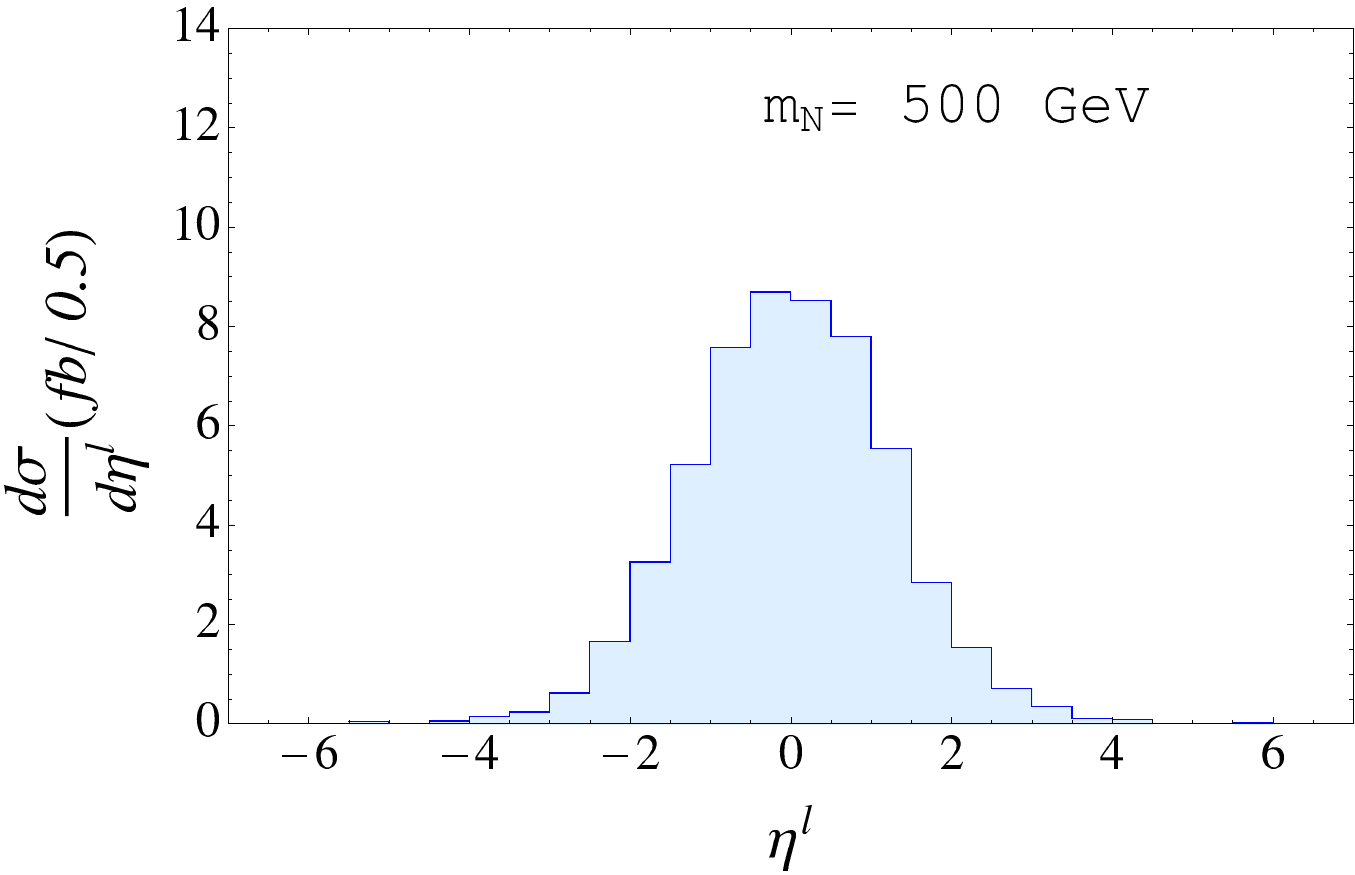}\\
\includegraphics[scale=0.55]{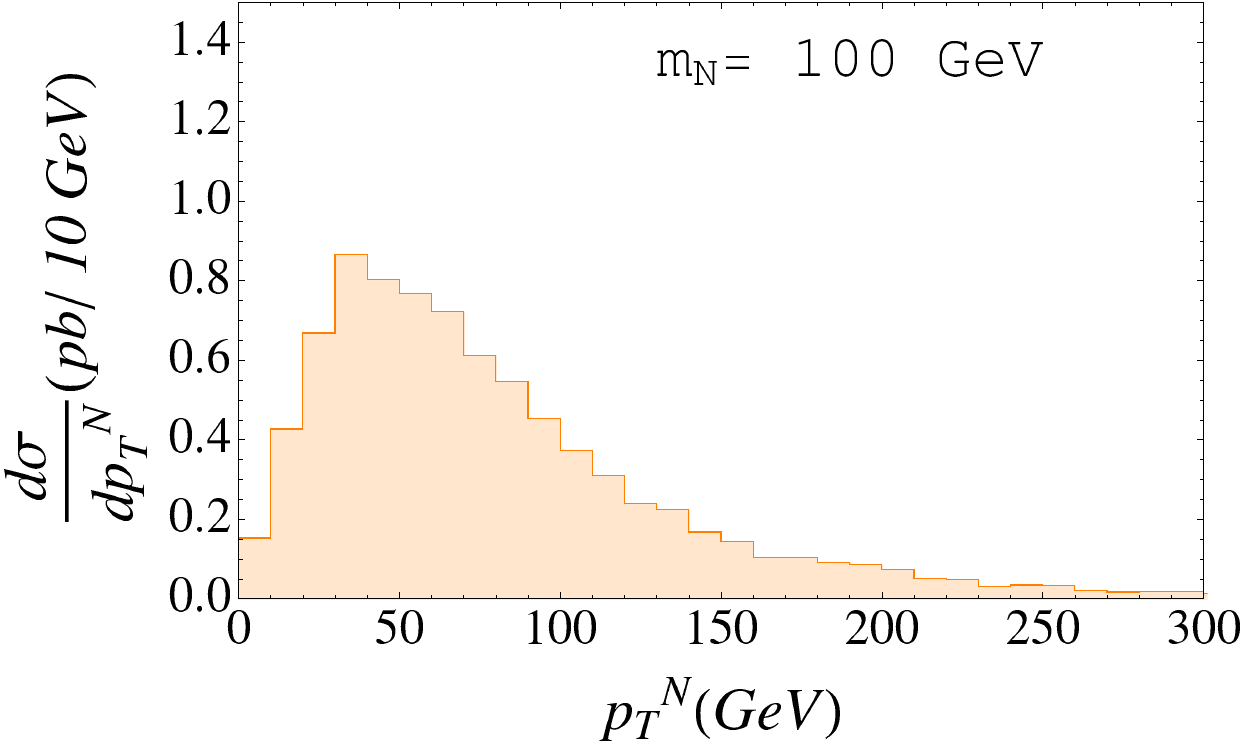}
\includegraphics[scale=0.51]{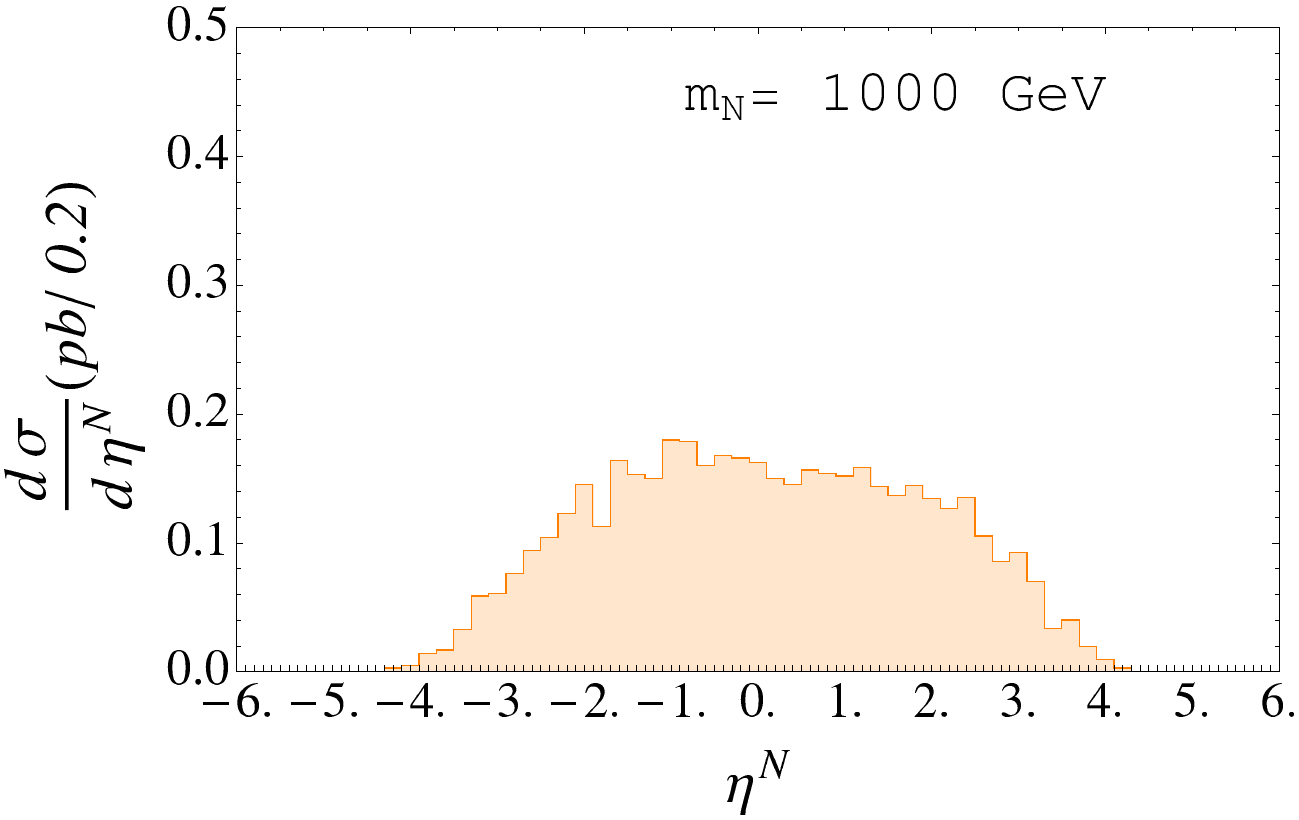}\\
\includegraphics[scale=0.55]{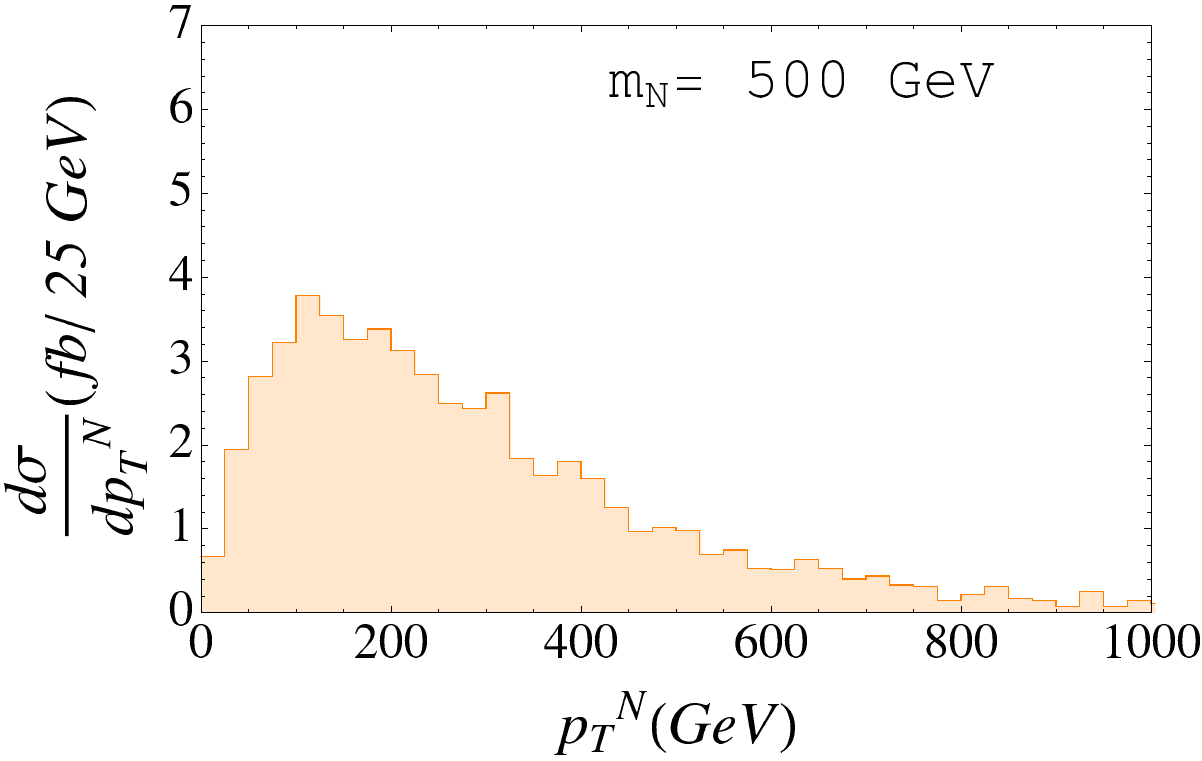}
\includegraphics[scale=0.63]{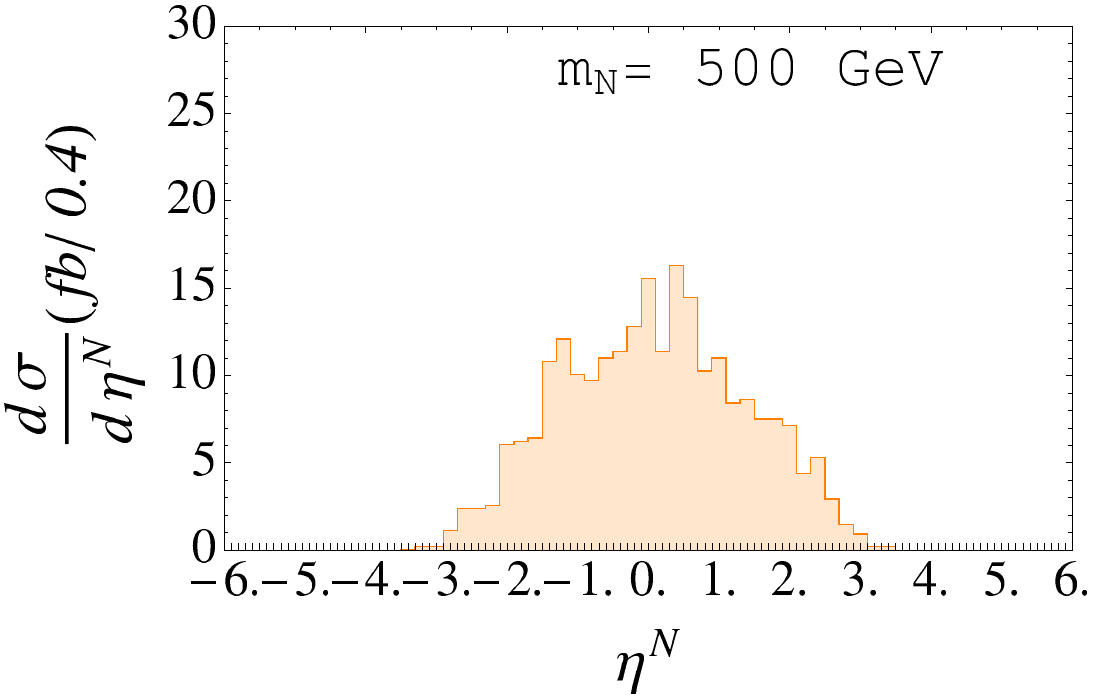}
\end{center}
\caption{The first and second rows show the $p_{T}$ and $\eta$ distributions of the lepton produced in the $N\ell jj$ final state from the $q\overline{q^{'}}$, $qg$ and $gg$ initial states. The third and fourth rows show the same for the heavy neutrino. The left column stands for $m_{N}=100$ GeV whereas the right one is for $m_{N}=500$ GeV.}
\label{pp2j1}
\end{figure}

\begin{figure}
\begin{center}
\includegraphics[scale=0.59]{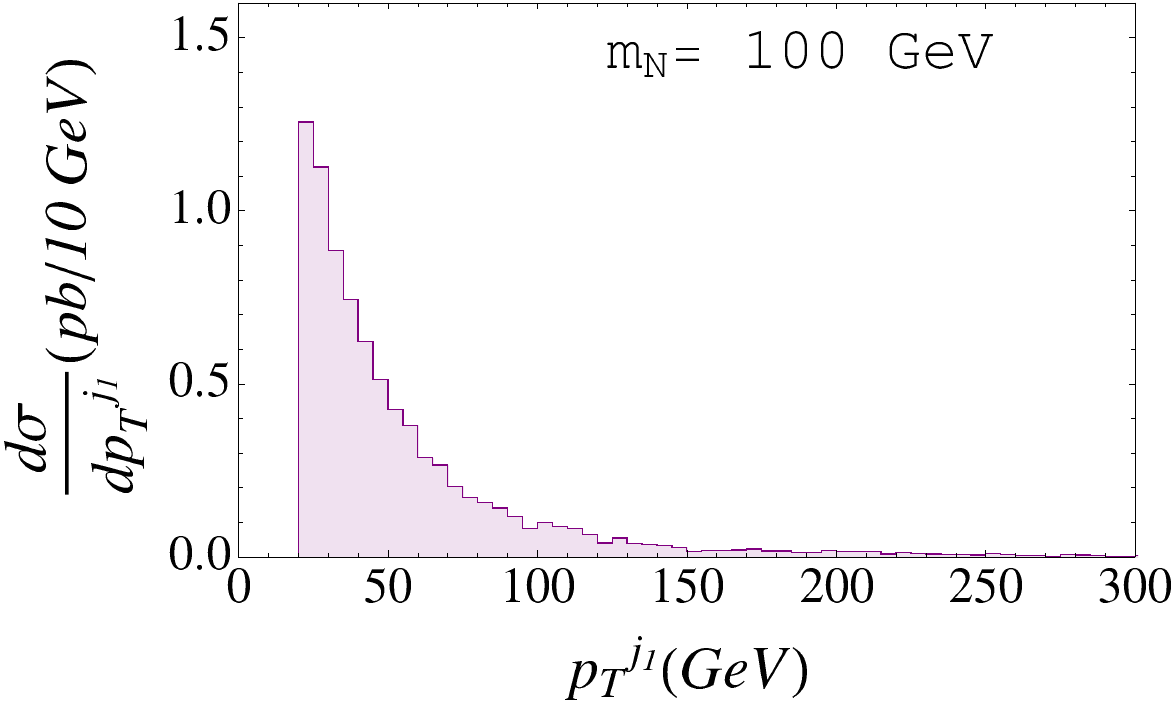}
\includegraphics[scale=0.57]{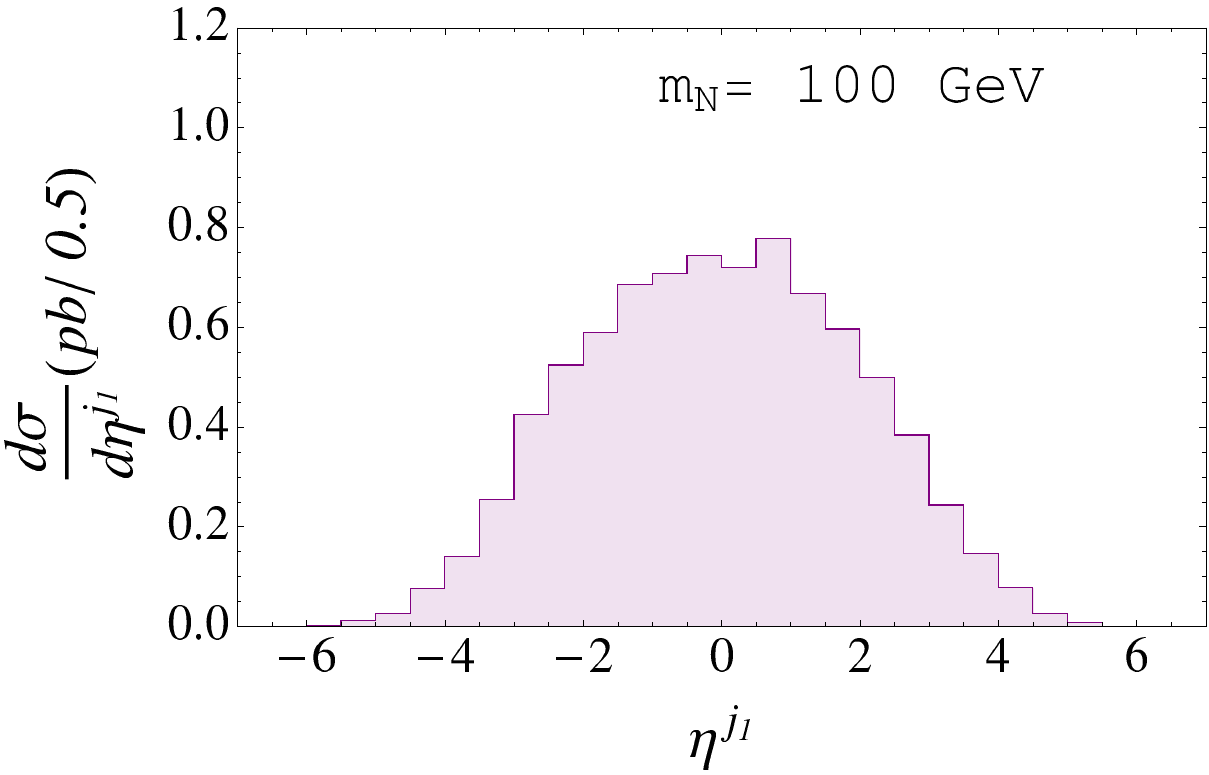}\\
\includegraphics[scale=0.59]{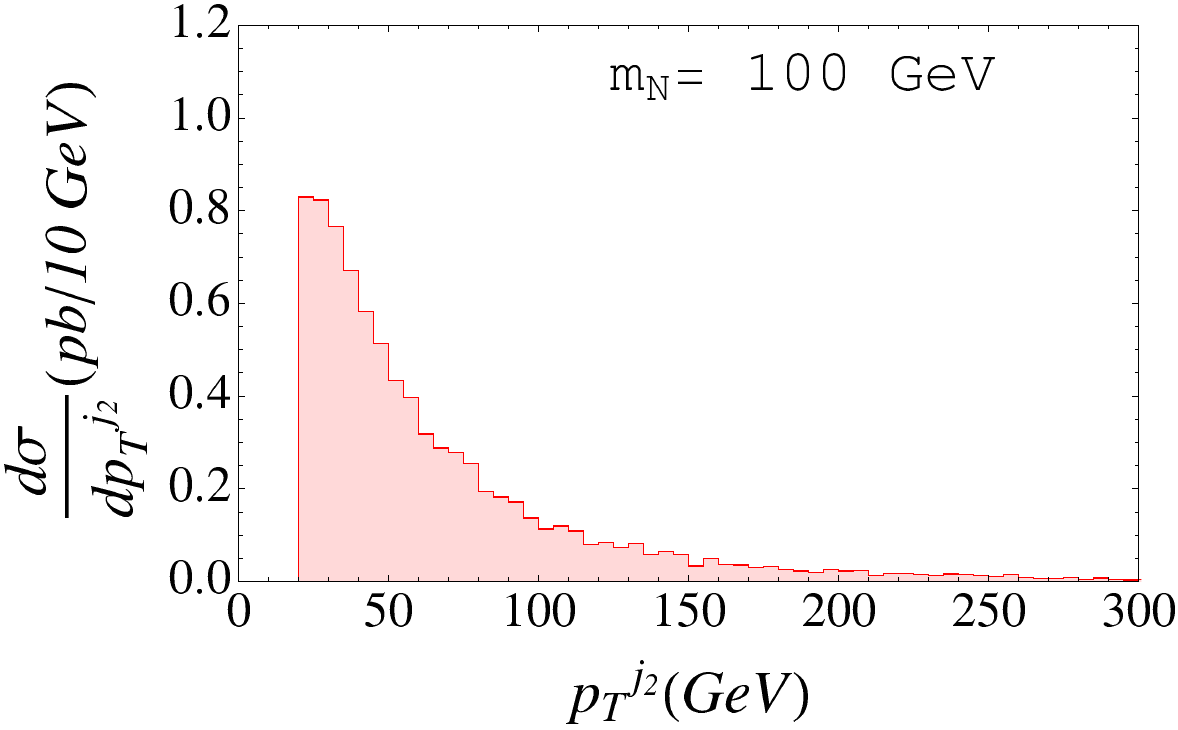}
\includegraphics[scale=0.61]{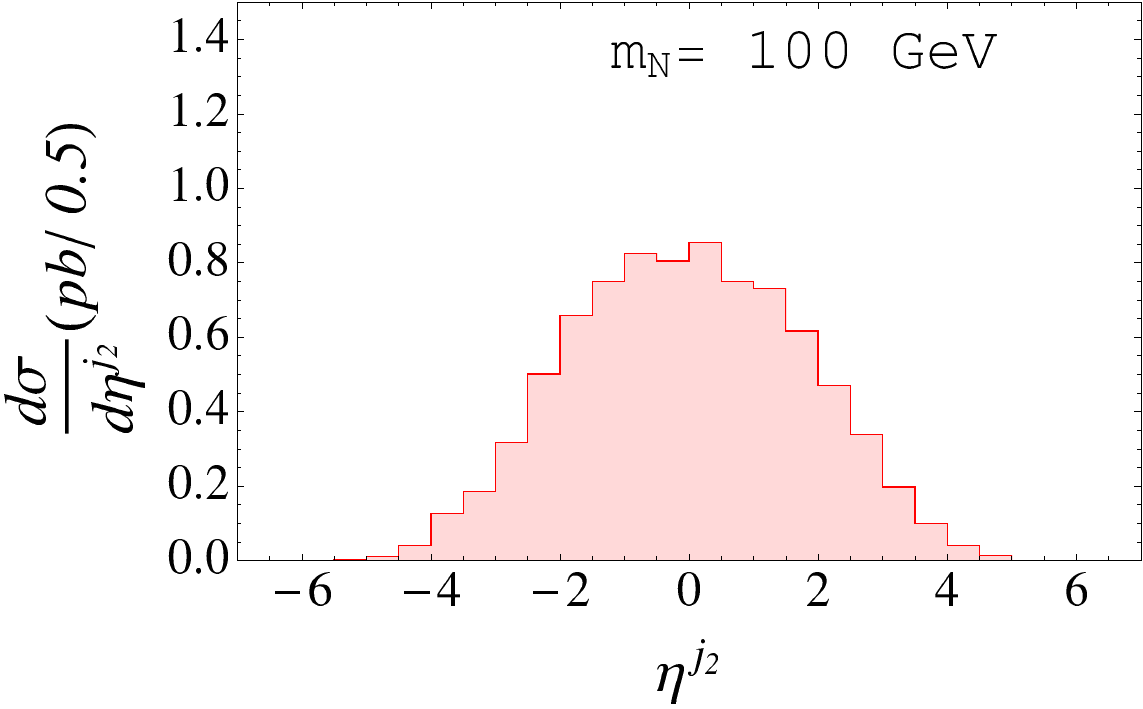}\\
\includegraphics[scale=0.65]{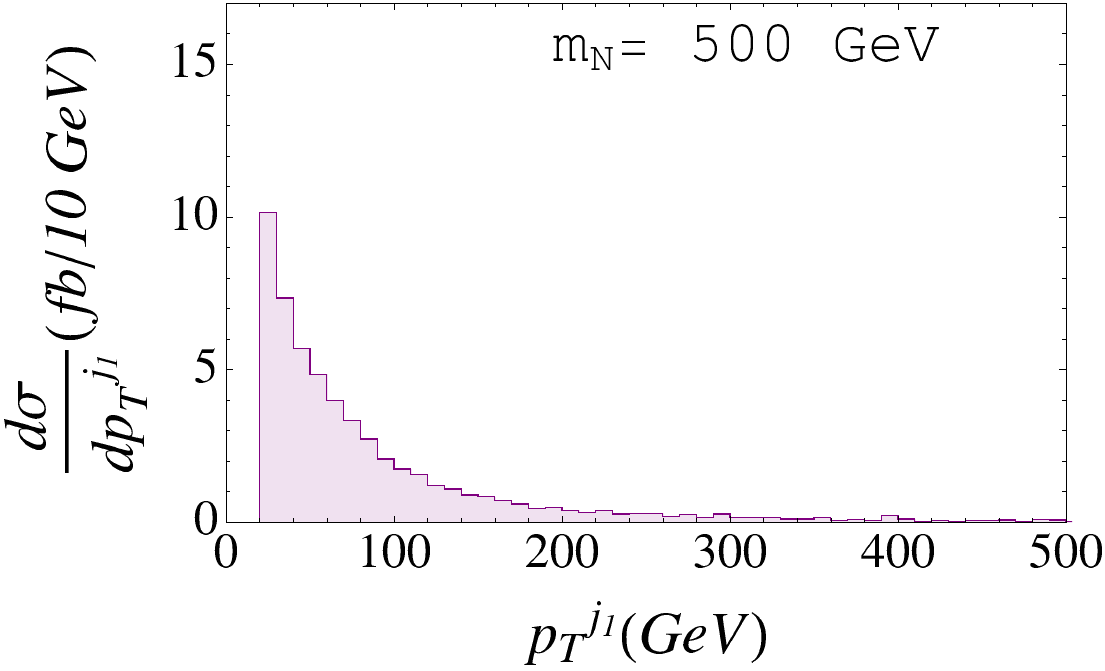}
\includegraphics[scale=0.63]{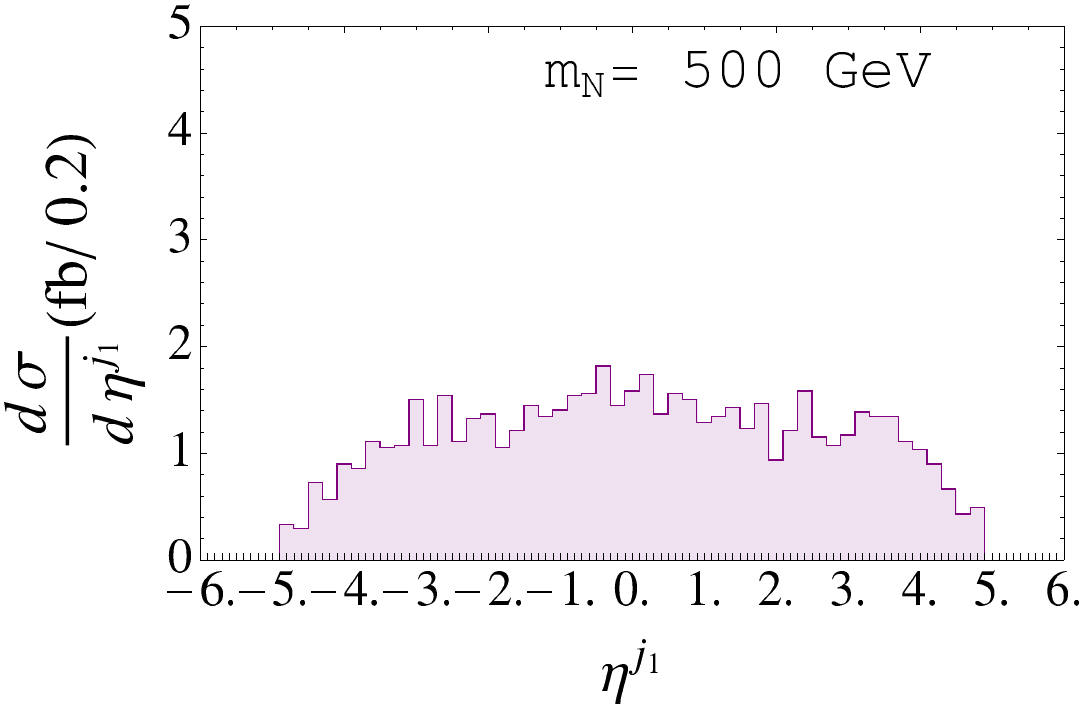}\\
\includegraphics[scale=0.67]{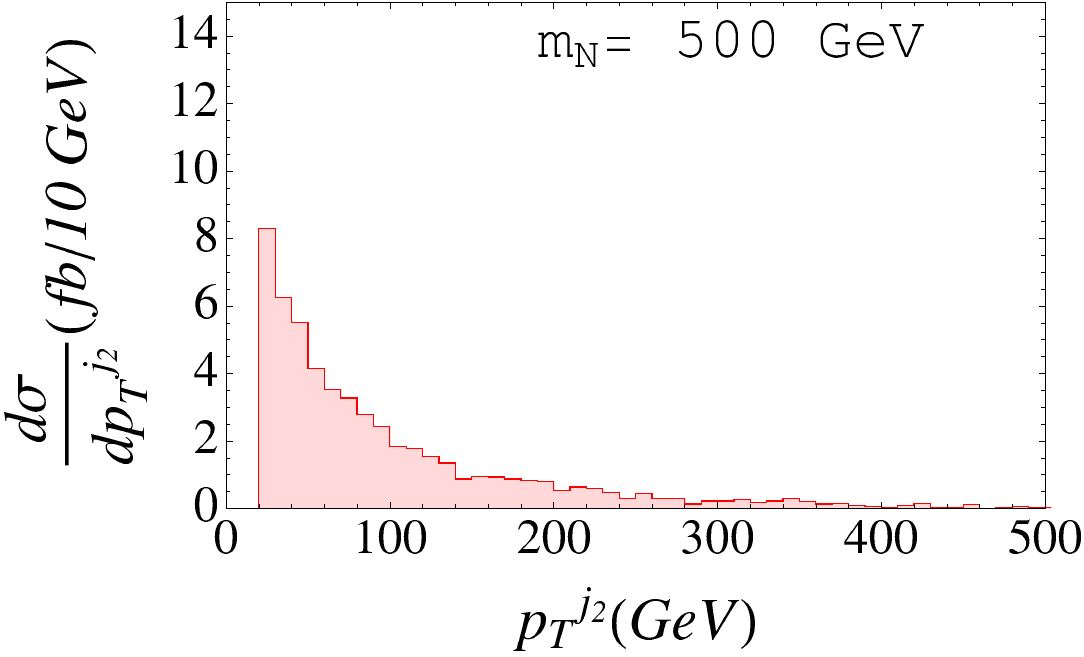}
\includegraphics[scale=0.57]{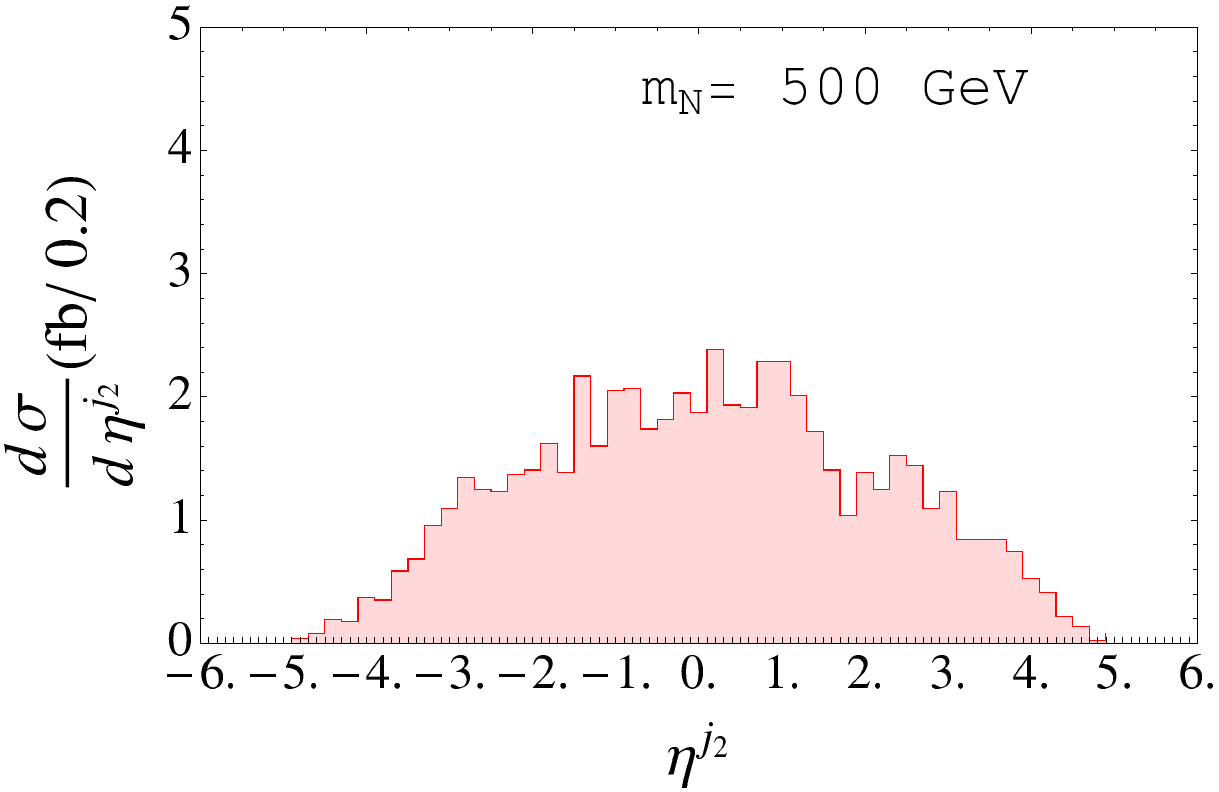}
\end{center}
\caption{The first and second rows show the $p_{T}$ and $\eta$ distributions of the non-leading jet produced in the $N\ell jj$ final state from the $q\overline{q^{'}}$, $qg$ and $gg$ initial states. The third and fourth rows show the same for the leading jet. The left column stands for $m_{N}=100$ GeV whereas the right one is for $m_{N}=500$ GeV.}
\label{pp2j2}
\end{figure}

\begin{figure}
\begin{center}
\includegraphics[scale=0.61]{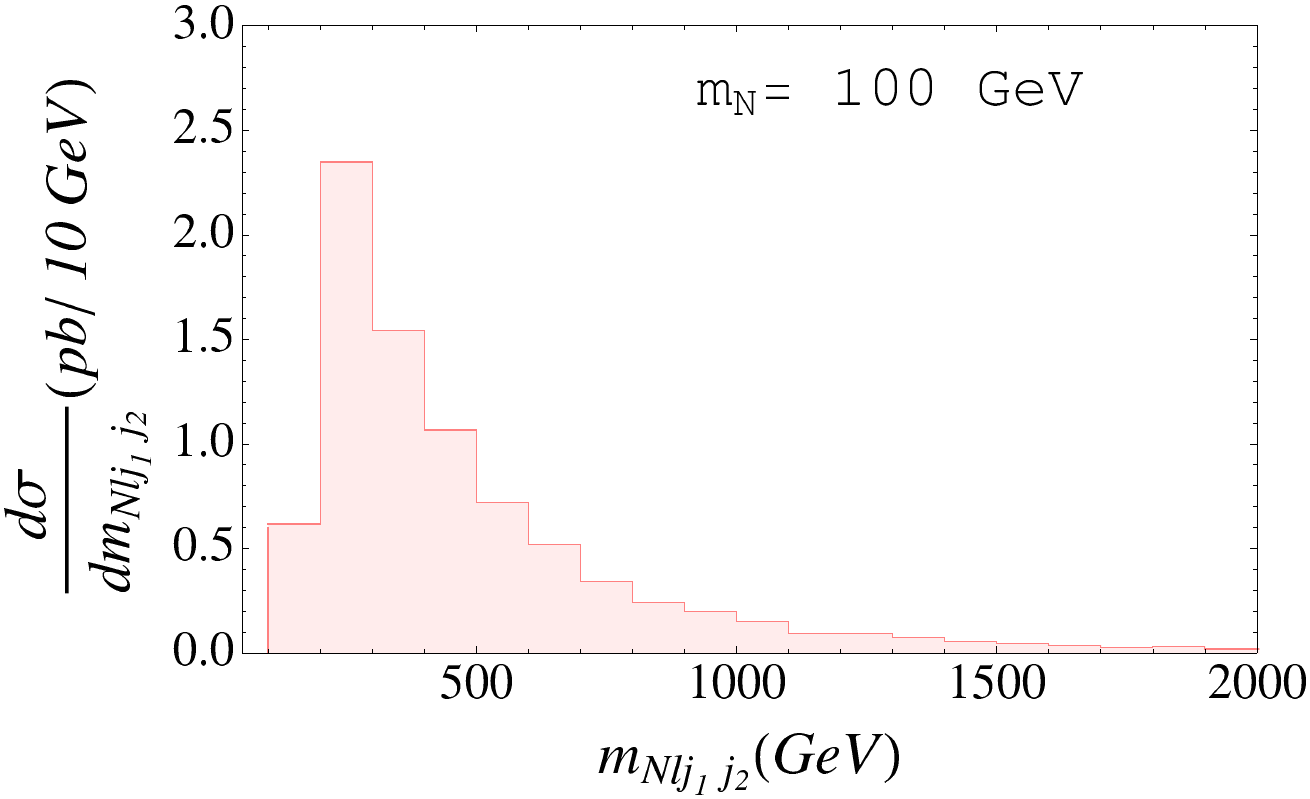}
\includegraphics[scale=0.69]{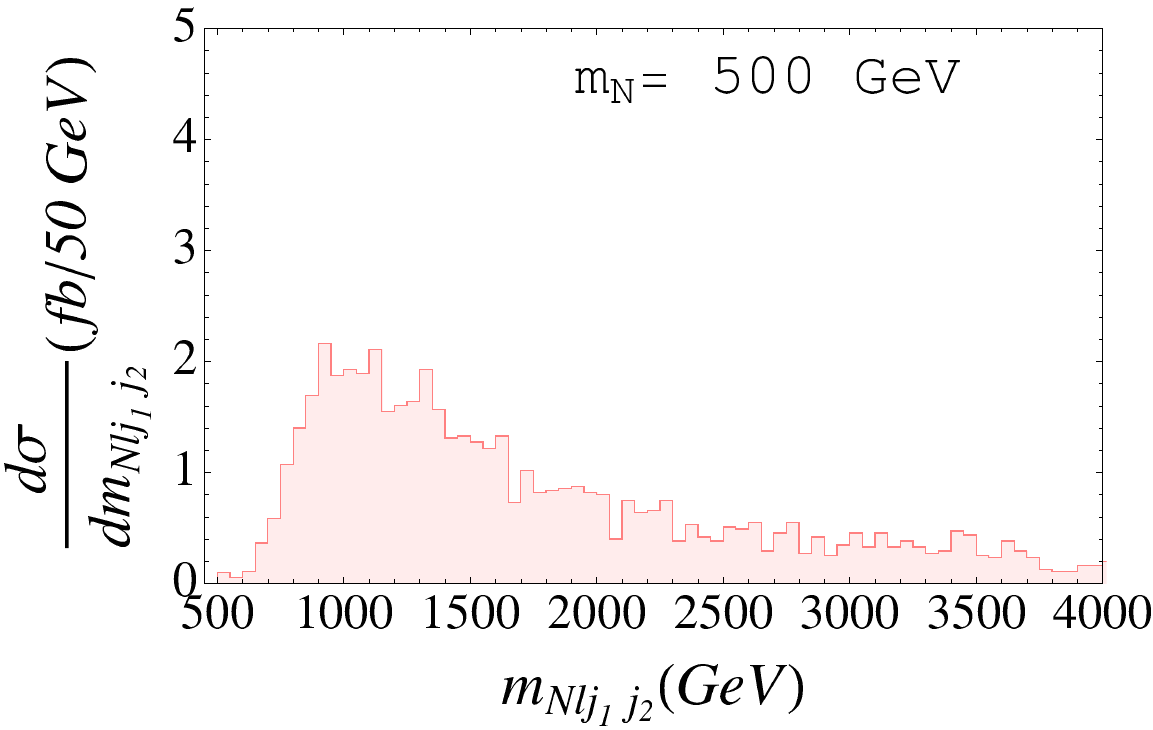}\\
\end{center}
\caption{The final state invariant mass $\left(m_{N\ell j_{1}j_{2}}\right)$ distributions produced in the $N\ell jj$ final state from the $q\overline{q^{'}}$, $qg$ and $gg$ initial states. The left panel shows $m_{N}=100$ GeV whereas the right one shows $m_{N}=500$ GeV.}
\label{pp2j3}
\end{figure}

The kinematics distributions for the final state $N\ell$ are shown in Fig.~\ref{pp0j}.
The $\eta^{\ell, N}$ distributions show peaks around $\eta=0$ and 
  become sharper with the increase in $m_{N}$.
The invariant mass distribution shows a sharp peak in the vicinity of the heavy neutrino mass threshold,
  which means most of the particles are produced almost at rest.
For the final state $N\ell j$, the kinetic distributions are shown in Figs.~\ref{pp1j1} and \ref{pp1j2}.
These distributions show similar behaviors to the results for the $N\ell$ final state.
Same for the final state $N\ell jj$ are shown in Figs.~\ref{pp2j1}-\ref{pp2j3}.
In Fig.~\ref{pp2j2} the leading jet is defined as $j_{2}$ while the $j_{1}$ is the non-leading jet 
 ($p_T^{j_{2}} > p_T^{j_{1}}$ for each event).
The $\eta^{j_{1, 2}}$ distributions are flatter than those in the $N\ell j$ case.
This is because many events include the radiated jets. 

\begin{figure}
\begin{center}
\includegraphics[scale=0.6]{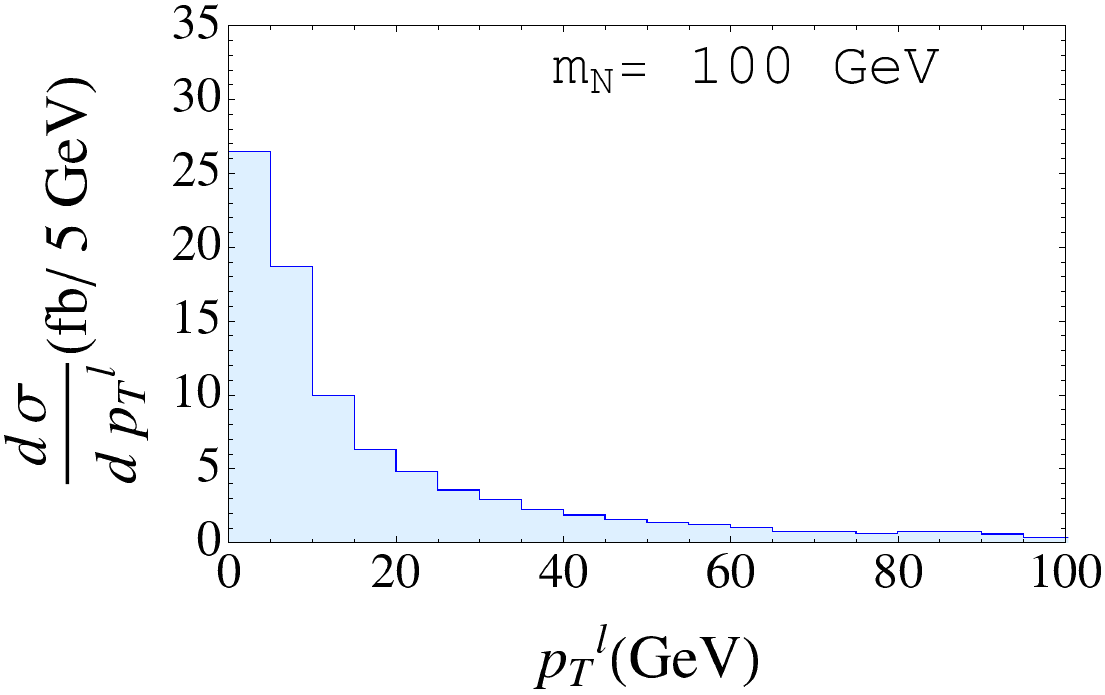}
\includegraphics[scale=0.63]{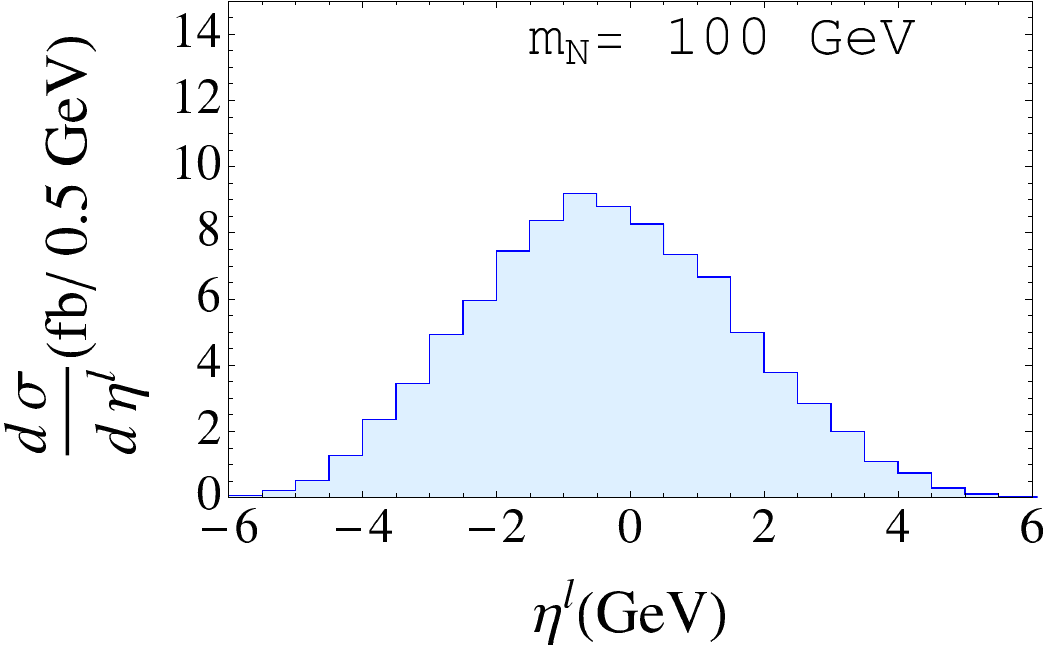}\\
\includegraphics[scale=0.57]{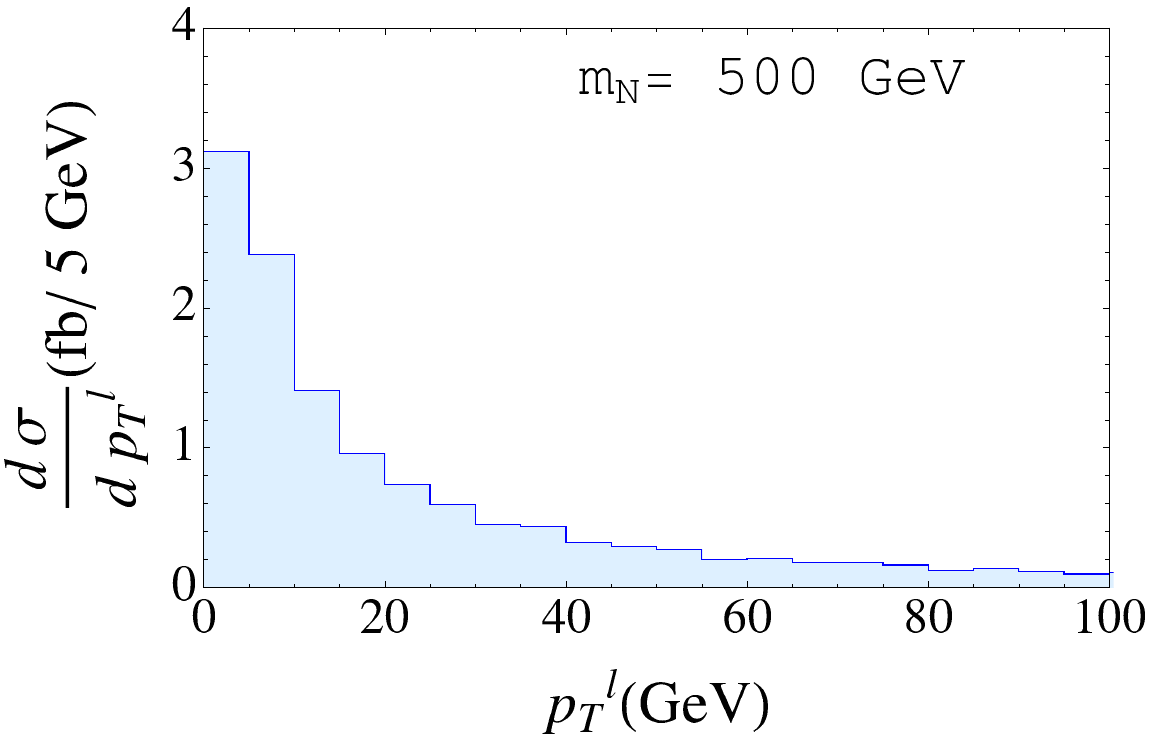}
\includegraphics[scale=0.55]{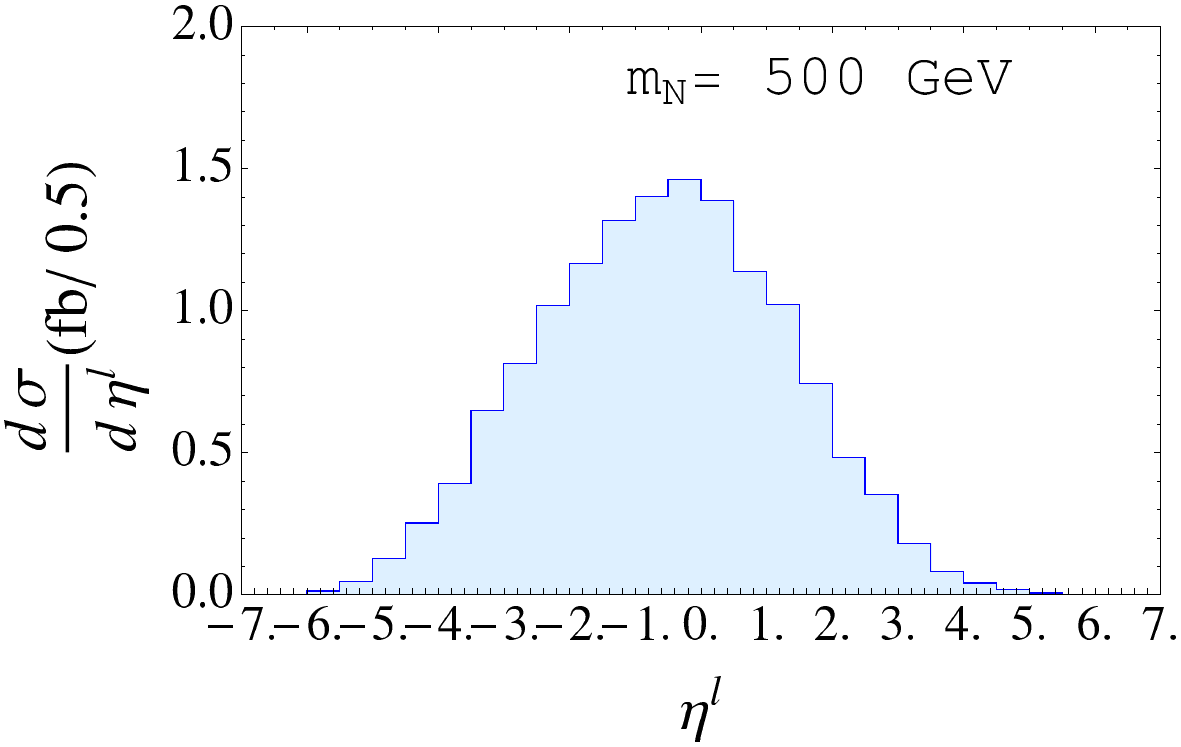}\\
\includegraphics[scale=0.56]{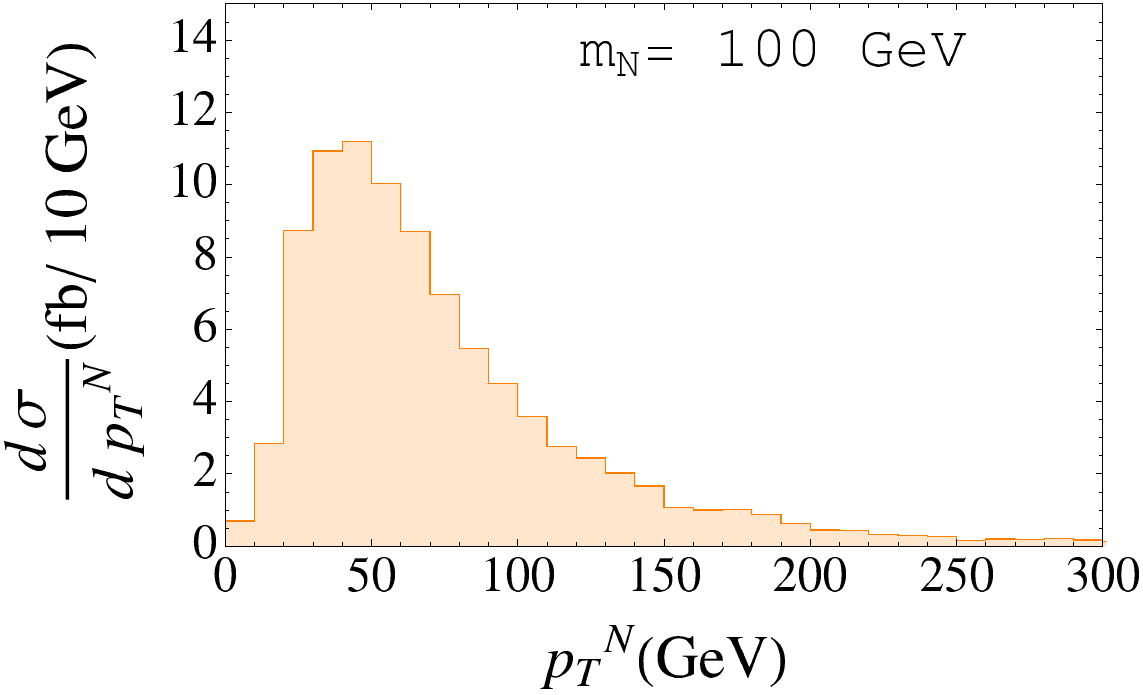}
\includegraphics[scale=0.55]{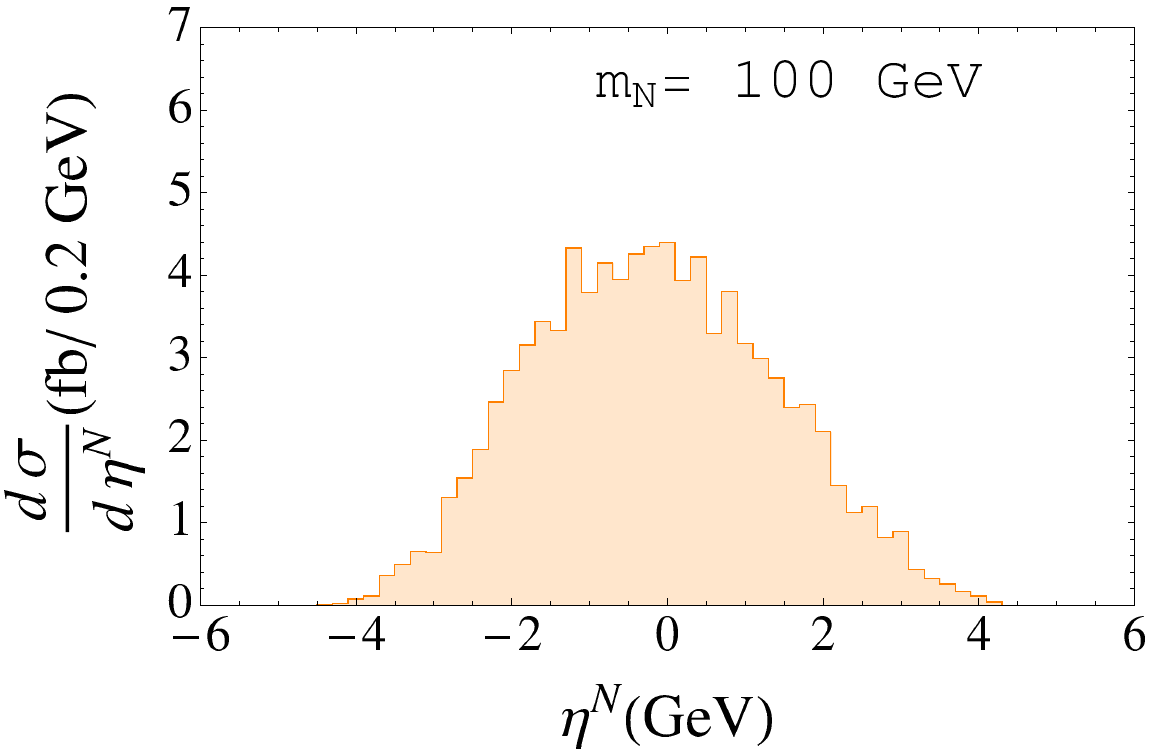}\\
\includegraphics[scale=0.51]{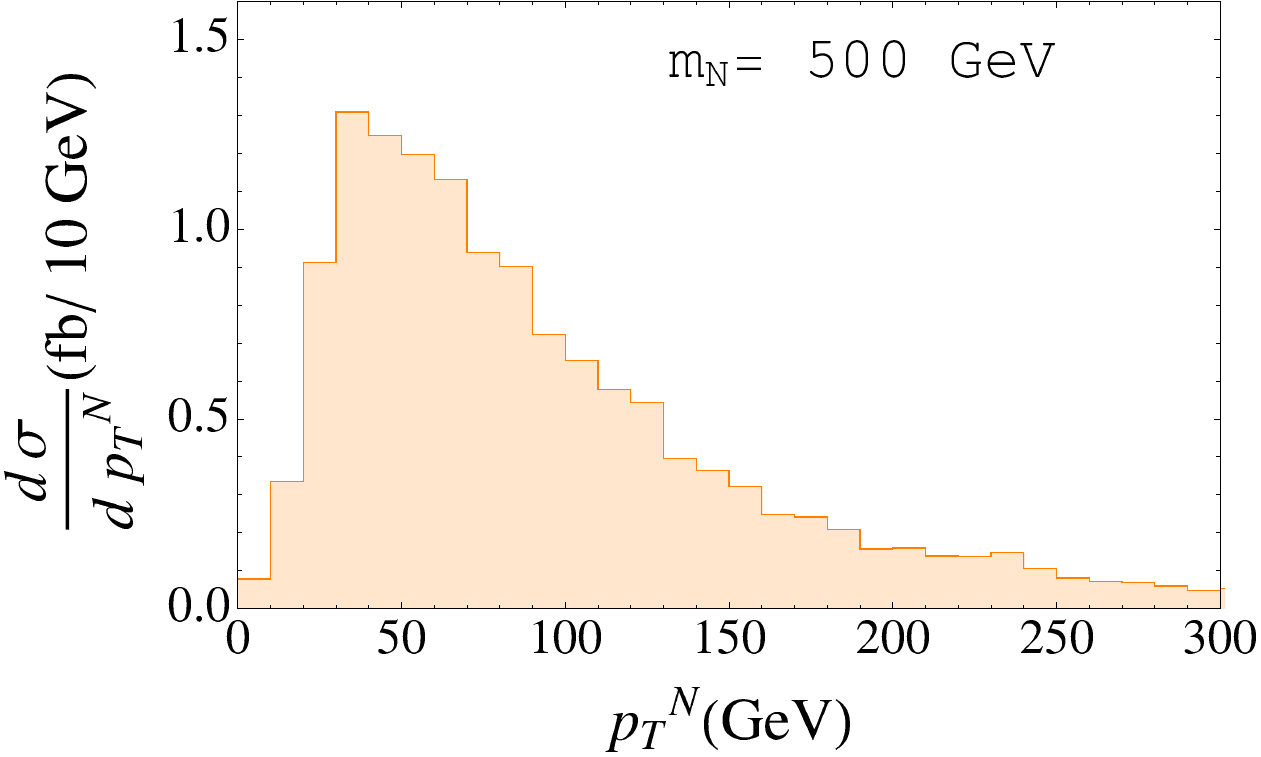}
\includegraphics[scale=0.45]{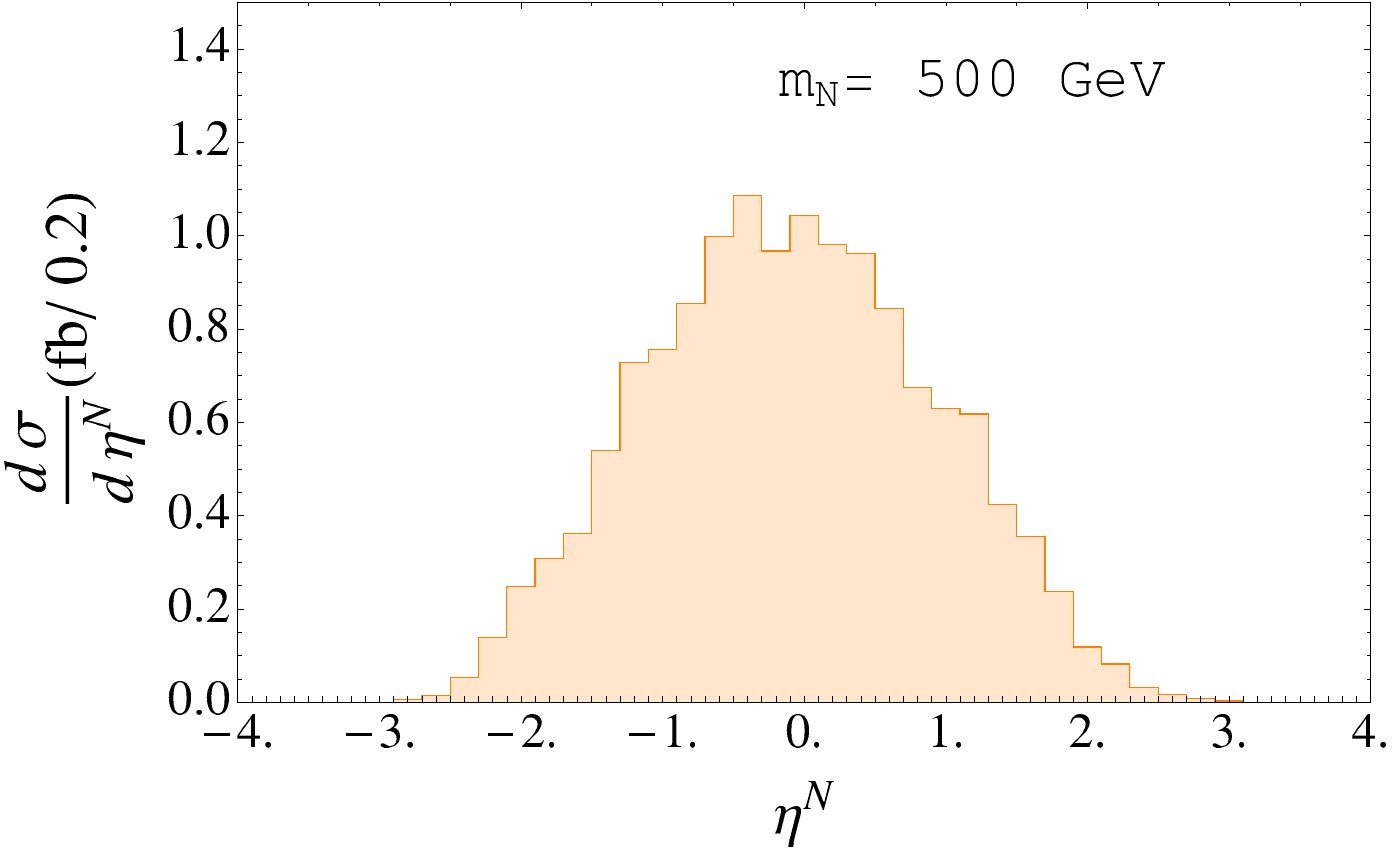}
\end{center}
\caption{The first and second rows show the $p_{T}$ and $\eta$ distributions of the lepton produced in the $N\ell j$ final state from the photon mediated elastic processes. The third and fourth rows show the same for the heavy neutrino. The left column stands for $m_{N}=100$ GeV whereas the right one is for $m_{N}=500$ GeV.}
\label{paelas1}
\end{figure}

\begin{figure}
\begin{center}
\includegraphics[scale=0.69]{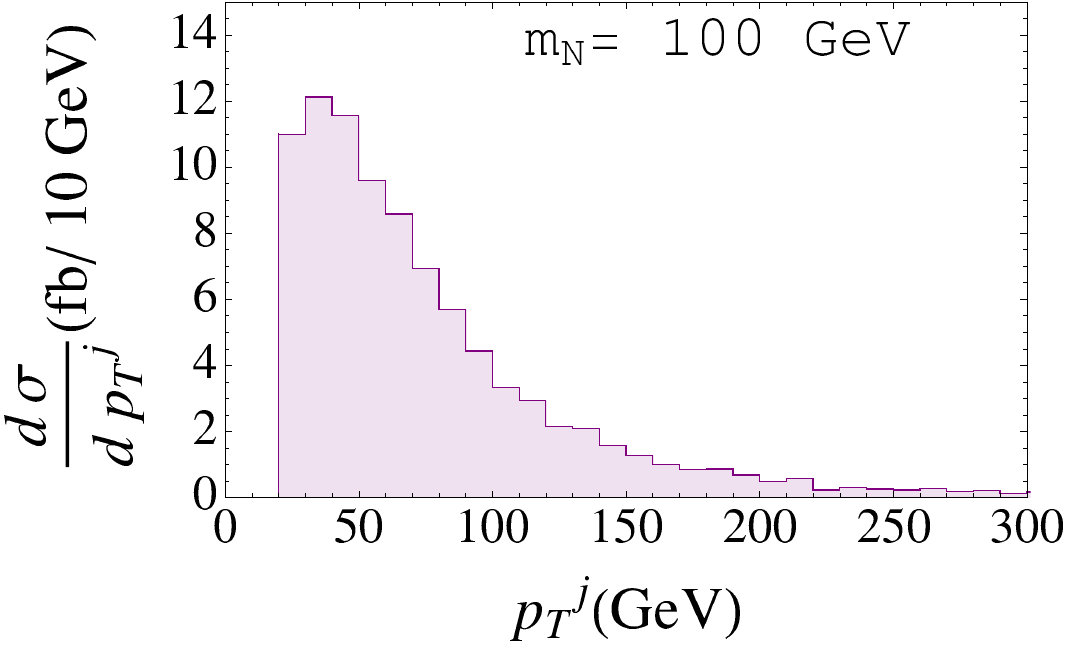}
\includegraphics[scale=0.53]{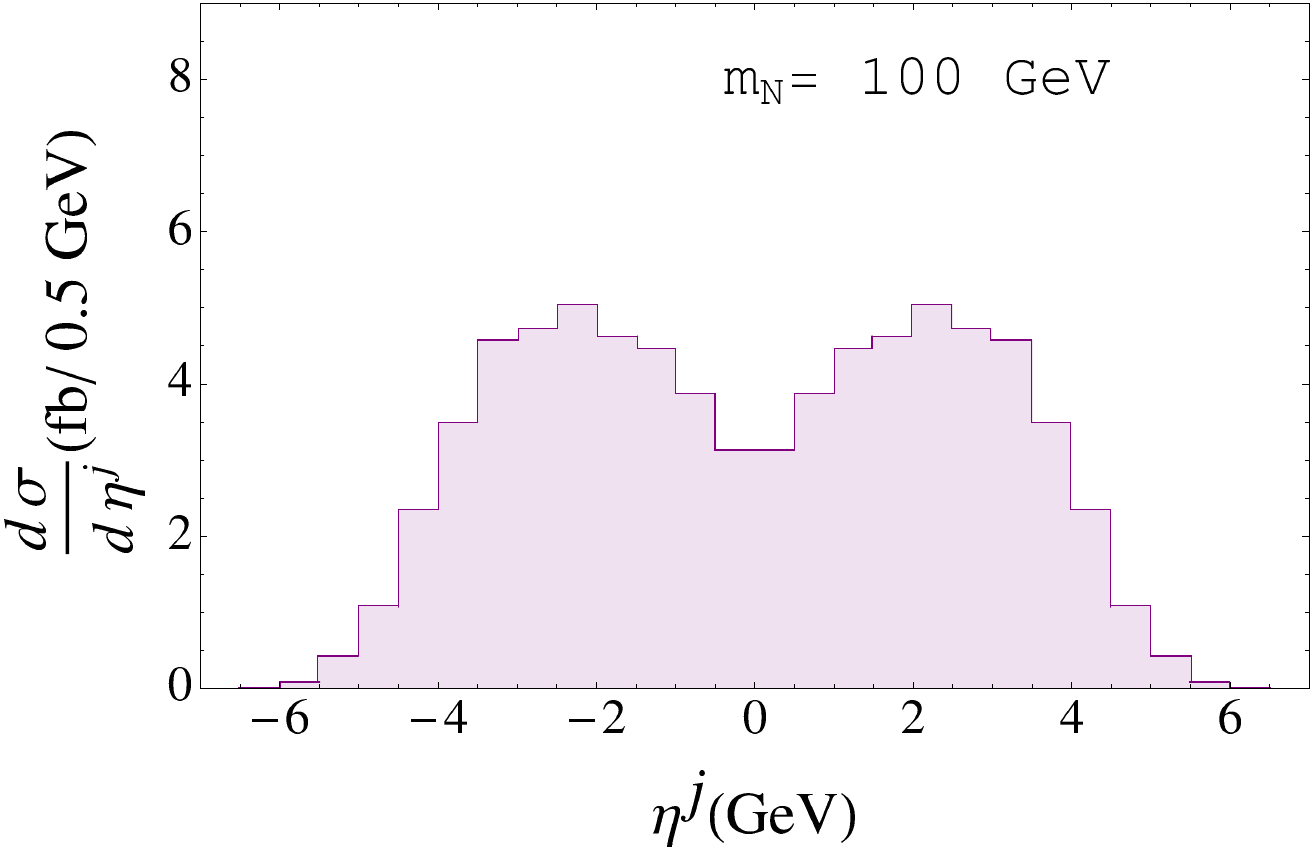}\\
\includegraphics[scale=0.53]{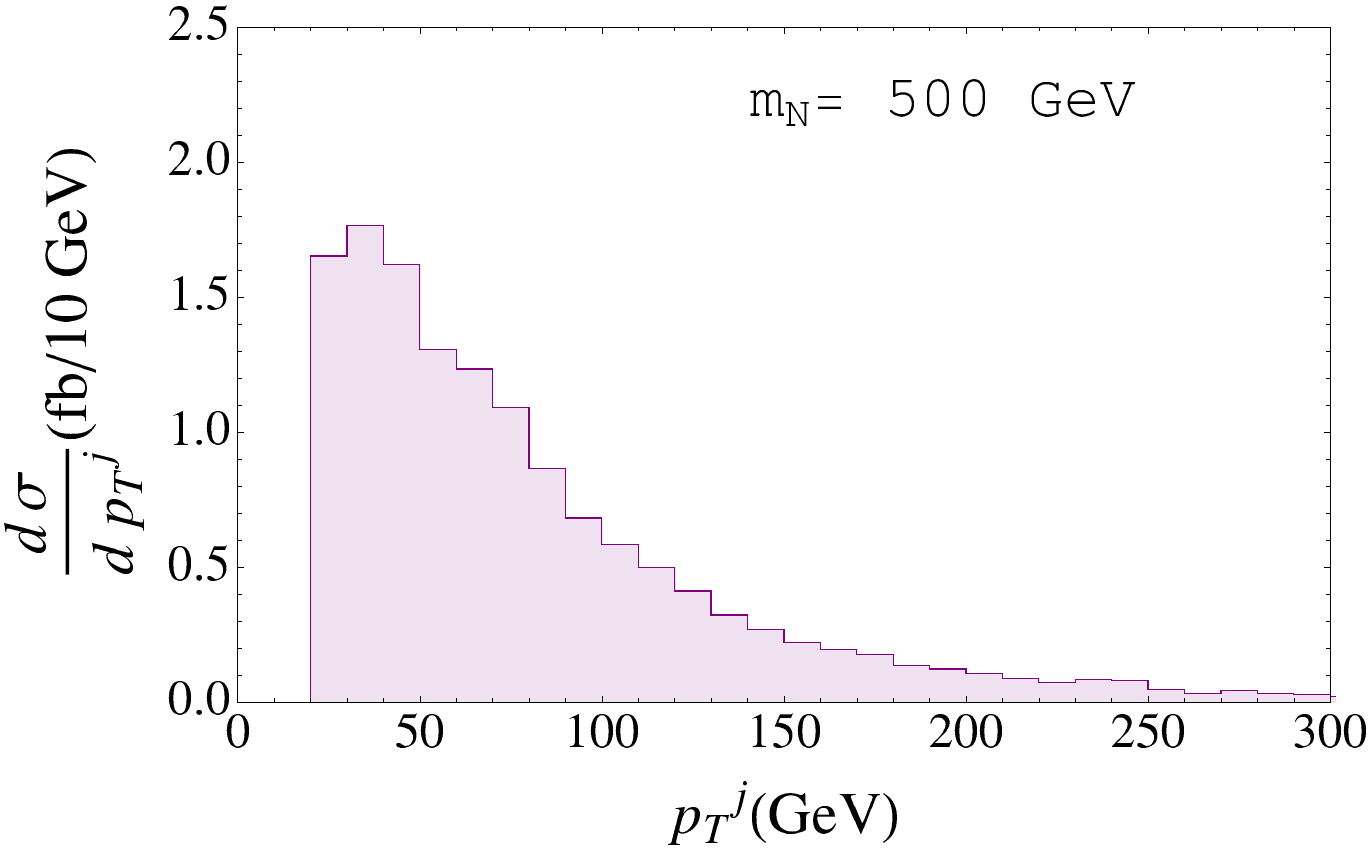}
\includegraphics[scale=0.57]{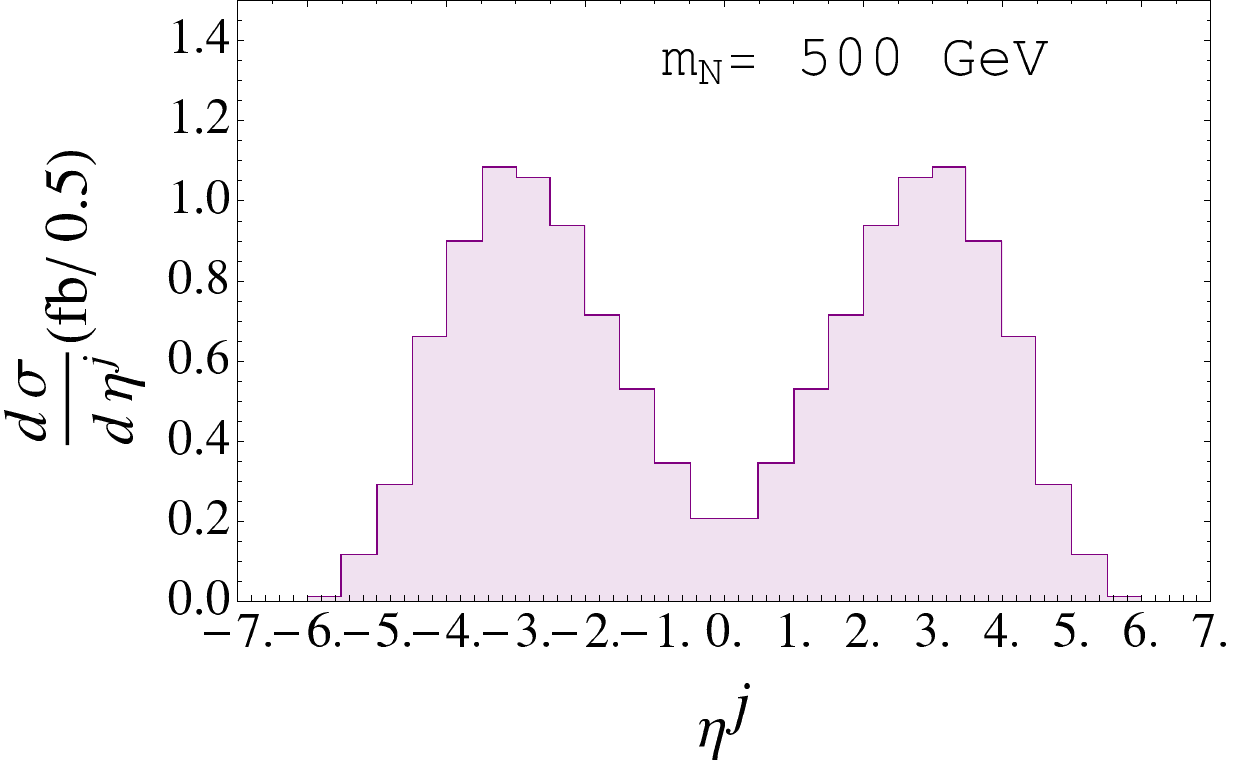}\\
\includegraphics[scale=0.53]{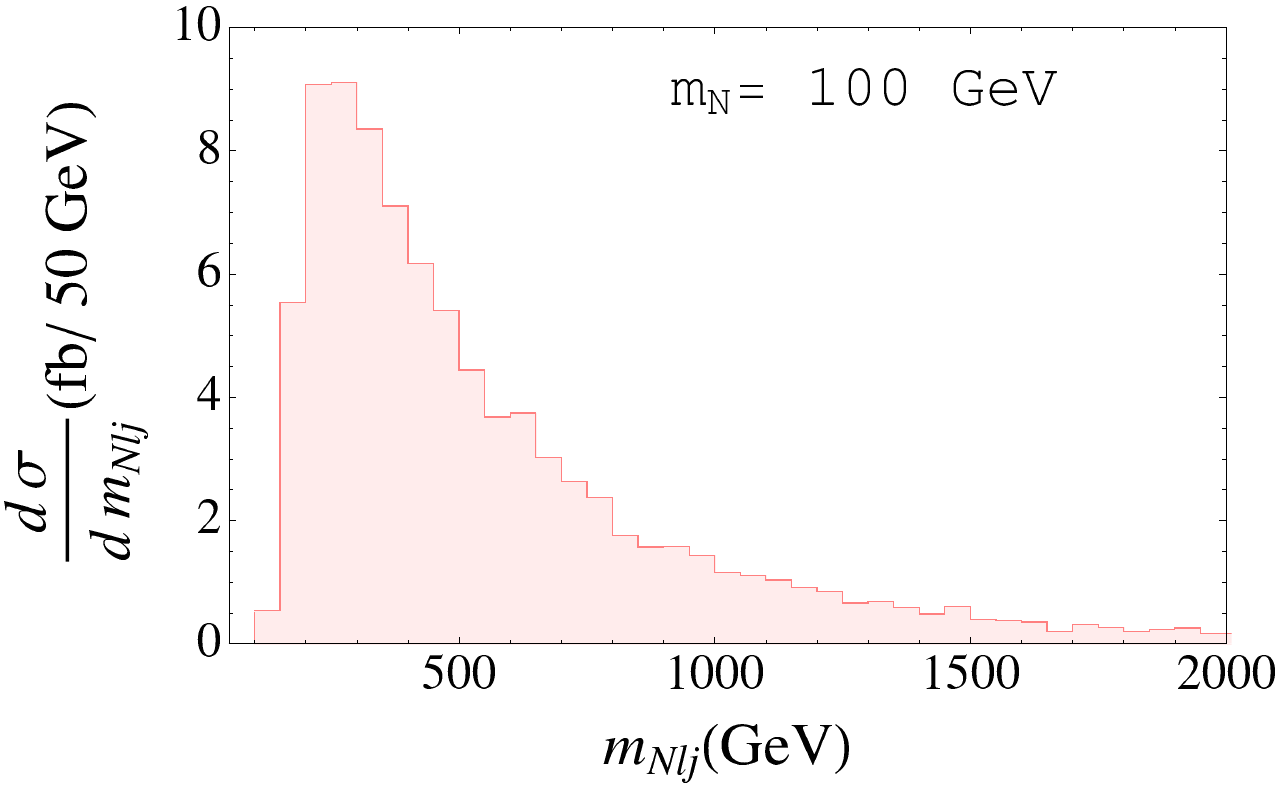}
\includegraphics[scale=0.63]{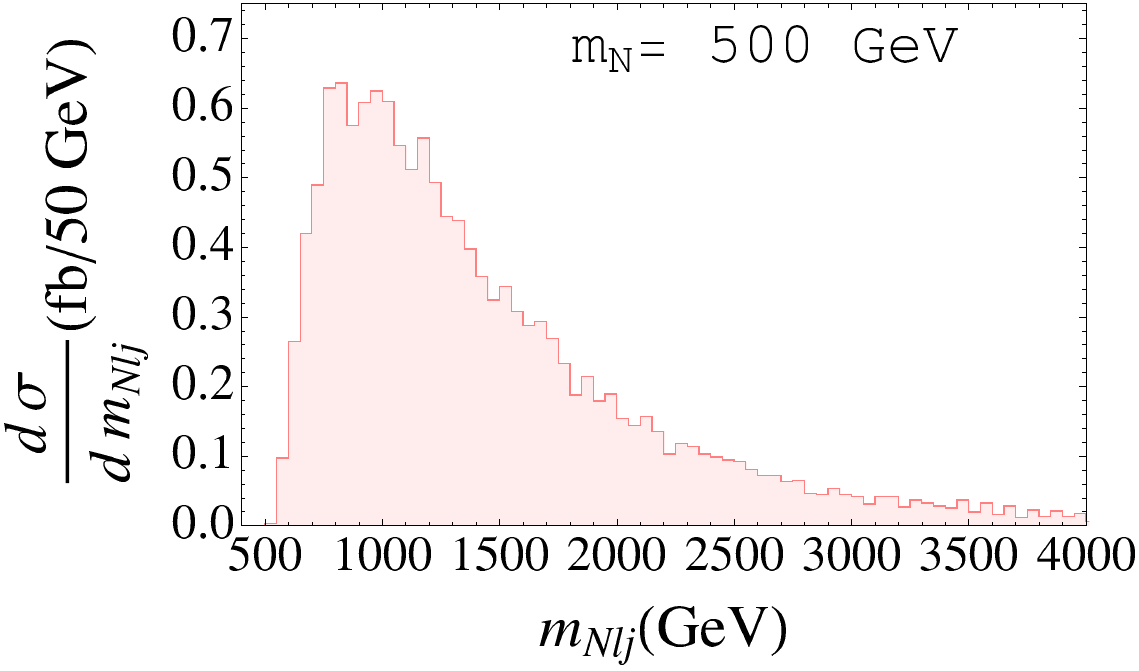}
\end{center}
\caption{The first and second rows show the $p_{T}$ and $\eta$ distributions of the jet produced in the $N\ell j$ final state from the photon mediated elastic processes. The third row shows the final state invariant mass $\left(m_{N\ell j}\right)$ distributions. The left column stands for $m_{N}=100$ GeV whereas the right one is for $m_{N}=500$ GeV.}
\label{paelas2}
\end{figure}

\begin{figure}
\begin{center}
\includegraphics[scale=0.6]{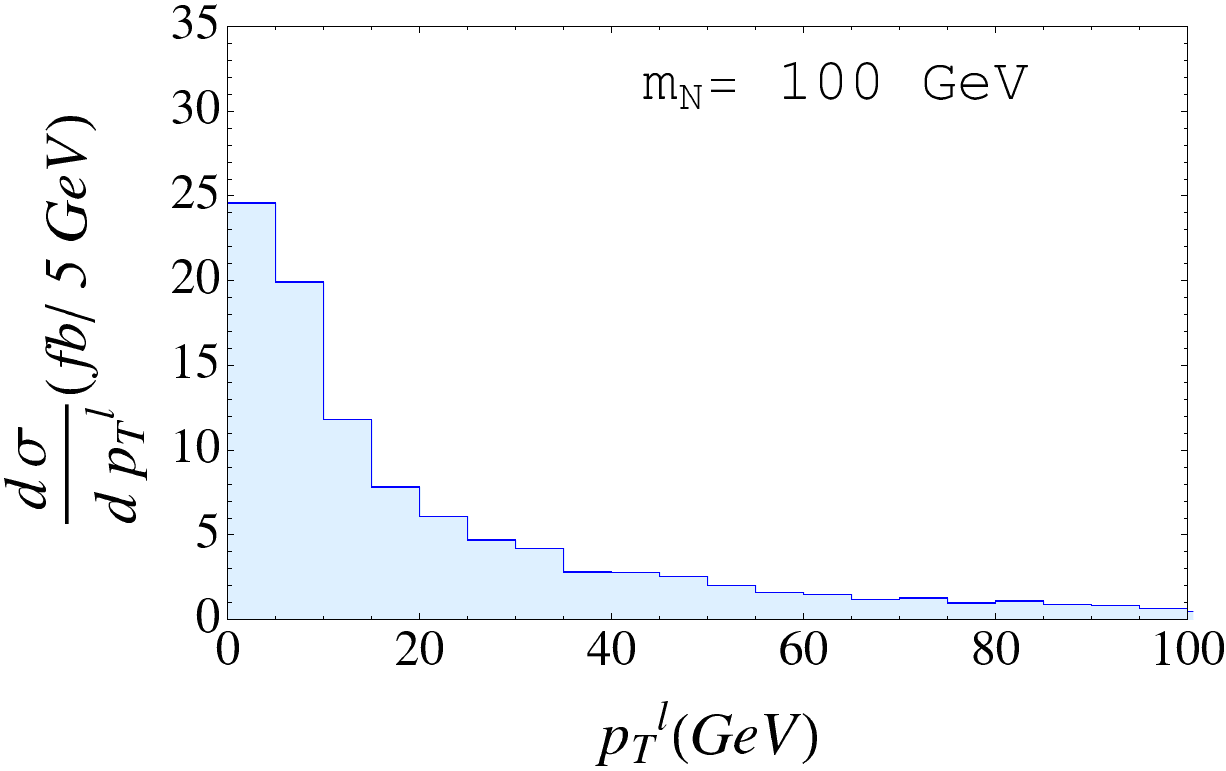}
\includegraphics[scale=0.65]{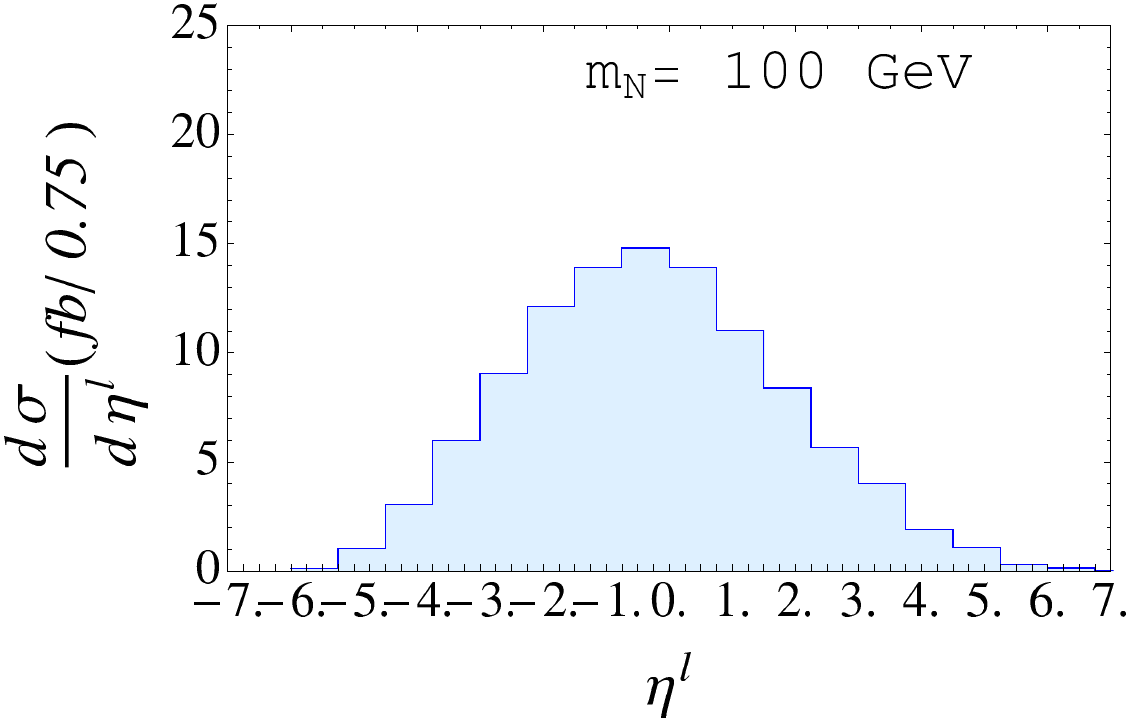}\\
\includegraphics[scale=0.73]{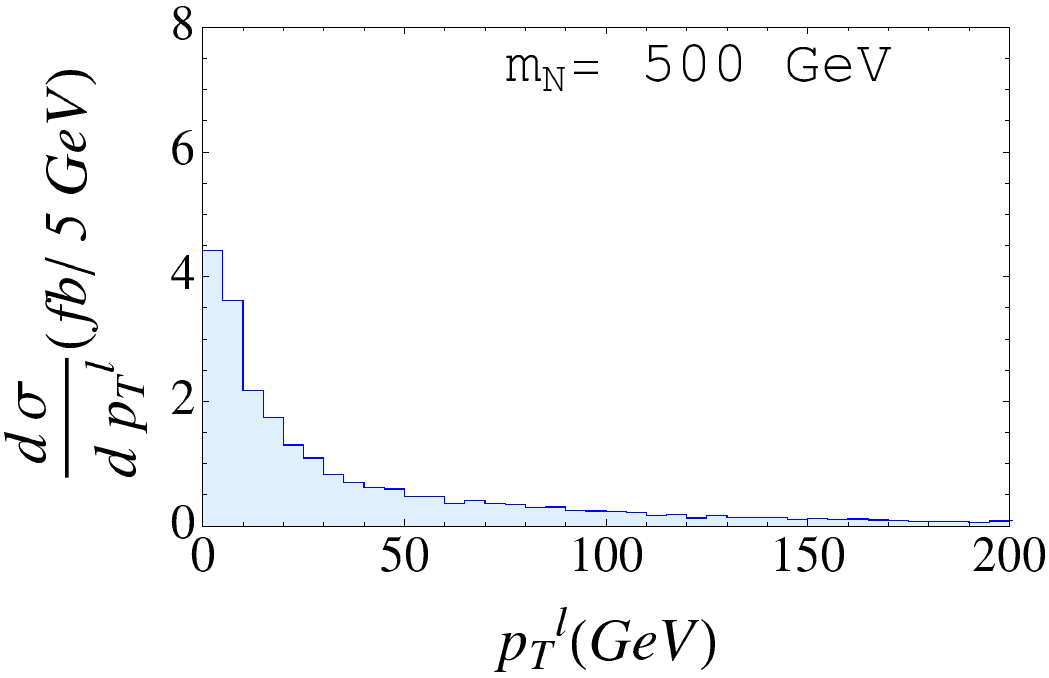}
\includegraphics[scale=0.71]{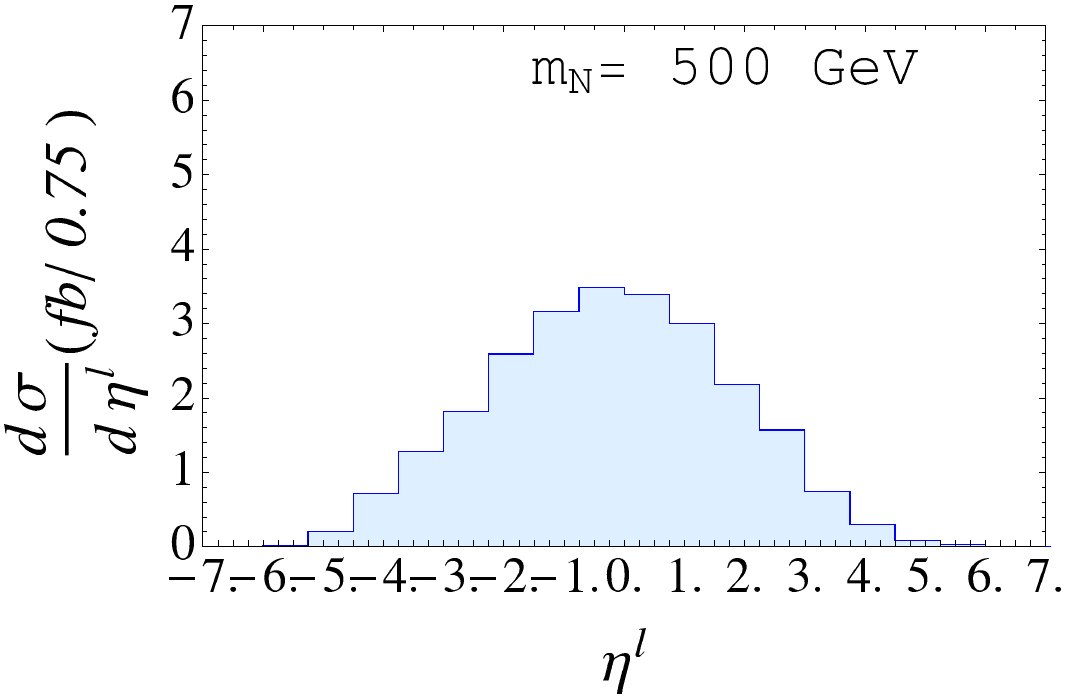}\\
\includegraphics[scale=0.57]{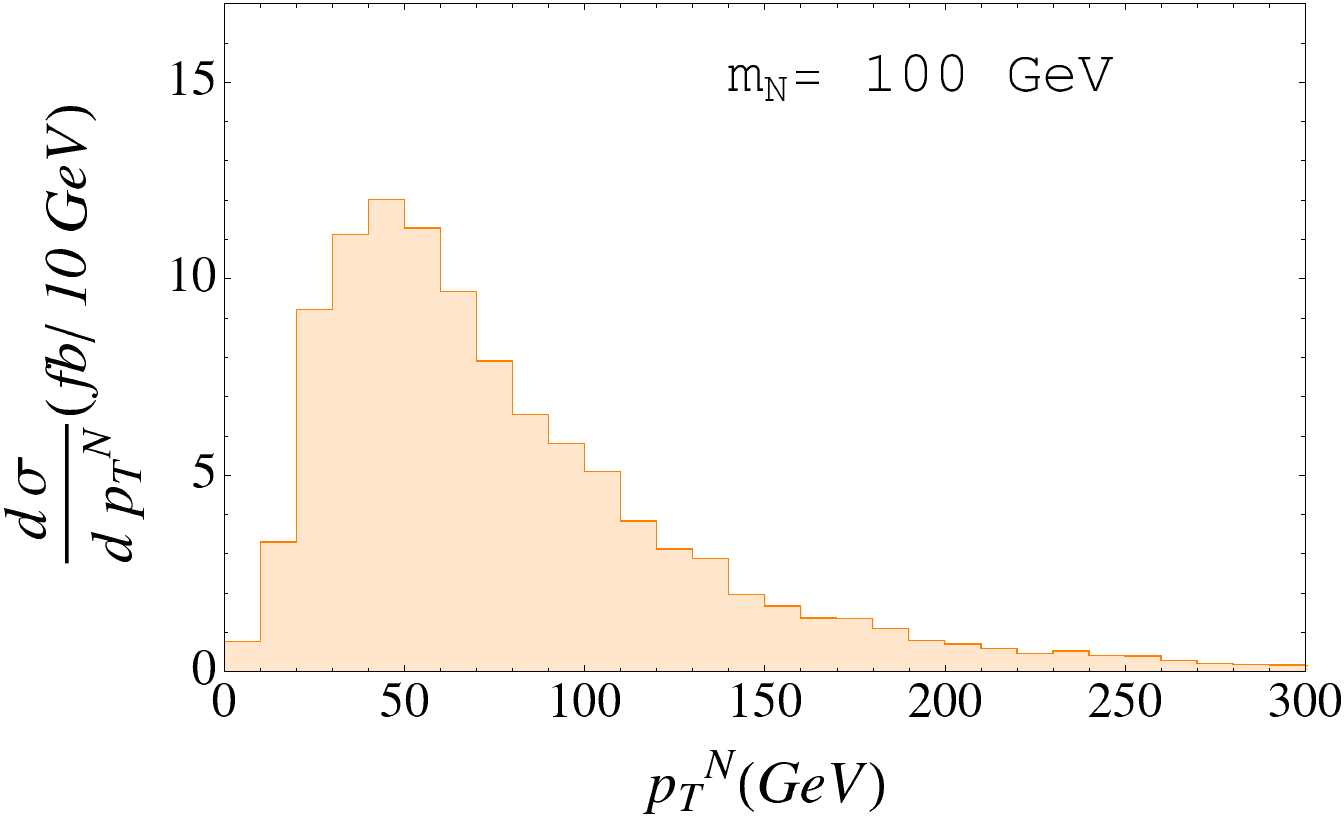}
\includegraphics[scale=0.62]{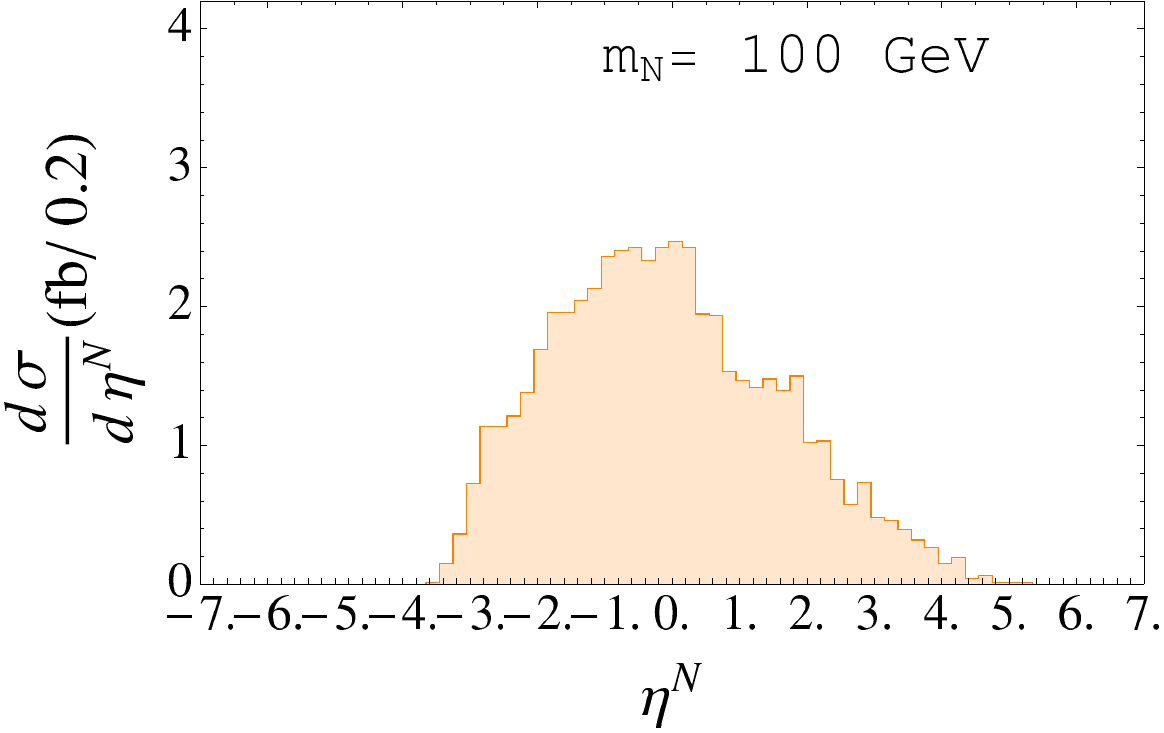}\\
\includegraphics[scale=0.59]{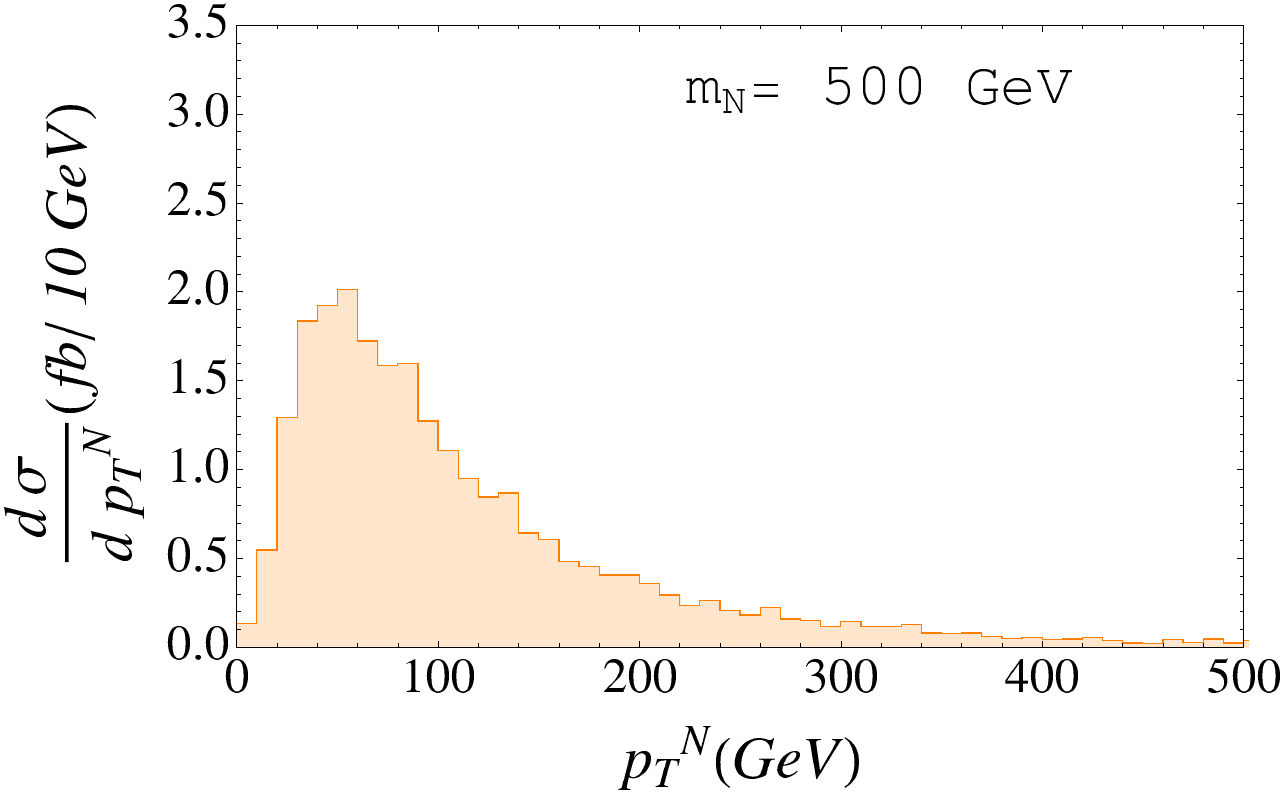}
\includegraphics[scale=0.65]{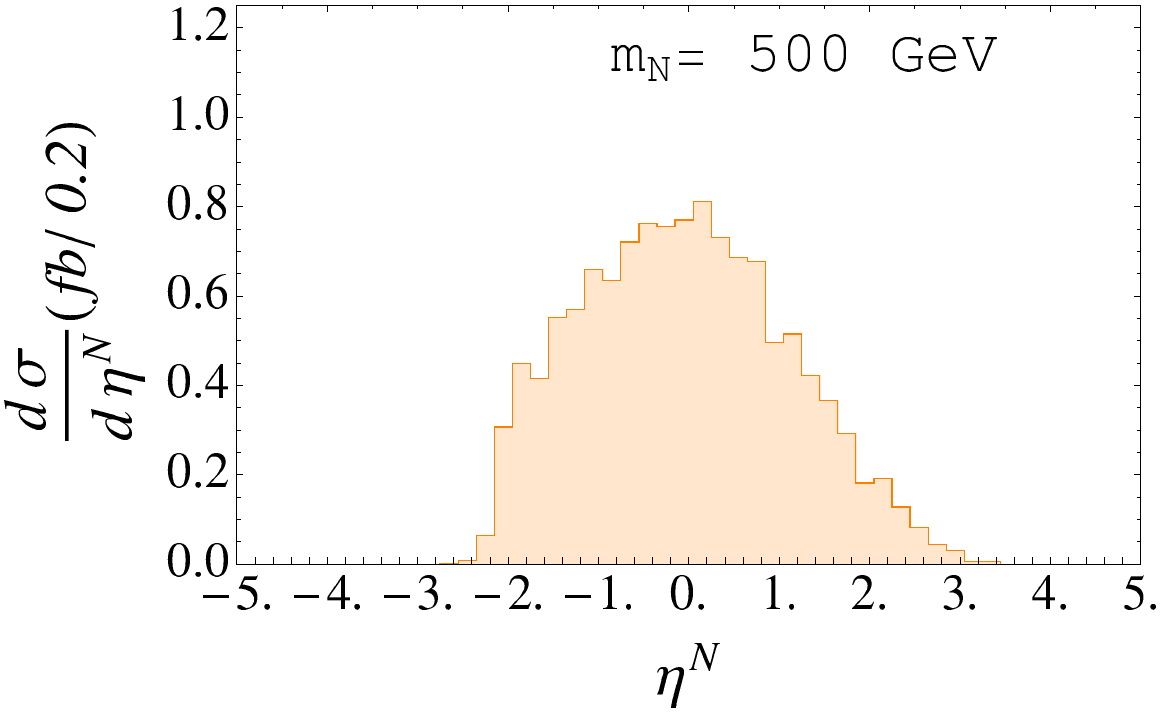}
\end{center}
\caption{The first and second rows show the $p_{T}$ and $\eta$ distributions of the lepton produced in the $N\ell j$ final state from the photon mediated inelastic processes. The third and fourth rows show the same for the heavy neutrino. The left column stands for $m_{N}=100$ GeV whereas the right one is for $m_{N}=500$ GeV.}
\label{painelas1}
\end{figure}

\begin{figure}
\begin{center}
\includegraphics[scale=0.71]{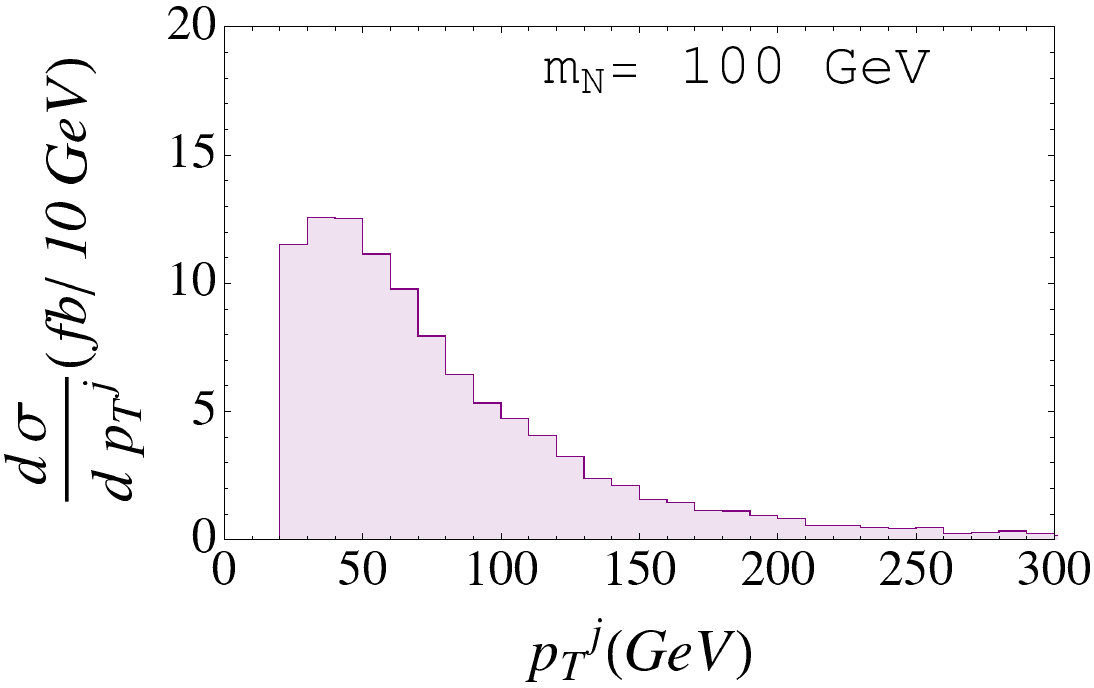}
\includegraphics[scale=0.65]{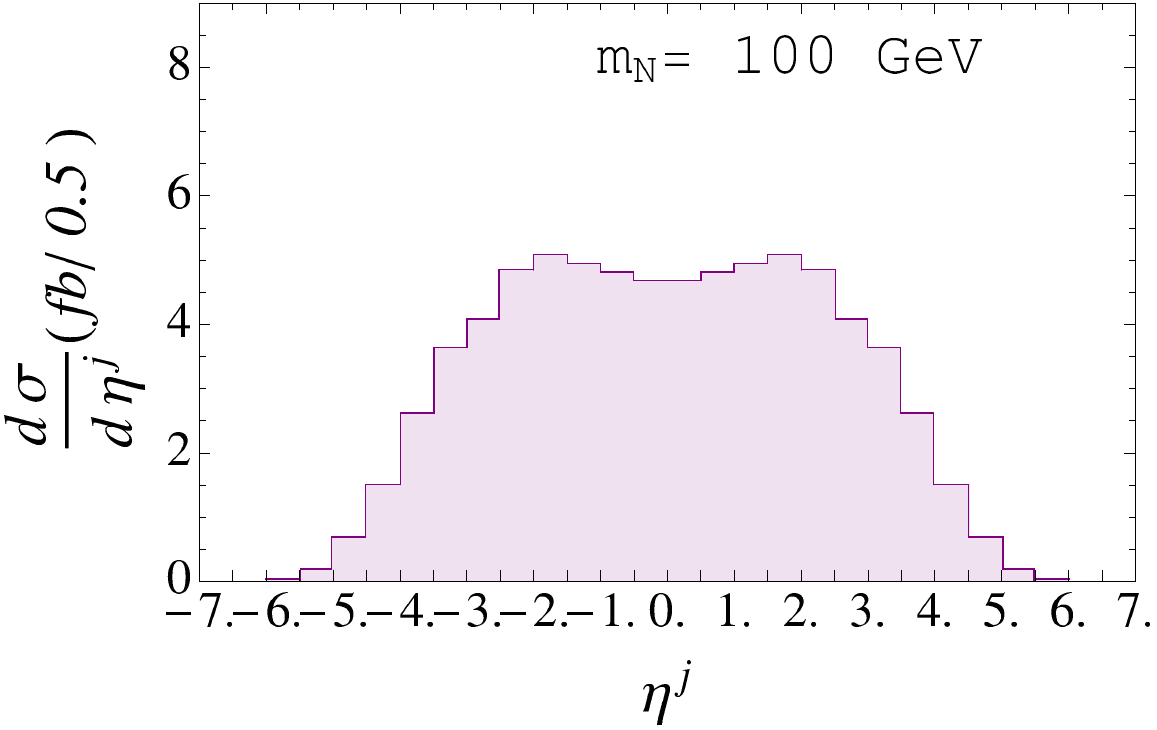}\\
\includegraphics[scale=0.77]{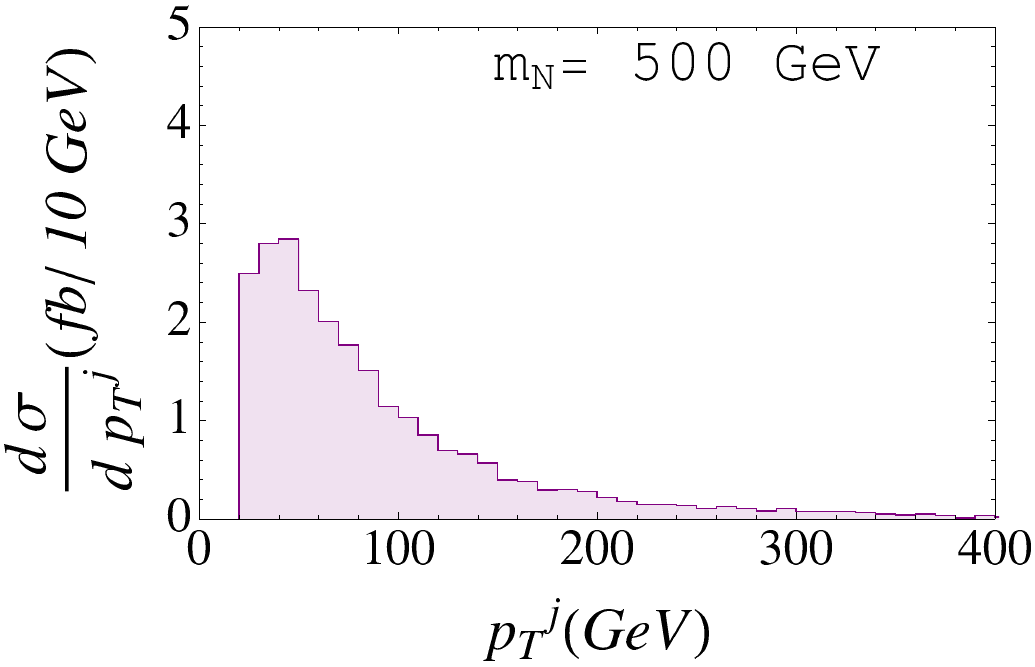}
\includegraphics[scale=0.67]{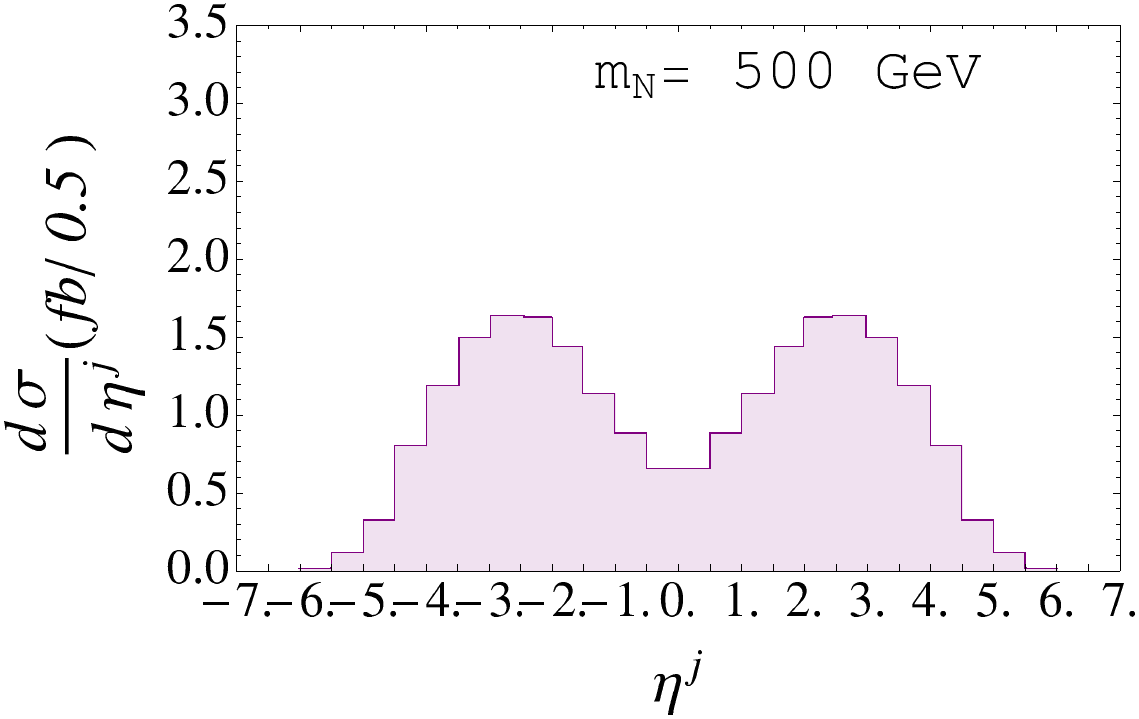}\\
\includegraphics[scale=0.70]{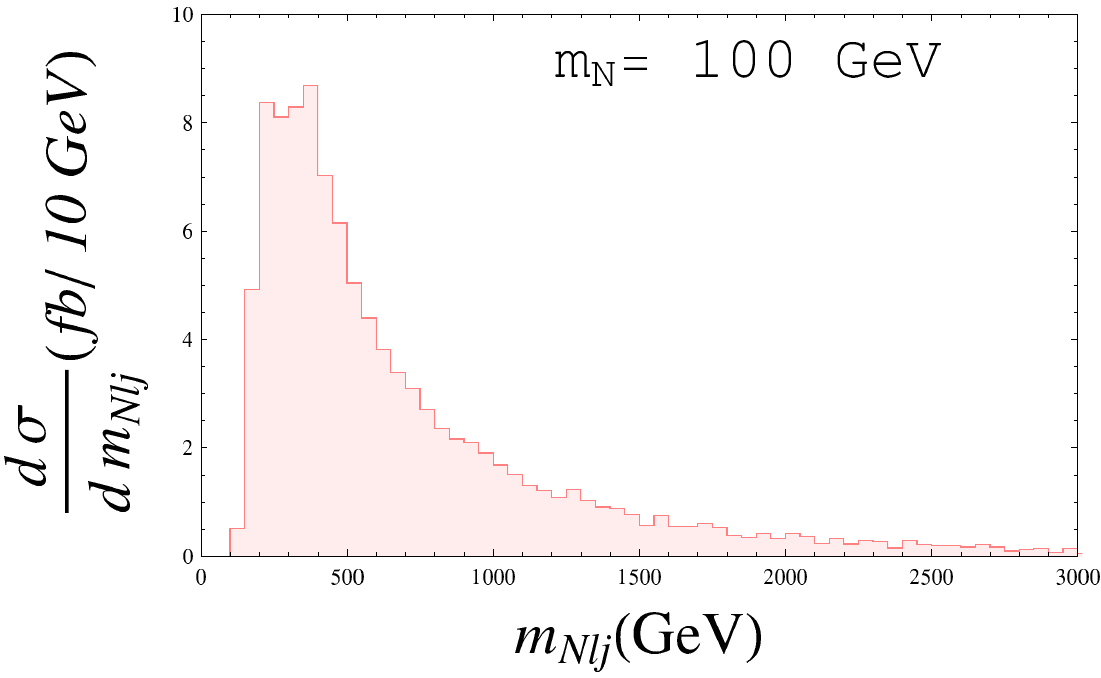}
\includegraphics[scale=0.69]{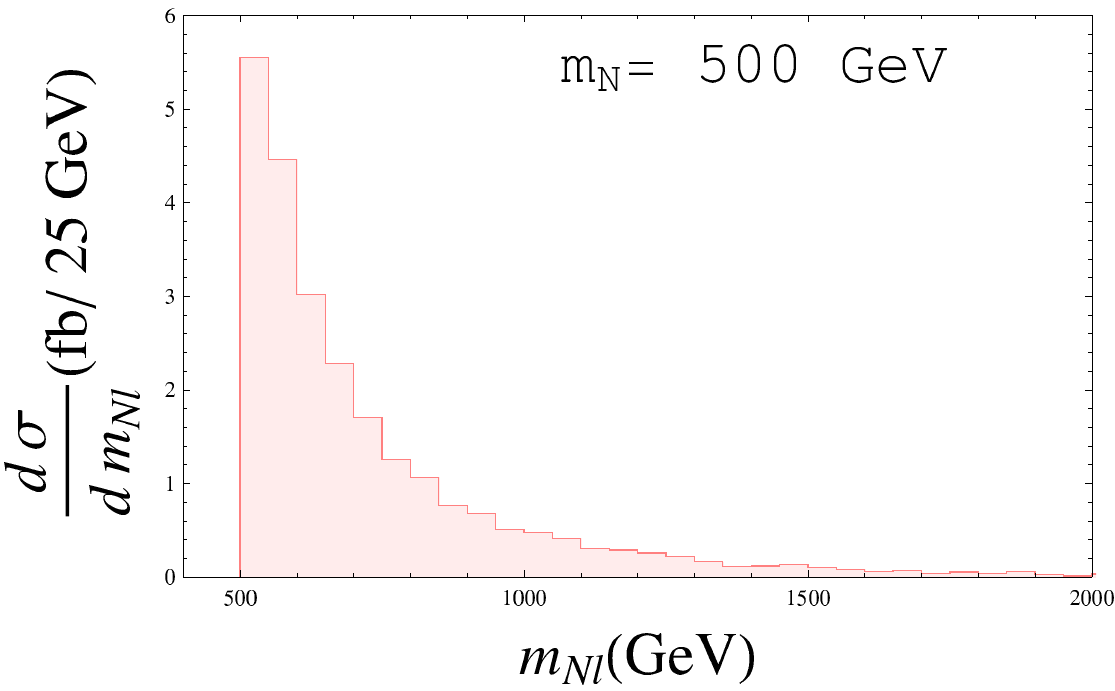}
\end{center}
\caption{The first and second rows show the $p_{T}$ and $\eta$ distributions of the jet produced in the $N\ell j$ final state from the photon mediated inelastic processes. The third row shows the final state invariant mass $\left(m_{N\ell}j\right)$ distributions. The left column stands for $m_{N}=100$ GeV whereas the right one is for $m_{N}=500$ GeV.}
\label{painelas2}
\end{figure}

\begin{figure}
\begin{center}
\includegraphics[scale=0.51]{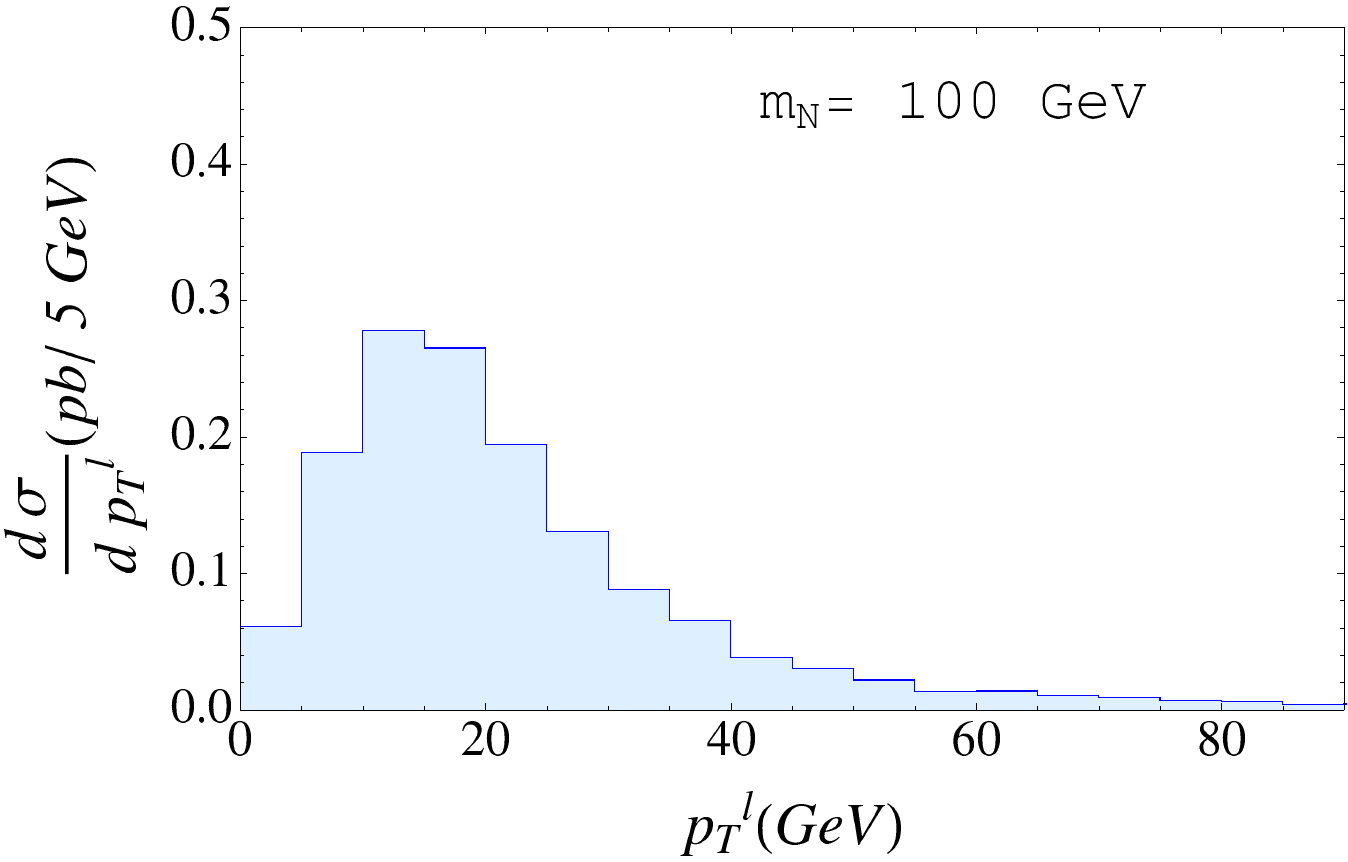}
\includegraphics[scale=0.49]{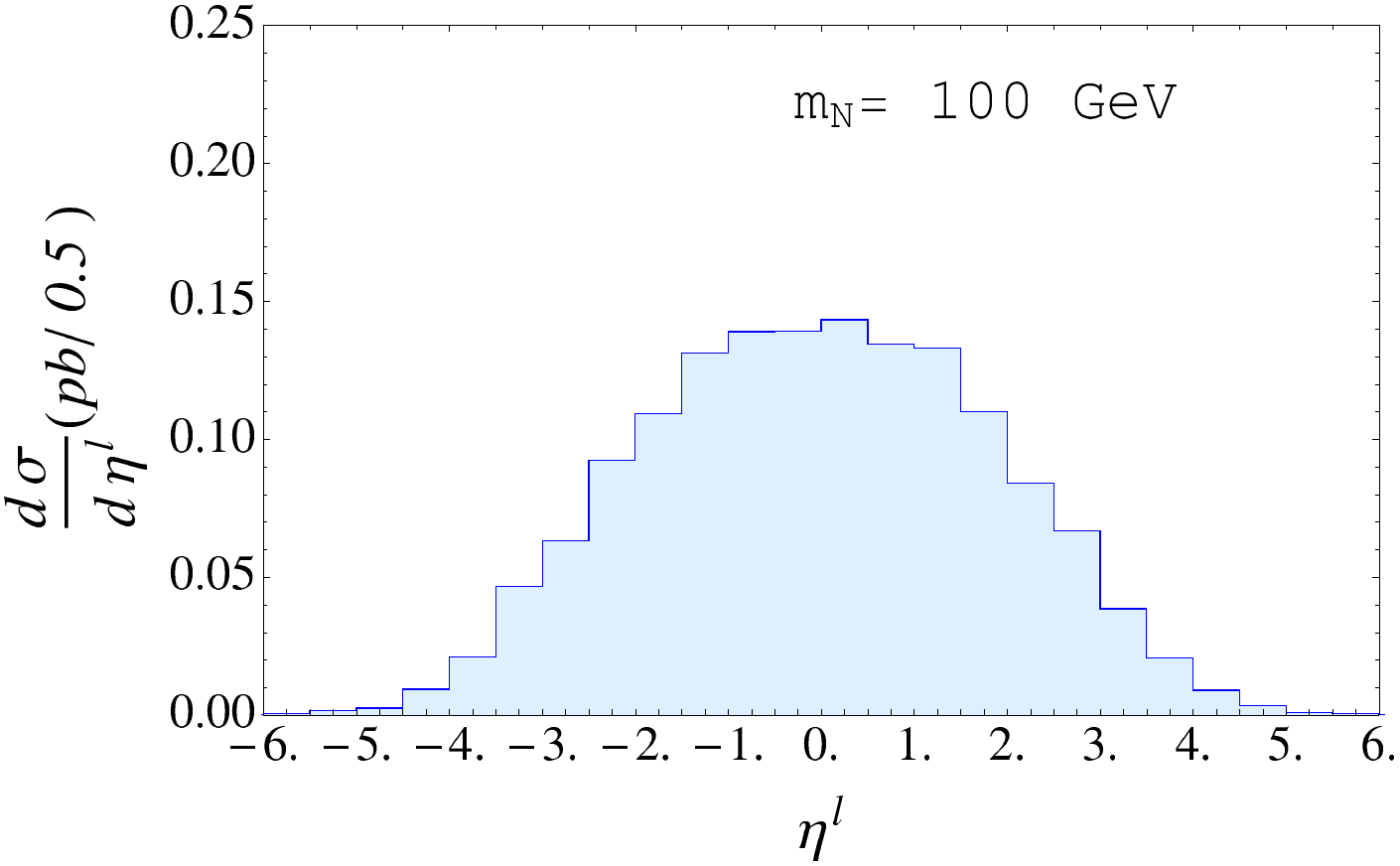}\\
\includegraphics[scale=0.61]{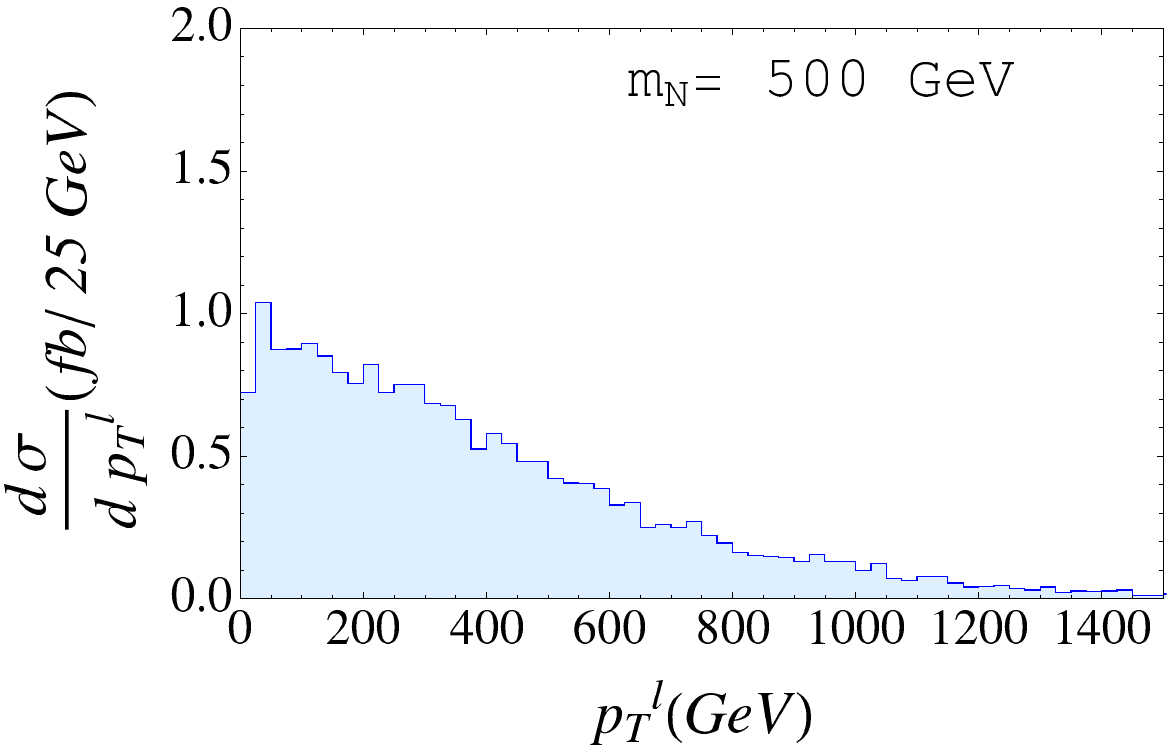}
\includegraphics[scale=0.60]{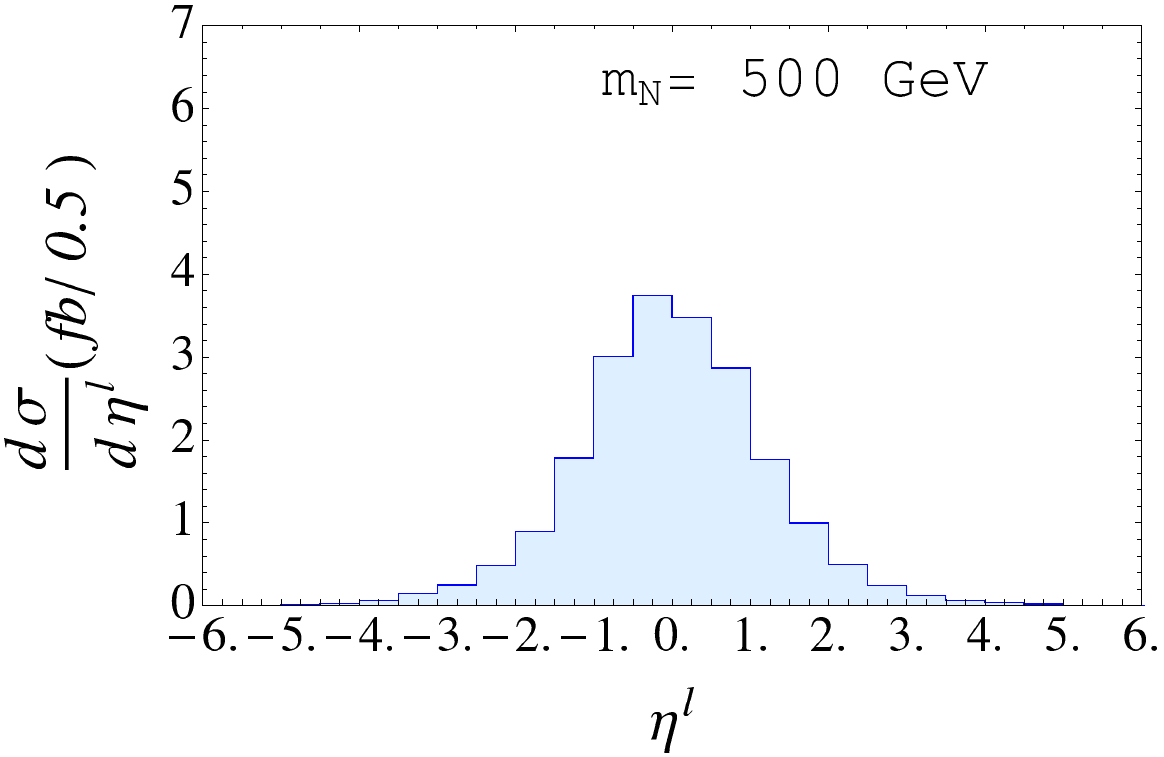}\\
\includegraphics[scale=0.54]{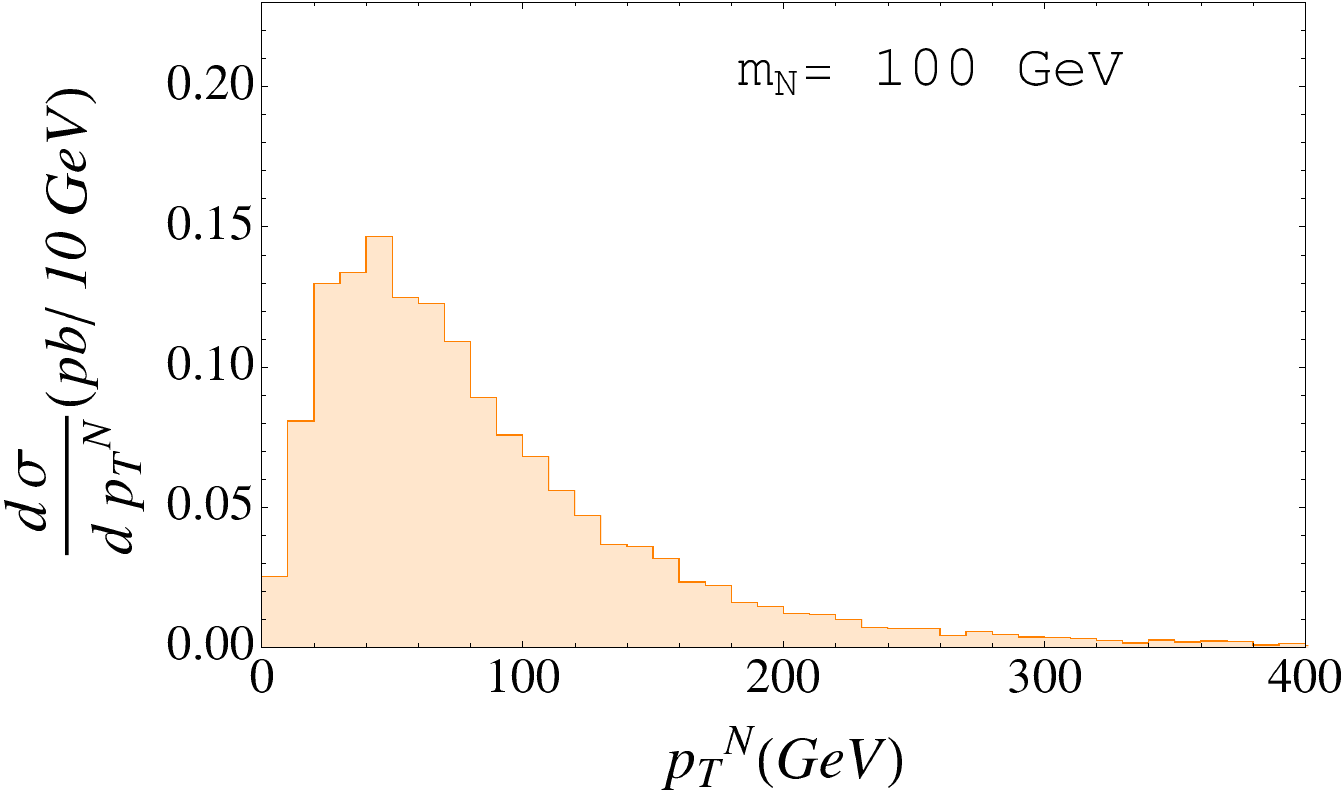}
\includegraphics[scale=0.64]{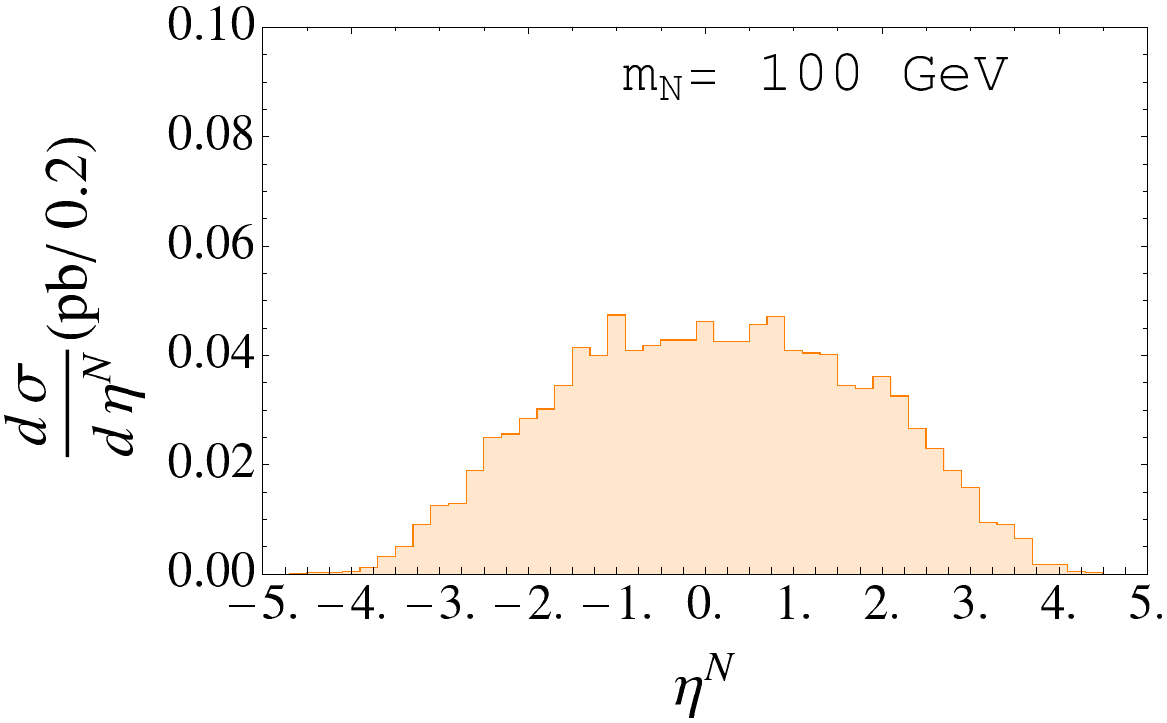}\\
\includegraphics[scale=0.53]{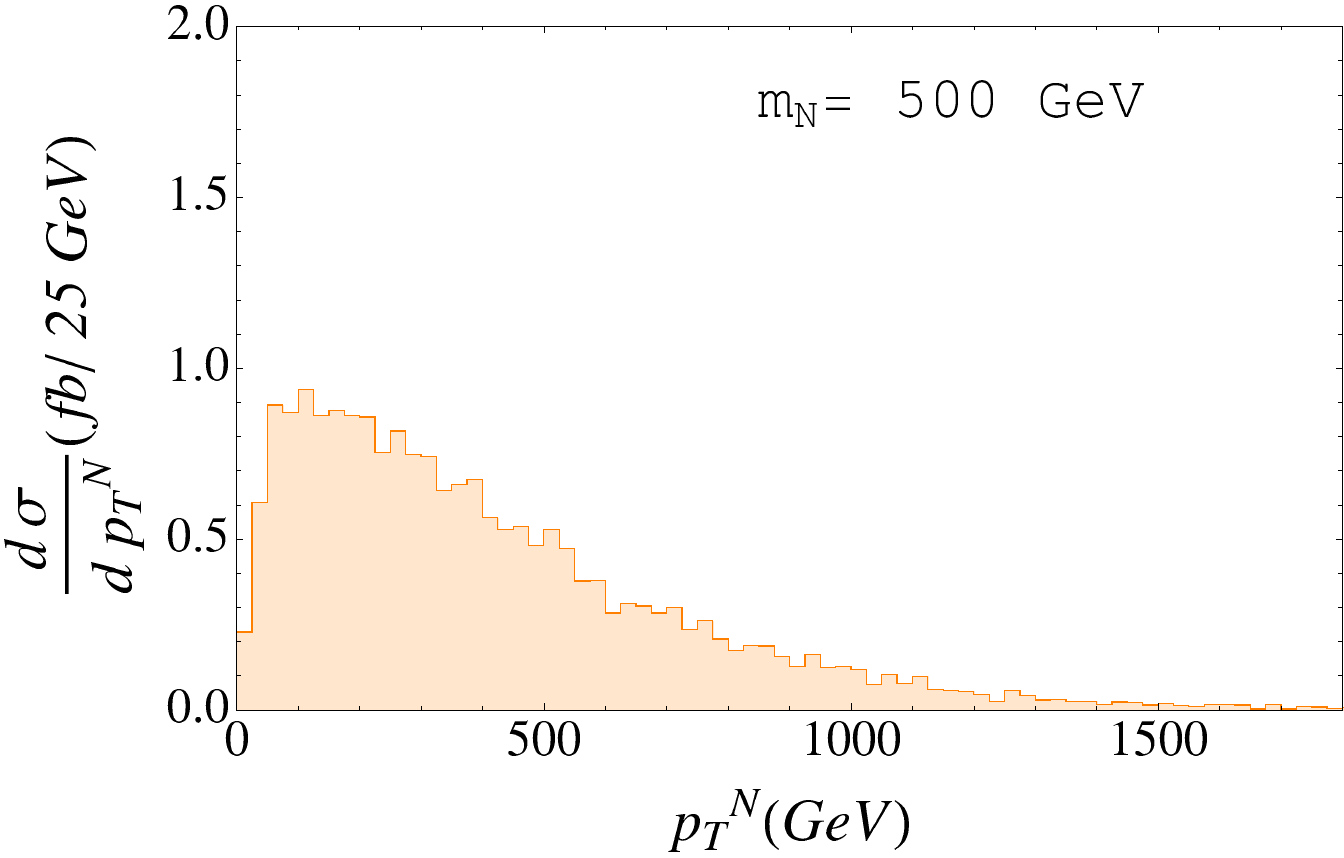}
\includegraphics[scale=0.47]{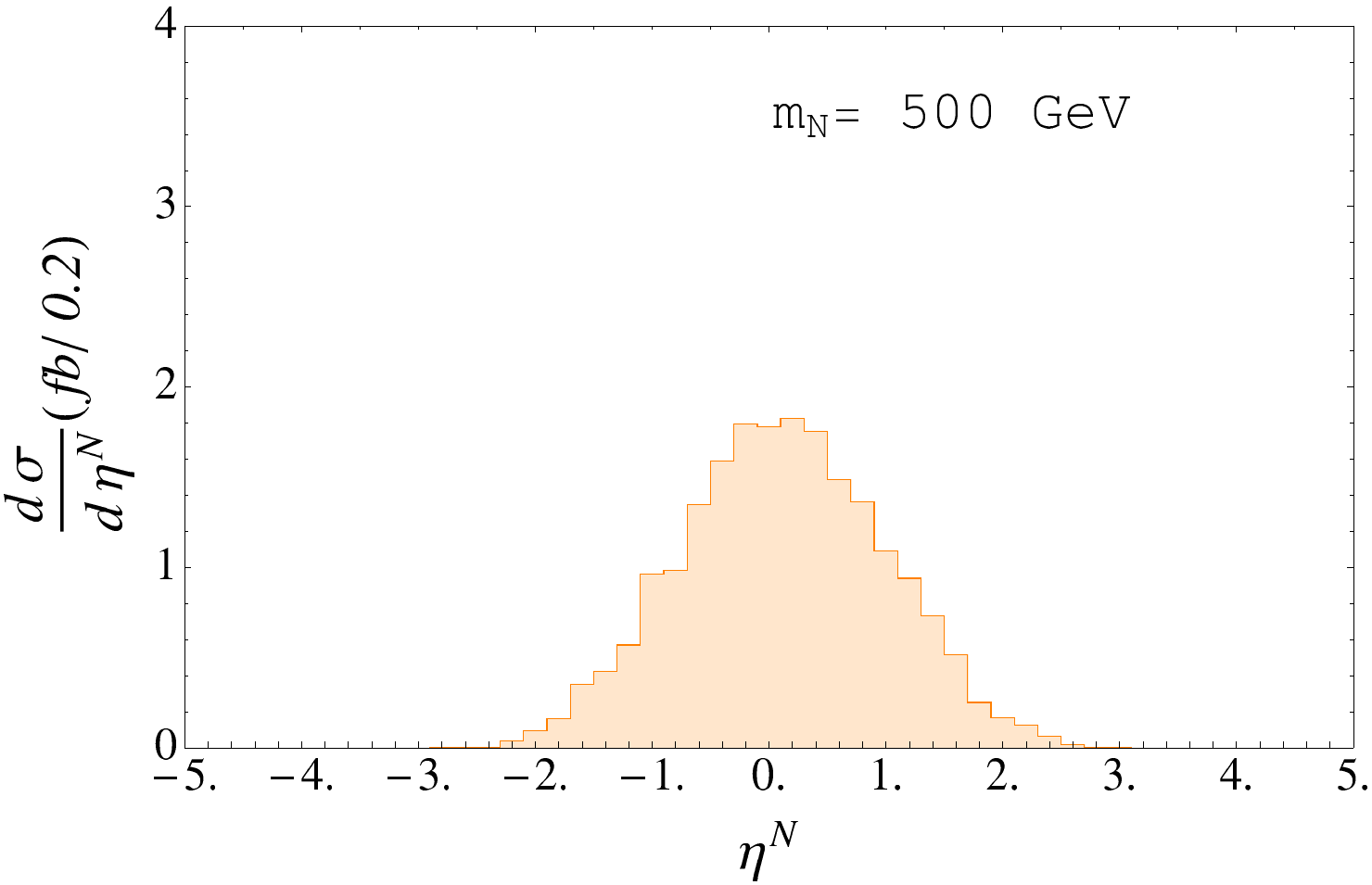}
\end{center}
\caption{The first and second rows show the $p_{T}$ and $\eta$ distributions of the lepton produced in the $N\ell jj$ final state from the photon mediated deep-inelastic processes. The third and fourth rows show the same for the heavy neutrino. The left column is for $m_{N}=100$ GeV whereas the right one stands for $m_{N}=500$ GeV.}
\label{ppDISj1}
\end{figure}

\begin{figure}
\begin{center}
\includegraphics[scale=0.61]{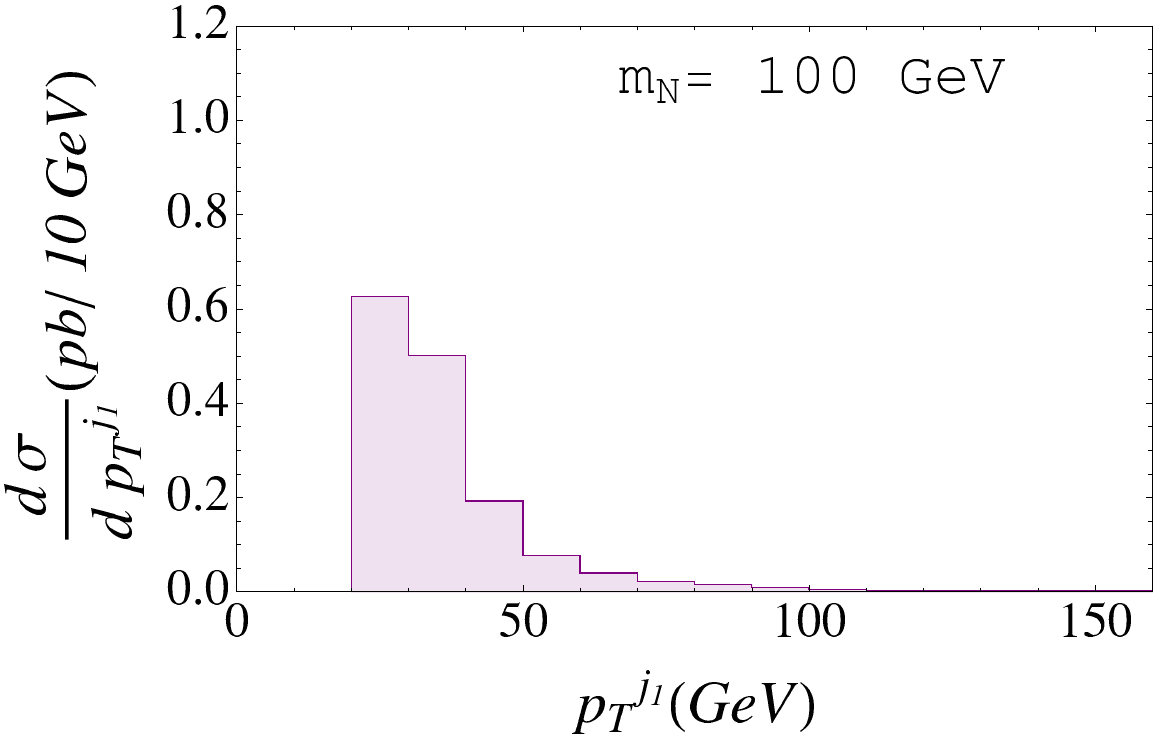}
\includegraphics[scale=0.63]{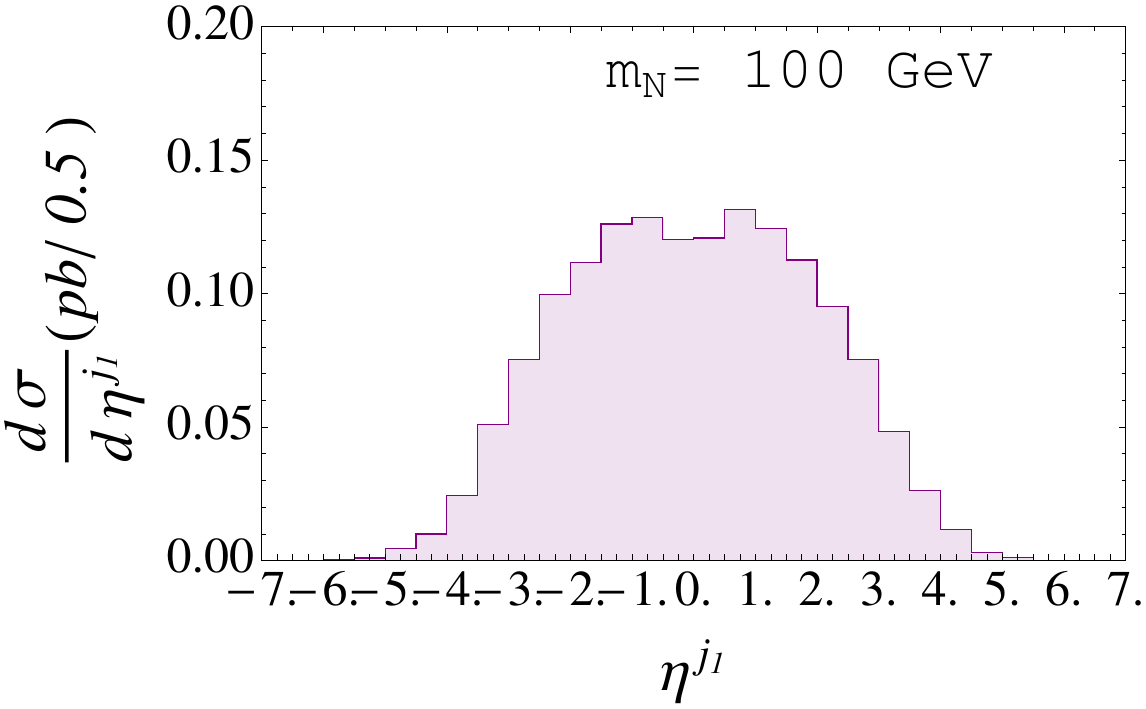}\\
\includegraphics[scale=0.57]{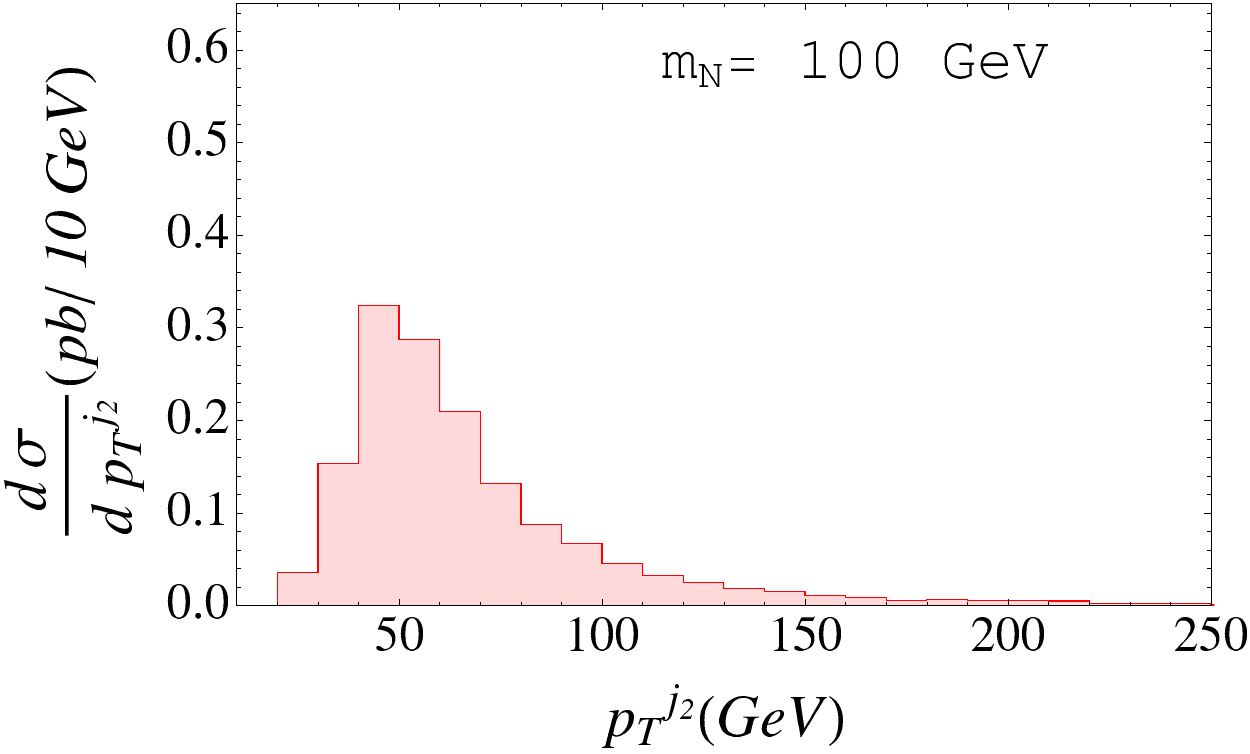}
\includegraphics[scale=0.55]{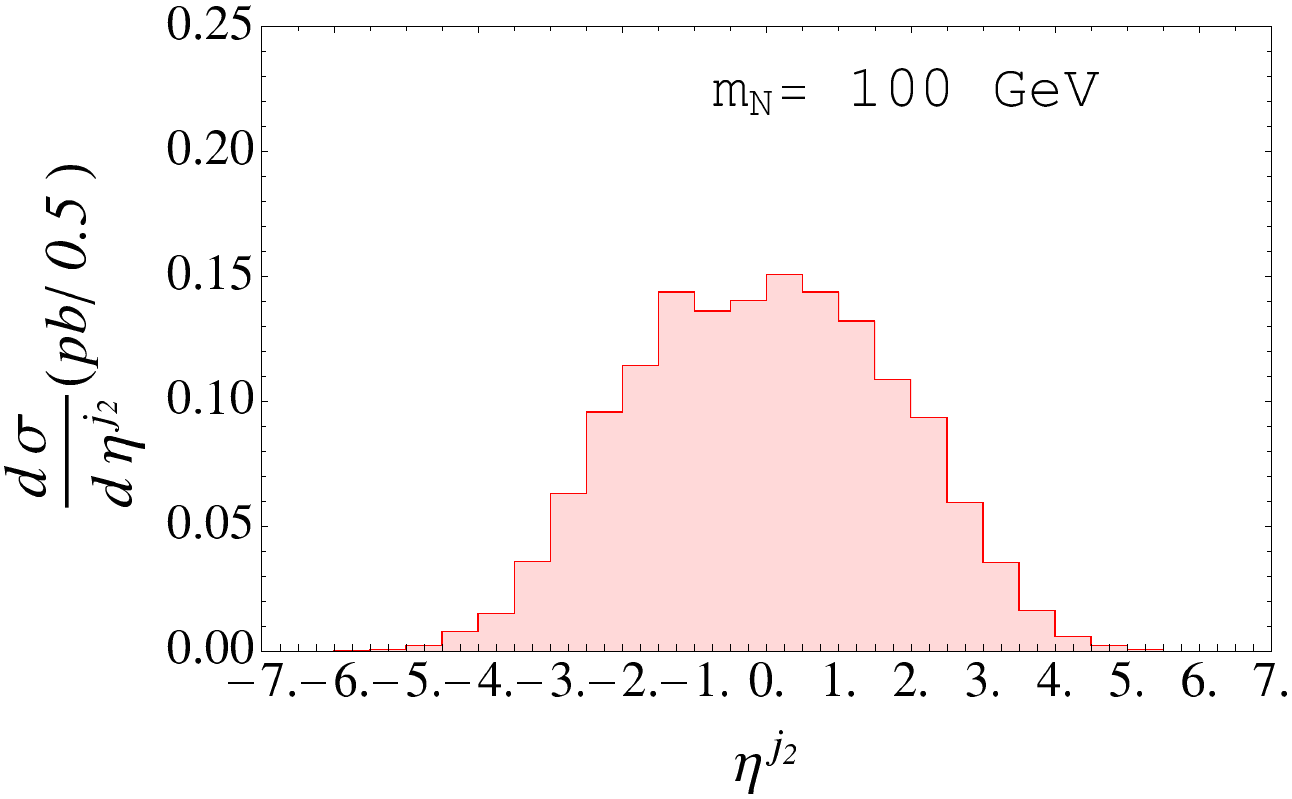}\\
\includegraphics[scale=0.55]{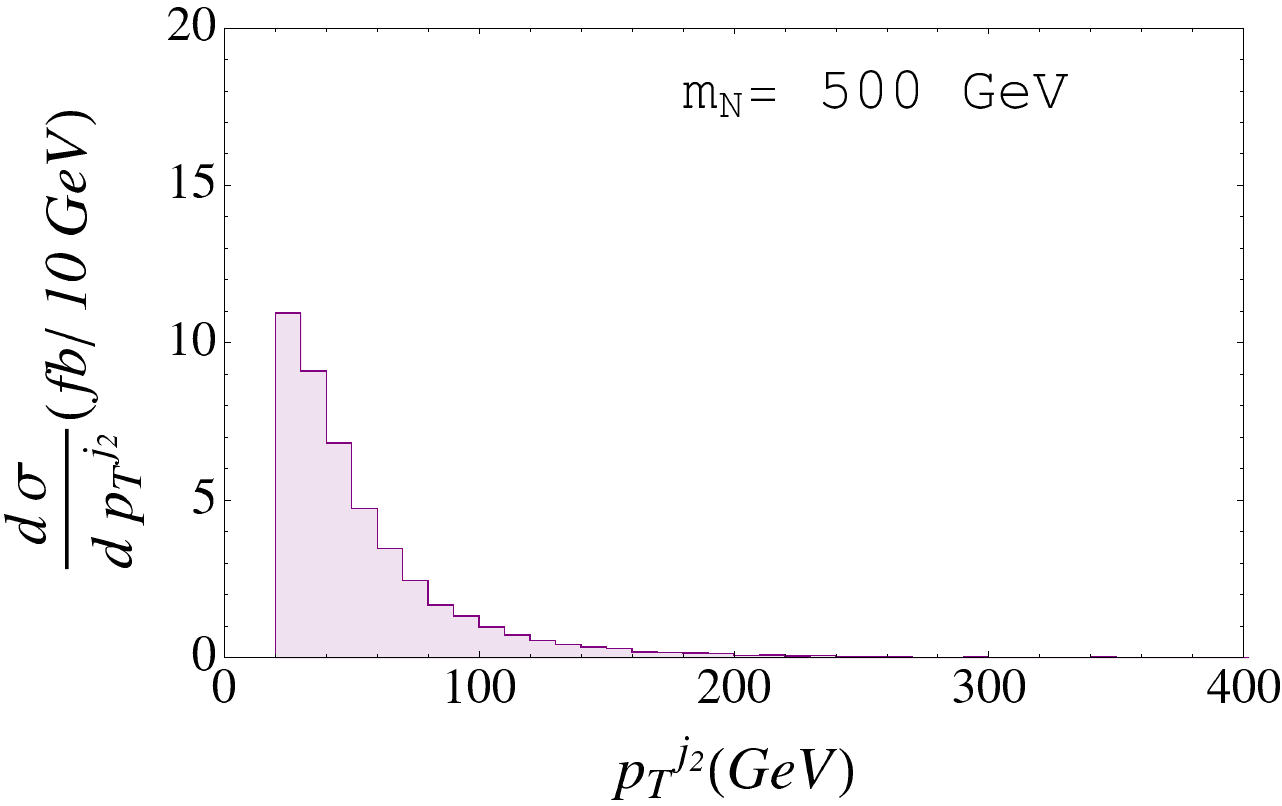}
\includegraphics[scale=0.48]{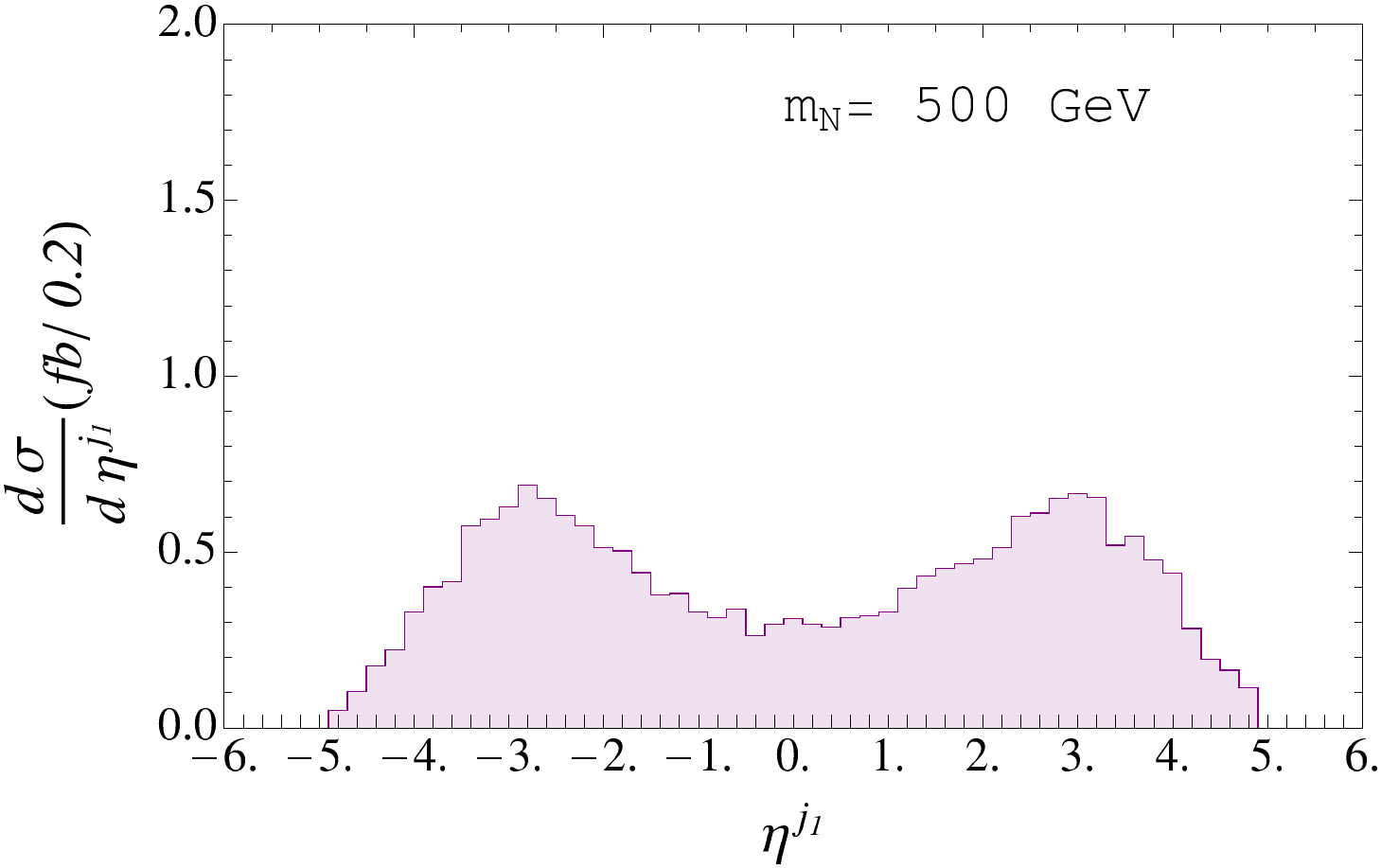}\\
\includegraphics[scale=0.53]{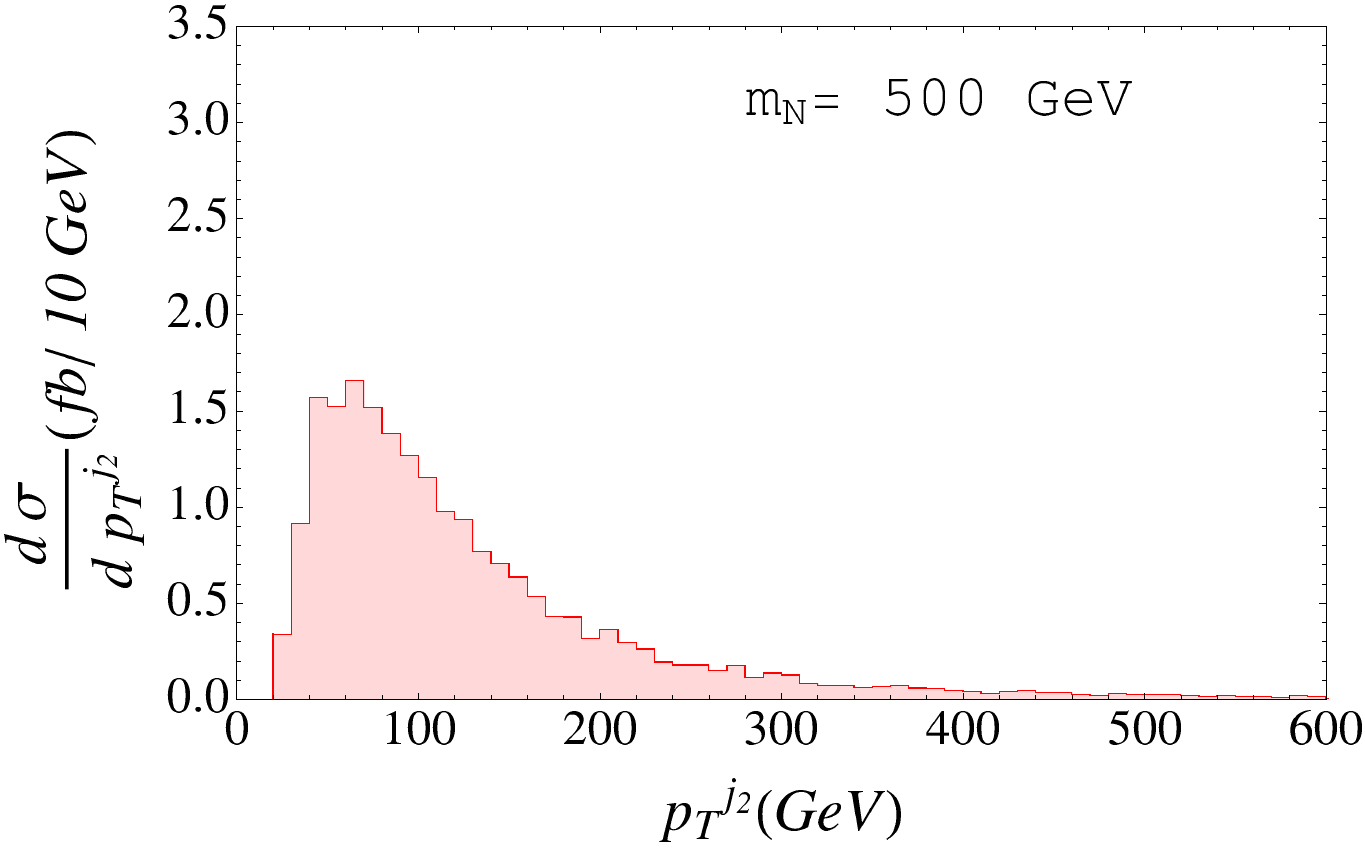}
\includegraphics[scale=0.45]{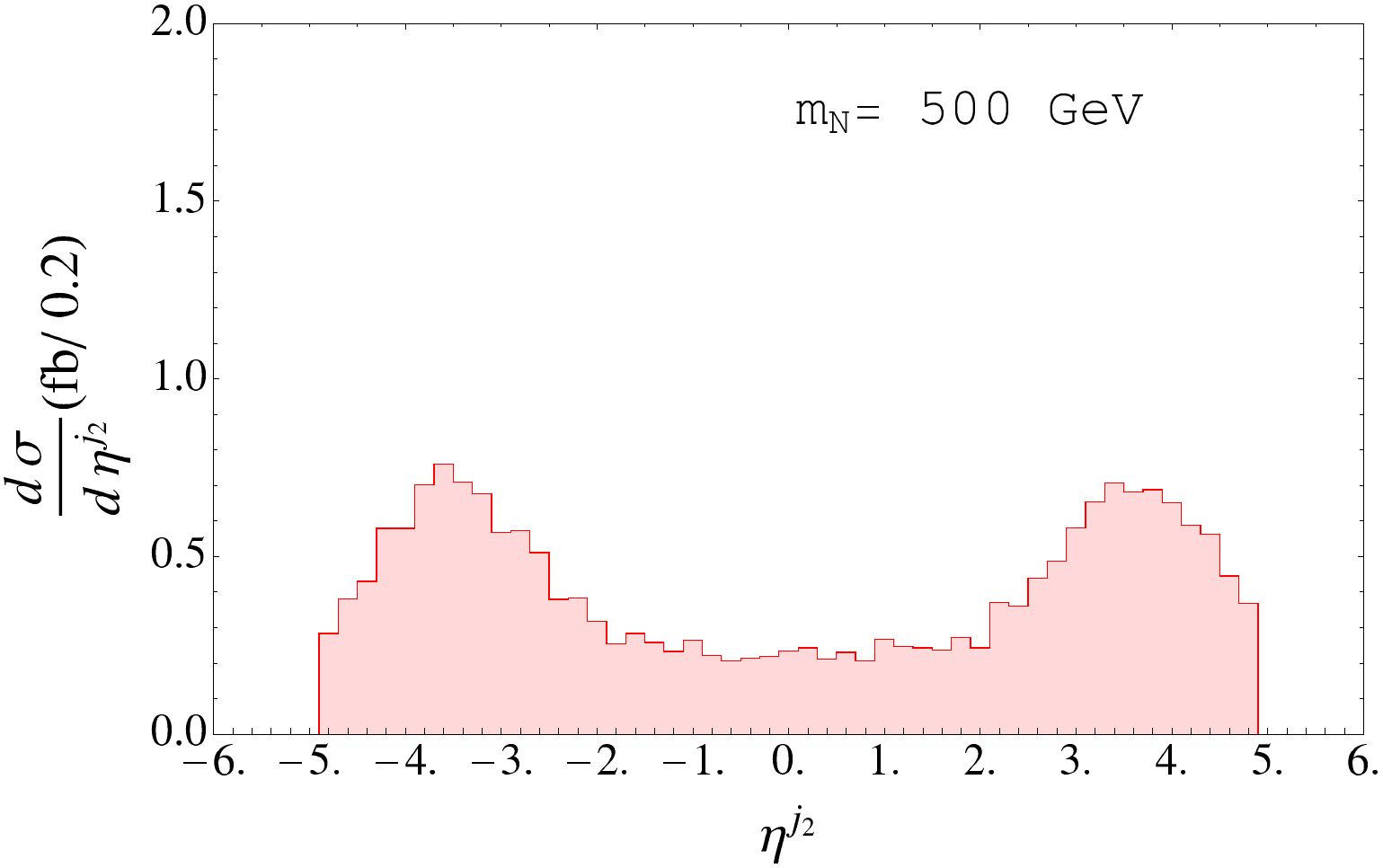}
\end{center}
\caption{The first and second rows show the $p_{T}$ and $\eta$ distributions of the non-leading jet produced in the $N\ell jj$ final state from photon mediated deep-inelastic processes. The third and fourth rows show the same for the leading jet. The left column is for $m_{N}=100$ GeV whereas the right one stands for $m_{N}=500$ GeV.}
\label{ppDISj2}
\end{figure}

\begin{figure}
\begin{center}
\includegraphics[scale=0.71]{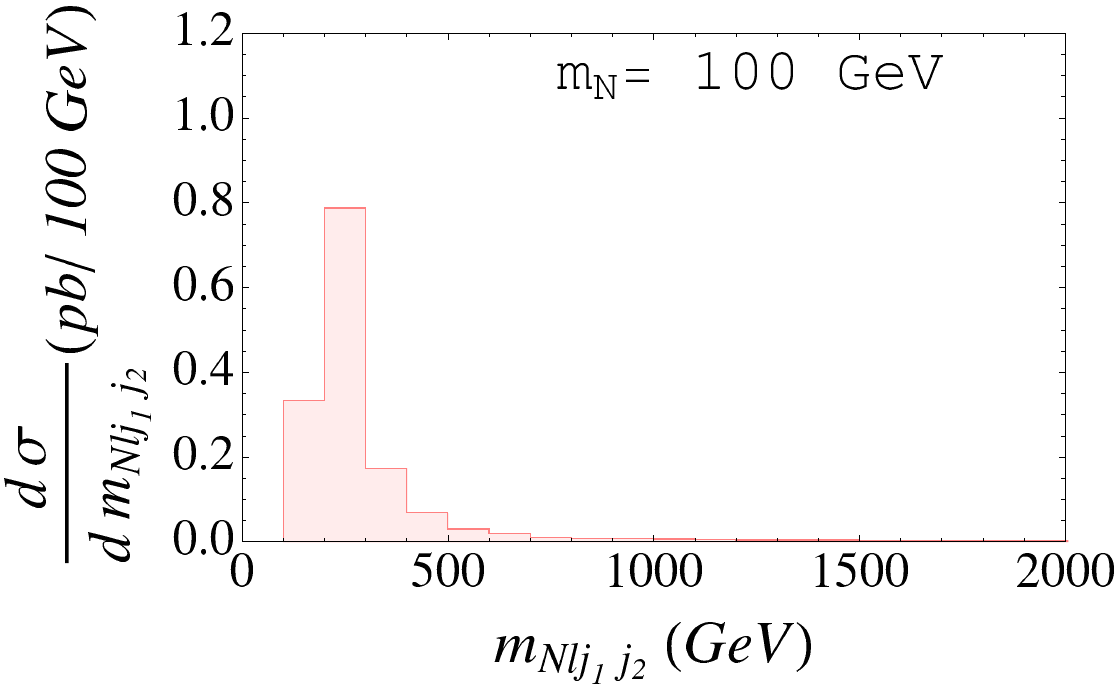}
\includegraphics[scale=0.65]{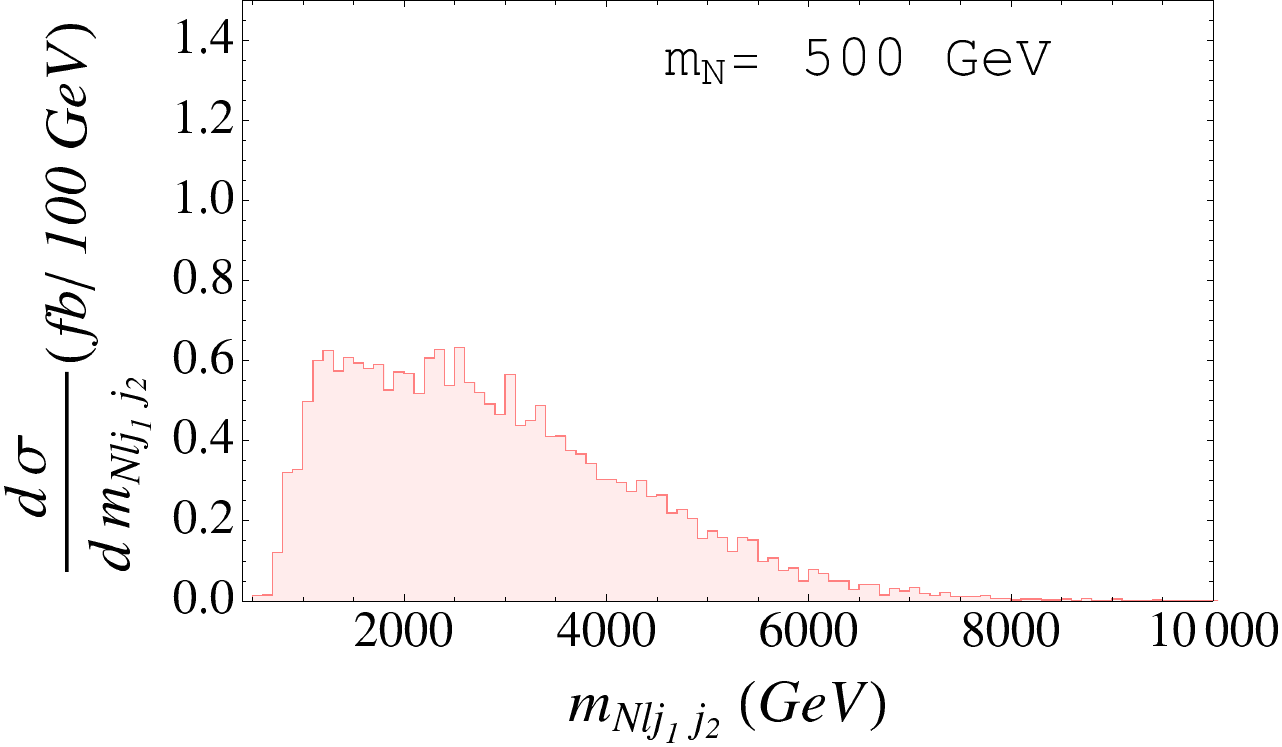}
\end{center}
\caption{Figure shows the final state invariant mass $\left(m_{N\ell j_{1}j_{2}}\right)$ distributions produced in the $N\ell jj$ final state from the photon mediated deep-inelastic processes. The left panel shows $m_{N}=100$ GeV whereas the right one is $m_{N}=500$ GeV.}
\label{ppDISj3}
\end{figure}

We also show the kinematic distributions from the photon mediated processes
  for the final states $N\ell j$ and $N\ell jj$. 
  The $N\ell j$ final state comes from the elastic and inelastic processes
  whereas the $N\ell jj$ comes from the deep inelastic process. 
Figs.~\ref{paelas1} and \ref{paelas2} show the kinematic distributions for the elastic process.
Corresponding results for the inelastic process are shown in Figs.~\ref{painelas1} and \ref{painelas2}.
Finally the results for the deep inelastic processes are shown in Figs~\ref{ppDISj1}-\ref{ppDISj3}.
The $\eta^{j_{1,2}}$ distributions show the double peaks in all three cases 
  which is a typical feature of the photon mediated processes \cite{ruiz}. 

\section{Data analysis and the upper bounds for the mixing angles}

The signal events at the parton level were generated by {\tt MadGraph5}. 
The showering and hadronization of the events were performed with {\tt PYTHIA6.4}~\cite{Pyth}
  and a fast detector simulation was done using {\tt DELPHES3}~\cite{DELPHES1, DELPHES2} bundled with {\tt MadGraph}. 
Hadrons were clustered into jets using the anti-$k_T$ algorithm as implemented in {\tt FastJet2}~\cite{fastjet1, fastjet2}
 with a distance parameter of 0.5. 
As our signal events are accompanied by $n$-jets, we incorporate the {\tt MLM} matching prescription 
  according to \cite{Matching}-\cite{Matching4} to avoid the double counting 
  of jets and use the matched cross section and 
  events for the analysis after the detector simulation. 
We consider two cases for the types of heavy neutrinos.
One is the Majorana case for which the signal process is
$pp\to N \ell_1^{\pm}/ N \ell_1^{\pm} j/ N \ell_1^{\pm} jj$, followed by $N\to \ell_2^{\pm} W^{\mp}$ and $W^{\mp} \to jj$. 
The other is the pseudo-Dirac case for which the signal process is
$pp\to N \ell_1^{+}/ N \ell_1^{+} j/ N l_1^{+} jj$, followed by $N\to \ell_2^{-} W^{+}$ and $W^{+} \to \ell_3^{+} \nu$ (
$pp\to \overline{N}\ell_1^{-}/ \overline{N} \ell_1^{-} j/ \overline{N} \ell_1^{-} jj$, followed by $\overline{N}\to \ell_2^{+} W^{-}$ and $W^{-} \to \ell_3^{-} \overline{\nu}$).

\subsection{\rm \textbf{Same-sign di-lepton plus di-jet signal}}
\begin{figure}
\begin{center}
\includegraphics[scale=0.9]{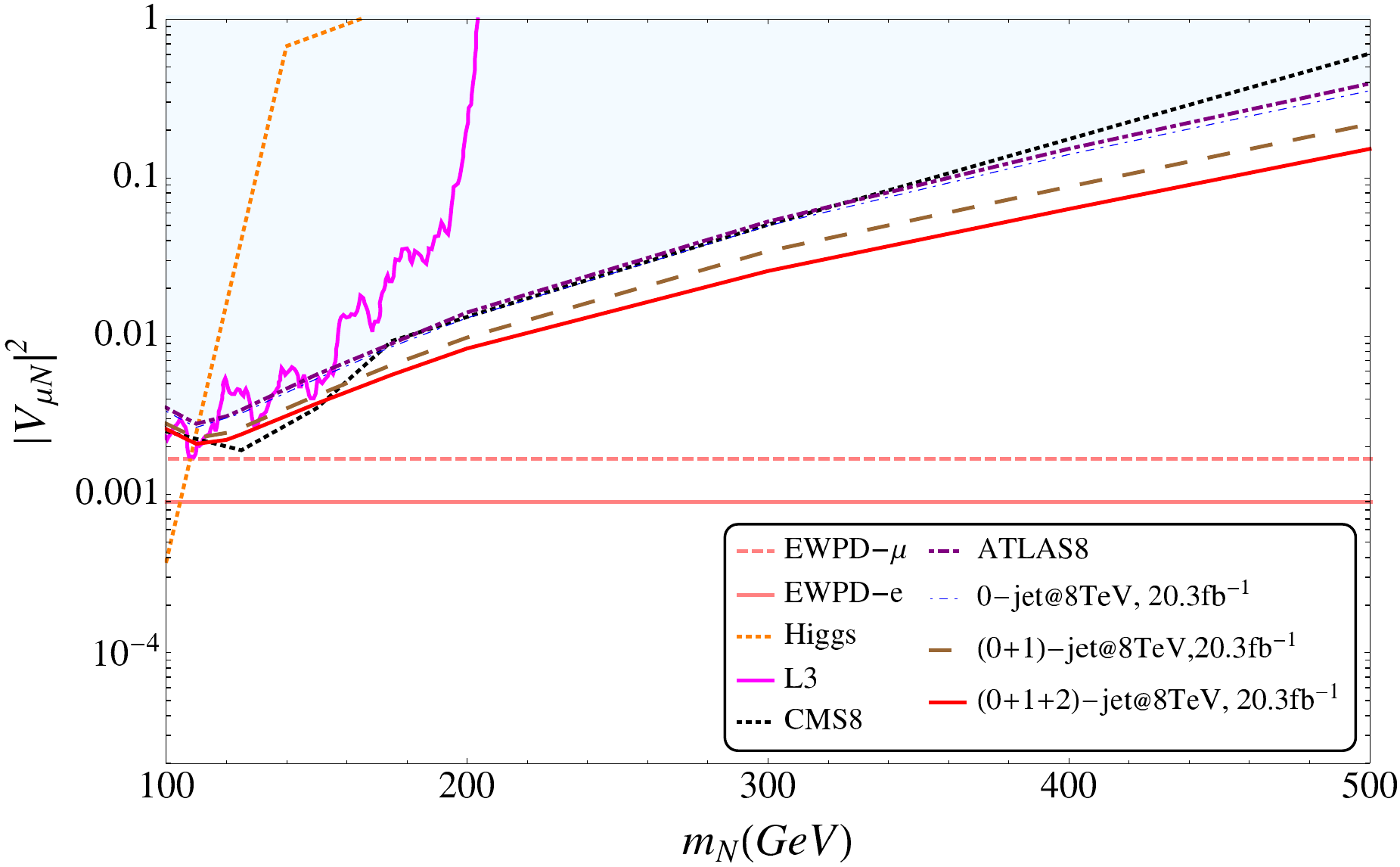}
\end{center}
\caption{Figure shows the upper bounds of square on the mixing angles as a function of $m_{N}$ using the ATLAS data at the 8 TeV \cite{ATLAS8}. The 0-jet (0-jet8, 20.3 fb$^{-1}$) result is compared with the ATLAS bounds (ATLAS8). The improved results with 1-jet ((dashed), (0+1)-jet@8TeV, 20.3 fb$^{-1}$) and 2-jet (thick, solid,(0+1+2)-jet@8, 20.3 fb$^{-1}$) are shown with respect to the bounds obtained by ATLAS \cite{ATLAS8}. These bounds are compared to (i) the $\chi^{2}$-fit to the LHC Higgs data \cite{LHCHiggs} (Higgs), (ii) from a direct search at LEP \cite{L3}(L3), valid only for the electron flavor, (iv) CMS limits from $\sqrt{s}=$8 TeV LHC data \cite{CMS8} (CMS8), for a heavy Majorana neutrino of the muon flavor and (v) indirect limit from a global fit to the electroweak precision data \cite{EWPD1, EWPD2, EWPD3} (EWPD), for both electron (solid, EWPD- $e$) and muon (dotted, EWPD- $\mu$) flavors. The shaded region is excluded.}
\label{figATLAS1}
\end{figure}

\begin{figure}
\begin{center}
\includegraphics[scale=0.85]{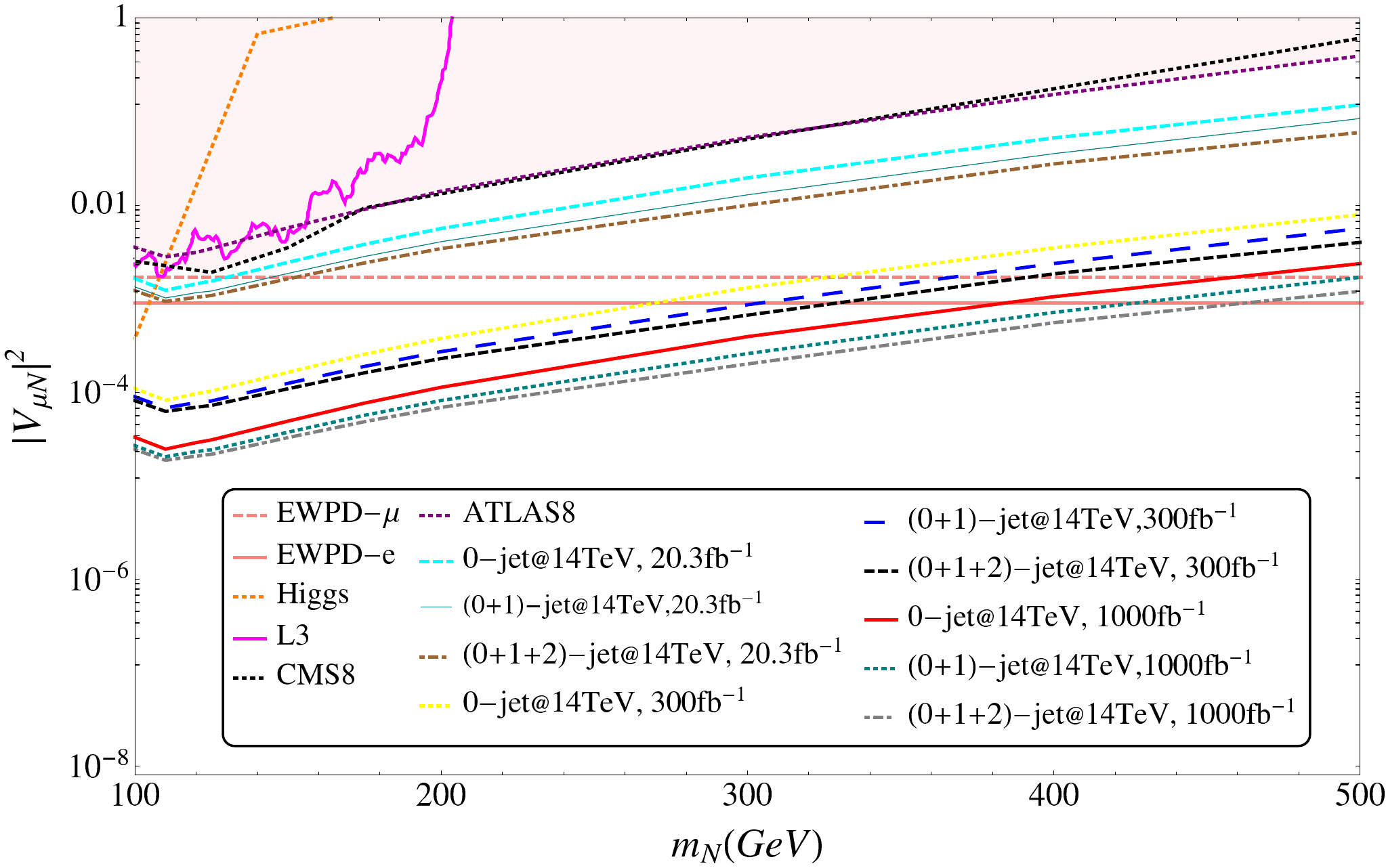}
\end{center}
\caption{Figure shows the search reach for the square of the mixing angles as a function of $m_{N}$ at 14 TeV LHC using the ATLAS data at the 8 TeV \cite{ATLAS8}. In this plot we give a prospective search reach for the 14 TeV. The results with 0-jet (0-jet@14TeV, 20.3 fb$^{-1}$), 1-jet ((0+1)-jet@14TeV, 20.3 fb$^{-1}$) and 2-jet ((0+1+2)-jet@14, 20.3 fb$^{-1}$) are shown at 20.3 fb $^{-1}$. The prospective results with 0-jet(0-jet@14TeV, 300 fb$^{-1}$), 1-jet ((0+ 1)-jet@14TeV, 300 fb$^{-1}$) and 2-jet ((0+1+2)-jet@14TeV, 300 fb$^{-1}$) are also plotted at 300 fb$^{-1}$ luminosity. The prospective results with 0-jet (0-jet@14teV, 1000 fb$^{-1}$), 1-jet ((0+ 1)-jet@14TeV, 1000fb$^{-1}$) and 2-jet ((0+1+2)-jet@14TeV, 1000 fb$^{-1}$) are plotted at 1000 fb$^{-1}$ and compared to (i) the $\chi^{2}$-fit to the LHC Higgs data \cite{LHCHiggs} (Higgs), (ii) from a direct search at LEP \cite{L3}(L3), valid only for the electron flavor, (iv) CMS limits from $\sqrt{s}=$8 TeV LHC data \cite{CMS8} (CMS8), for a heavy Majorana neutrino of the muon flavor and (v) indirect limit from a global fit to the electroweak precision data \cite{EWPD1, EWPD2, EWPD3} (EWPD), for both electron (solid, EWPD- $e$) and muon (dotted, EWPD- $\mu$) flavors. The shaded region is excluded.}
\label{figATLAS2}
\end{figure}

For simplicity we consider the case that only one generation of the heavy neutrino
  is light and accessible to the LHC which couples to only the second generation of the lepton flavor. 
To generate the events in the {\tt MadGraph} we use the {\tt NN23LO1 PDF} \cite{NNPDF1}-\cite{NNPDF4}
with {\tt Xqcut}$=p_{T}^{j}=20$ GeV and {\tt QCUT}$=25$ GeV.
We calculate the cross sections for the processes $\sigma_{0-j}=\sigma(pp\to N \mu^{\pm} \to\mu^{\pm} \mu^{\pm} jj)$, 
  $\sigma_{1-j}=\sigma(pp\to N \mu^{\pm} j \to\mu^{\pm} j \mu^{\pm} jj)$
  and $\sigma_{2-j}=\sigma(pp\to N \mu^{\pm} jj\to\mu^{\pm}jj \mu^{\pm} jj)$ as functions of $m_{N}$.
Comparing our generated events with the recent ATLAS results \cite{ATLAS8} at the $8$ TeV LHC with the luminosity $20.3$ fb$^{-1}$,
  we obtain an upper limit on the mixing angles between the Majorana type 
  heavy neutrino and the SM leptons as a function of $m_{N}$.
In the ATLAS analysis the upper bound of the production cross section ($\sigma^{ATLAS}$) is obtained for the final state with
  the same-sign di-muon plus di-jet as a function of $m_{N}$.
Using these cross sections we obtain the upper bounds on the mixing angles as follows
\bea
|V_{\ell N}|^{2}_{0-j} \lesssim \frac{\sigma^{ATLAS}}{\sigma_{0-j}},
\label{mixqq}
\eea
\bea
|V_{\ell N}|^{2}_{1-j} \lesssim \frac{\sigma^{ATLAS}}{\sigma_{0-j}+ \sigma_{1-j}},
\label{mix1j}
\eea
and 
\bea
|V_{\ell N}|^{2}_{2-j} \lesssim \frac{\sigma^{ATLAS}}{\sigma_{0-j}+ \sigma_{1-j}+  \sigma_{2-j}}.
\label{mix2j}
\eea
Our resultant upper bounds on the mixing angles are shown in Fig.~\ref{figATLAS1}, along with the bounds from
  ATLAS \cite{ATLAS8}, CMS \cite{CMS8}, LEP (L3) \cite{L3}, electroweak precision data for electron (EWPD-e)
  and muon (EWPD-$\mu$) \cite{EWPD1}-\cite{EWPD3}, and finally LHC Higgs data (Higgs) \cite{LHCHiggs}.
The bound obtained from $\sigma_{0-j}$ is consistent with the bound presented in \cite{ATLAS8}. 
We can see that a significant improvement on the bounds by adding the $\sigma_{1-j}$ and $\sigma_{2-j}$ cross sections.
We have also calculated the cross sections, $\sigma_{0-j}$, $\sigma_{1-j}$ and $\sigma_{2-j}$, for the 14 TeV LHC. 
Applying the ATLAS bound at 8 TeV with $20.3$ fb$^{-1}$ luminosity, 
  we put a prospective upper bound on the mixing angles.
We scaled the bound for the luminosities of $300$ fb$^{-1}$ and  $1000$ fb$^{-1}$. 
The results are shown in Fig~\ref{figATLAS2}. 

\begin{figure}

\begin{center}
\includegraphics[scale=0.95]{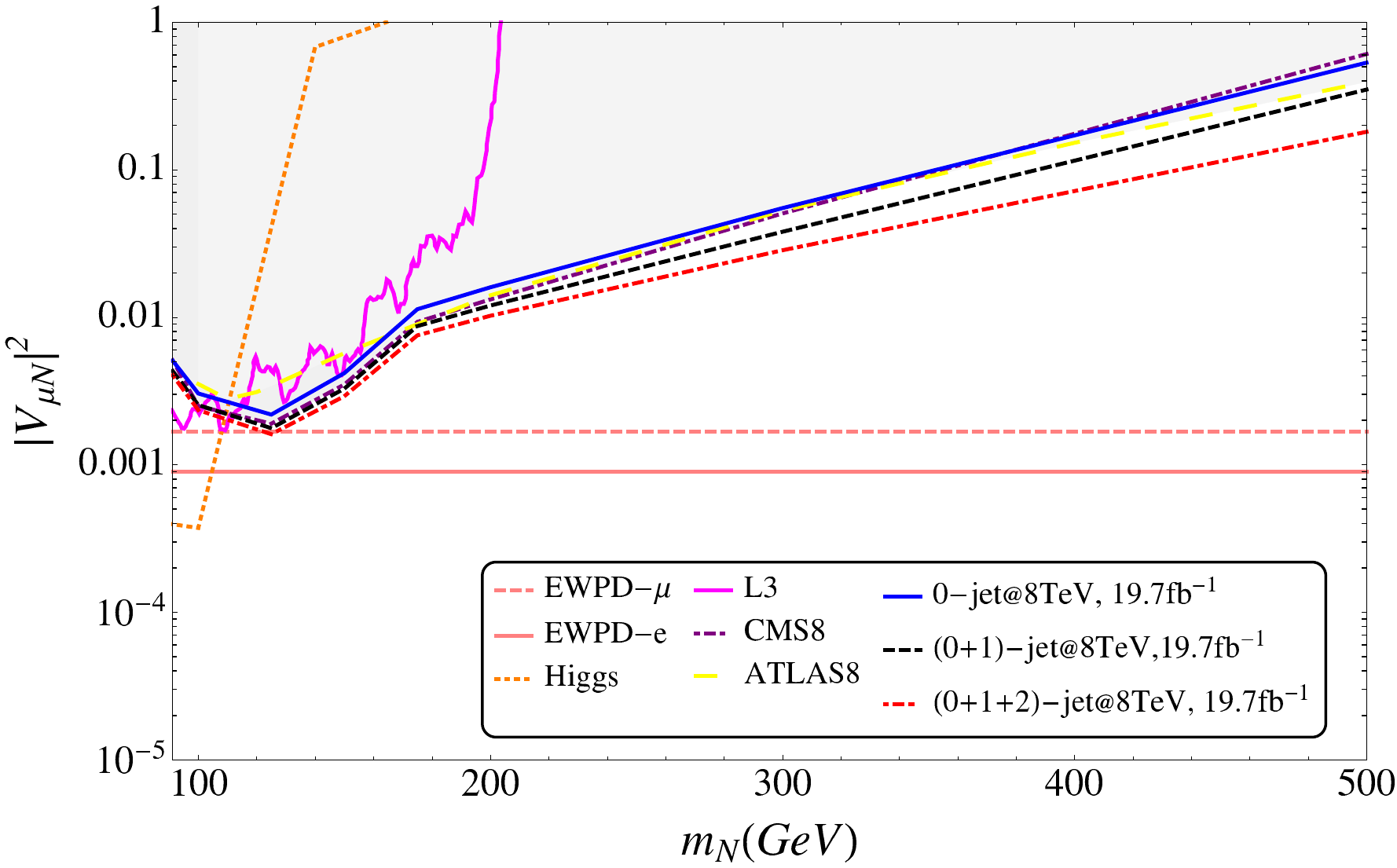}
\end{center}
\caption{Figure shows the upper bounds on square of the mixing angles as a function of $m_{N}$ using the CMS data at the 8 TeV \cite{CMS8}. The 0-jet (0-jet@8TeV, 19.7 fb$^{-1}$) result is compared with the CMS bounds (CMS8). The improved results with 1-jet ((0+ 1)- jet@8TeV, 19.7 fb$^{-1}$) and 2-jet ((0+ 1+ 2)-jet@8TeV, 19.7 fb$^{-1}$) are shown with respect to the CMS data from \cite{CMS8}. The bounds are compared to (i) the $\chi^{2}$-fit to the LHC Higgs data \cite{LHCHiggs} (Higgs), (ii) from a direct search at LEP \cite{L3}(L3), valid only for the electron flavor, (iv) ATLAS limits from $\sqrt{s}=$8 TeV LHC data \cite{ATLAS8} (ATLAS 8), for a heavy Majorana neutrino of the muon flavor and (v) indirect limit from a global fit to the electroweak precision data \cite{EWPD1, EWPD2, EWPD3} (EWPD), for both electron (solid, EWPD- $e$) and muon (dotted, EWPD- $\mu$) flavors. The shaded region is excluded.}
\label{figCMSdilep1}
\end{figure}

\begin{figure}
\begin{center}
\includegraphics[scale=0.85]{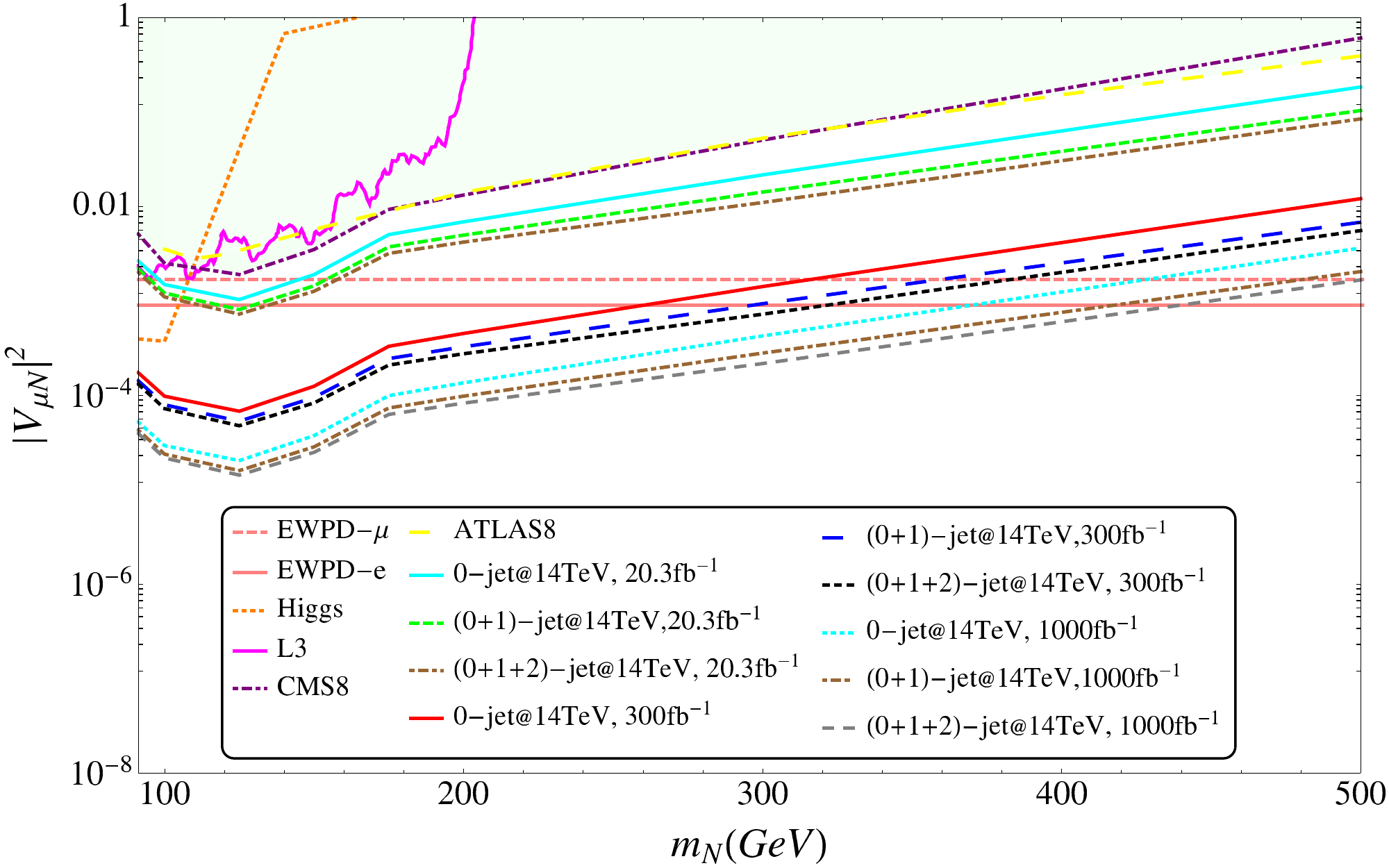}
\end{center}
\caption{Figure shows the prospective search reach for square of the mixing angles as a function of $m_{N}$ at the 14 TeV LHC using the CMS data at the 8 TeV \cite{CMS8}.
The 0-jet (0-jet@14TeV, 20.3 fb$^{-1}$), 1-jet ((0+1)-jet@14TeV, 20.3 fb$^{-1}$)and 2-jet ((0+1+2)-jet@14TeV, 20.3 fb$^{-1}$) are shown at the 20.3 fb$^{-1}$. 
The prospective results with 0-jet(0-jet@14TeV, 300 fb$^{-1}$), 1-jet ((0+ 1)-jet@14TeV, 300 fb$^{-1}$) and 2-jet ((0+1+2)-jet@14TeV, 300 fb$^{-1}$) are also plotted at 300 fb$^{-1}$ luminosity. The prospective results with 0-jet (0-jet@14TeV, 1000 fb$^{-1}$), 1-jet ((0+ 1)-jet@14TeV, 1000fb$^{-1}$) and 2-jet ((0+1+2)-jet@14TeV, 1000 fb$^{-1}$) are plotted at 1000 fb$^{-1}$ and compared to (i) the $\chi^{2}$-fit to the LHC Higgs data \cite{LHCHiggs} (Higgs), (ii) from a direct search at LEP \cite{L3}(L3), valid only for the electron flavor, (iv) CMS limits from $\sqrt{s}=$8 TeV LHC data \cite{CMS8} (CMS8), for a heavy Majorana neutrino of the muon flavor and (v) indirect limit from a global fit to the electroweak precision data \cite{EWPD1, EWPD2, EWPD3} (EWPD), for both electron (solid, EWPD- $e$) and muon (dotted, EWPD- $\mu$) flavors. The shaded region is excluded.}
\label{figCMSdilep2}
\end{figure}

Recently the CMS has performed the same-sign di-lepton plus di-jet search \cite{CMS8}.
Using this reslt and adopting the same procedure for the ATLAS result we calculate 
  the upper bound on the mixing angles at the 8 TeV LHC.
The results are shown in Fig.~\ref{figCMSdilep1}.
A significant improvement for the upper bound on the mixing angle is obtained by adding the
   $\sigma_{1-j}$ and $\sigma_{2-j}$ results to the  $\sigma_{0-j}$ case.
We can see that at the low mass region the bound obtained by the CMS analysis is comparable to our 
  $0$-jet plus $1$-jet case and in the high mass region that is comparable to our 0-jet case.
Using the 8 TeV CMS result \cite{CMS8} we obtain prospective bound at the 14 TeV for the luminosities 
  of  $20.3$ fb$^{-1}$, $300$ fb$^{-1}$ and  $1000$ fb$^{-1}$.
The results are shown in Fig.~\ref{figCMSdilep2}.

\subsection{\rm \textbf{Tri-lepton signal}}

In this analysis we consider two cases. 
One is the Flavor Diagonal case (FD), where
  we introduce three generations of the degenerate heavy neutrinos 
  and each generation couples with the single, corresponding lepton flavor.
The other one is the Single Flavor case (SF) where only one heavy neutrino
is light and accessible to the LHC which couples to 
only the first or second generation of the lepton flavor.
In this analysis we use the {\tt CTEQ6L PDF} \cite{CTEQ}.
We have calculated the bound on the mixing angle for the tri-lepton case
  using the $N\ell j$ final state in \cite{DDO1}.
In the following we extend this analysis by adding the photon mediated processes
  to the $N\ell j$ final state. 
We also include the $N\ell jj$ final state.

We generate the 1-jet and the 2-jet parton level processes separately in {\tt MadGraph} and 
  then gradually hadronize and perform detector simulations with {\tt Xqcut}$=p_{T}^{j}=30$ GeV and 
  {\tt QCUT}$=36$ GeV for the hadronization. 
In our analysis we use the matched cross section after the detector level analysis. 
After the signal events are generated we adopt the following basic criteria, 
  used in the CMS tri-lepton analysis \cite{CMS-trilep}: 
\begin{itemize}
\item [(i)] The transverse momentum of each lepton: $p^\ell_T > 10$ GeV.
\item [(ii)] The transverse momentum of at least one lepton: $p^{\ell,{\rm leading}}_{T} > 20$ GeV.
\item [(iii)]  The jet transverse momentum: $p_T^j > 30$ GeV. 
\item [(iv)] The pseudo-rapidity of leptons: $|\eta^\ell| < 2.4$ and of jets: $|\eta^j| < 2.5$.
\item [(v)] The lepton-lepton separation: $\Delta R_{\ell \ell} > 0.1$ and the lepton-jet separation: $\Delta R_{\ell j} > 0.3$. 
\item [(vi)] The invariant mass of each OSSF (opposite-sign same flavor) lepton pair: $m_{\ell^+ \ell^-}< 75$  GeV or $> 105$ GeV to avoid the on-$Z$ region which was excluded from the CMS search. Events with $m_{\ell^+ \ell^-}< 12$ GeV are rejected to eliminate background from low-mass Drell-Yan processes and hadronic decays. 
\item [(vii)]  The scalar sum of the jet transverse momenta: $H_{T} < 200$ GeV. 
\item [(viii)] The missing transverse energy: $\slashed{E}_{T} < 50$ GeV. 
\end{itemize}
The additional tri-lepton contributions come from $N \rightarrow Z \nu, h \nu$, followed by the $Z$, $h$ decays into a pair of OSSF leptons. 
However, the $Z$ contribution is rejected after the implementation of the invariant mass cut for the OSSF leptons to suppress the SM background and 
  the $h$ contribution is suppressed due to very small Yukawa couplings of electrons and muons. 
Using different values of $\slashed{E}_{T}$ and $H_{T}$, the CMS analysis provides different number of observed events and 
  the corresponding SM background expectations. 
For our analysis the set of cuts listed above are the most efficient ones as implemented by the CMS analysis \cite{CMS-trilep}.
To derive the limits on $|V_{\ell N}|^{2}$, we calculate the signal cross section normalized by the 
  square of the mixing angle as a function of the heavy neutrino mass $m_{N}$
  for both SF and FD cases, by imposing the CMS selection criteria listed above. 
The corresponding number of signal events passing all the cuts 
  is then compared with the observed number of events at the 19.5 fb$^{-1}$ luminosity \cite{CMS-trilep}.
For the selection criteria listed above, the CMS experiment observed:
\begin{itemize}
\item[(a)] 510 events with the SM background expectation of 560$\pm$87 events for $m_{\ell^{+}\ell^{-}} <$ 75 GeV.
\item[(b)] 178 events with the SM background expectation of 200$\pm$35 events for $m_{\ell^{+}\ell^{-}} >$ 105 GeV. 
\end{itemize}
In case (a) we have an upper limit of 37 signal events, while in 
  case (b) leads to an upper limit of 13 signal events. 
Using these limits, we can set an upper bound on $|V_{\ell N}|^{2}$ for a given value of $m_{N}$.

\begin{figure}
\begin{center}
\includegraphics[scale=0.8]{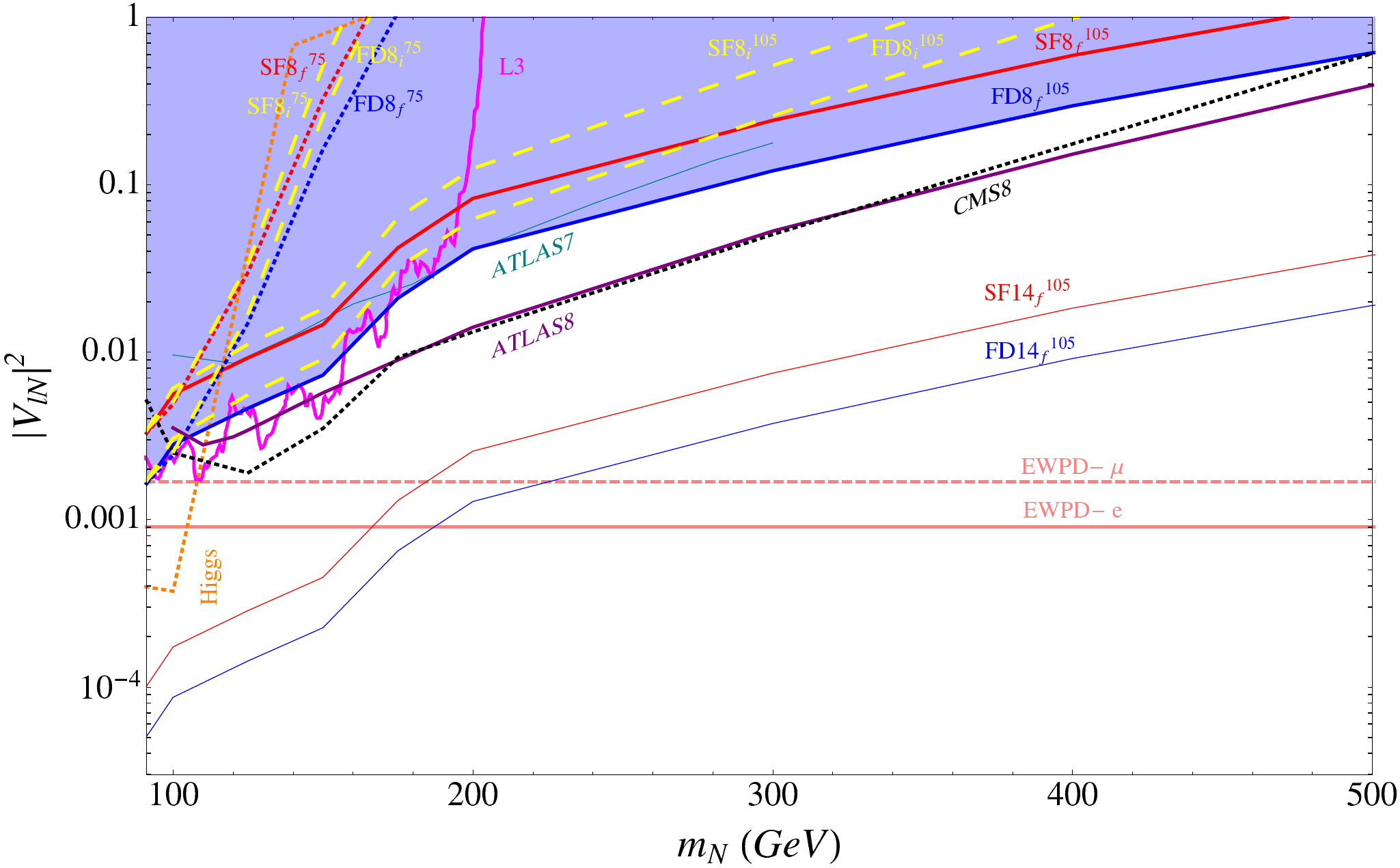}
\end{center}
\caption{The 95 $\%$ CL upper limits on the light-heavy neutrino mixing angles $|V_{\ell N}|^2$ as a function of the heavy Dirac neutrino mass $m_{N}$, derived from the CMS tri-lepton data at $\sqrt{s}=$8 TeV LHC for 19.5 fb$^{-1}$ luminosity \cite{CMS-trilep}. The exclusion (shaded) regions are shown for two benchmark scenarios: (i) single flavor (SF) and (ii) flavor diagonal (FD), with two choices of the selection cut $m_{l^{+}l^{-}}\textless$75 GeV (thick dotted) and $\textgreater$105 GeV (thick solid). The previous results with the selection cut $m_{l^{+}l^{-}}\textless$75 GeV (thick dot-dashed) and $\textgreater$105 GeV (thick large-dashing) at the 1-jet level are shown in this context from \cite{DDO1}. The corresponding conservative projected limits from $\sqrt{s}=$14 TeV LHC data with 300 fb$^{-1}$ integrated luminosity are shown by thin solid lines ($SF14_{f}^{105}$ and $FD14_{f}^{105}$) above the $Z$-pole. Some relevant existing upper limits (all at 95$\%$ CL) are also shown for comparison: (i) from a $\chi^{2}$-fit to the LHC Higgs data \cite{LHCHiggs} (Higgs), (ii) from a direct search at LEP \cite{L3}(L3), valid only for the electron flavor, (iii) ATLAS limits from $\sqrt{s}=$7 TeV LHC data \cite{ATLAS7} (ATLAS7) and $\sqrt{s}=$8 TeV LHC data \cite{ATLAS8} (ATLAS8), valid for a heavy Majorana neutrino of the muon flavor, (iv) CMS limits from $\sqrt{s}=$8 TeV LHC data \cite{CMS8} (CMS8), for a heavy Majorana neutrino of the muon flavor and (v) indirect limit from a global fit to the electroweak precision data \cite{EWPD1, EWPD2, EWPD3} (EWPD), for both electron (solid) and muon (dotted) flavors. The data for 1-jet case from \cite{DDO1} are shown by $SF_{i}^{75}$, $SF_{i}^{105}$, $FD_{i}^{75}$and $FD_{i}^{105}$ (large dashed lines). The shaded region is excluded.}
\label{figtrilep}
\end{figure}

In Fig.~\ref{figtrilep} we plot our results of the upper bound for the SF and FD cases.
The thick red dashed line ($SF8_{f}^{75}$) stands for the results of the SF case by employing the CMS observation (a) whereas 
  the thick red solid line ($SF8_{f}^{105}$) stands for the results of the SF case by employing the CMS observation (b). 
The corresponding results for the FD case are represented by the thick blue dashed line ($FD8_{f}^{75}$) and
  the thick blue solid line ($FD8_{f}^{105}$), respectively.
The results of the previous analysis \cite{DDO1} are shown by the yellow dashed lines ($SF8_{i}^{75}$, $SF8_{i}^{105}$, $FD8_{i}^{75}$ and $FD8_{i}^{105}$).
Our new results have improved the upper bound on the mixing angles. 
This improvement becomes more significant in the high $m_{N}$ region because the photon mediated process dominates 
  the $N$ production cross section (compare Figs.~\ref{fig4}, \ref{fig5a} and \ref{fig7} each other).
We have also calculated the cross sections at the 14 TeV LHC. 
Comparing the cross sections at 14 TeV to those at 8 TeV we have obtained the prospective upper bounds
  at the 14 TeV LHC, which are shown as the thin solid lines ($SF14_{f}^{105}$ and $FD14_{f}^{105}$) above the $Z$-pole.

\section{Conclusion}

We have studied the productions of heavy Majorana and pseudo-Dirac neutrinos 
  from the type-I and the inverse seesaw mechanisms, respectively.  
The heavy neutrinos can be produced at the LHC through the
  sizable mixings with the light neutrinos in the SM.
For the heavy neutrino production process, we have
  considered various initial states 
  such as $q\overline{q^{'}}$, $qg$ and $gg$ as well as photon mediated processes
  to produce the final states $N\ell$, $N\ell j$ and $N\ell jj$.
We have found that the cross section to produce $N\ell j$ from the 
  $q\overline{q^{'}}$ initial state dominates over the $qg$ for $m_{N} \gtrsim 150$ GeV, 
  while the $qg$ initial state takes over for $m_{N} \lesssim 150$ GeV at the 8 TeV LHC.
This happens because at low Bjorken scaling parameter the gluon PDF
  dominates over the quark PDF.
For the final state $N\ell jj$ the cross section obtained from $qg$ 
  dominates over that from the $q\overline{q^{'}}$ for 
  $m_{N} \lesssim 200$ GeV, while 
  for $m_{N} \gtrsim 200$ GeV the $q\overline{q^{'}}$ dominates.
The gluon fusion channel follows the cross sections from the 
  $qg$ and $q\overline{q^{'}}$ initial states.
The gluon fusion channel can not dominate 
  for the $m_{N}$ values we studied, because in order to
  produce the heavy neutrino the Bjorken scaling parameters for 
  the two gluons can not be small simultaneously.
We have also studied the photon mediated processes such as elastic, 
  inelastic and the deep-inelastic processes.
For $m_{N} \lesssim 400$ $(300)$ GeV, the deep-inelastic
  process dominates over the other photon mediated processes
  for $p_{T}^{j} \gtrsim 10$ $(20)$ GeV.
The inelastic process dominates over the deep-inelastic 
  process for $m_{N} \gtrsim 400$ $(300)$ for $p_{T}^{j} \gtrsim 10$ $(20)$ GeV.
The elastic process always yields the lowest cross section.

We have showed the kinematic distributions of the heavy neutrino
  produced with the associated leptons and jets for $m_{N}=100$ GeV and $m_{N}=500$ GeV.
We have plotted the differential scattering cross sections as a function of
  $p_{T}$ and $\eta$ of the heavy neutrino, associated lepton and the jets,
and the invariant mass distributions for the final states. 
The $p_{T}^{\ell}$ distribution has a peak around $\frac{m_{N}}{5}$ 
  for the $q\overline{q^{'}}$, $qg$ and $gg$ initial states,
  while for the photon mediated processes the distribution 
  is found to be almost independent of $m_{N}$.
The $\eta^{\ell}$ distributions of the leptons 
  are peaked around $0$ for all the initial states and
  become sharper with the increase in $m_{N}$.
The $p_{T}$ and $\eta$ distributions of the heavy neutrinos 
  are similar to those of the leptons irrespective of the 
  initial states.
The $p_{T}$ distributions of the jets start from cut values used
  in the analysis for final states with single and two-jets.
The $p_{T}^{j}$ distributions for the $N\ell j$ final state from the 
  $q\overline{q^{'}}$ and $qg$ initial states sharply drop with respect to 
  those for the $N\ell j j$ final state from the 
  $q\overline{q^{'}}$, $qg$ and $gg$ initial states.
In case of the photon mediated processes the $p_{T}$ distributions 
  of the jets are observed to have peaks around $\frac{m_{W}}{2}$
  for the $N\ell j$ final state from the elastic, inelastic cases and
  the leading jet from the $N\ell jj$ final state from the deep-inelastic process.
The $\eta$ distributions of the leading and non-leading jets 
  of the photon mediated processes show the distinct double peaks
  at the forward backward regions.
This is a typical feature of the photon mediated process.
We notice that the invariant mass distributions of the different 
  final states sharply peak at the vicinity of the heavy neutrino 
  mass, referring to the production of the heavy neutrino almost at rest.

We have studied the collider signatures of Majorana and pseudo-Dirac heavy neutrinos.
From the Majorana heavy neutrino we consider the same-sign di-lepton plus di-jet in the final state.
We have generated the signal cross sections by using {\tt MadGraph}. 
We compare our results with the recent ATLAS \cite{ATLAS8} and CMS \cite{CMS8} data at $\sqrt{s}=$ 8 TeV with 
  20.3 fb$^{-1}$ and 19.7 fb$^{-1}$ luminosities, respectively, we have obtained the upper bound on the mixing angles. 
Using a variety of the initial states we have improved the upper bounds previously obtained in \cite{ATLAS8} and \cite{CMS8}. 
For the pseudo-Dirac heavy neutrino we have generated the tri-lepton plus missing 
  energy as the signal events using {\tt MadGraph} and performed the
  detector simulation for various initial states.
Comparing our simulation results with the recent anomalous multi-lepton search results by
  CMS \cite{CMS-trilep} at $\sqrt{s}=$ 8 TeV with 19.5 fb$^{-1}$ luminosity
  we have obtained the upper bound on the mixing angles, 
  which improves the previously given result in \cite{DDO1}.
Finally we have scaled our results at the 8 TeV LHC to obtain a prospective
  search reach at the 14 TeV LHC with higher luminosities.

\section*{Acknowledgements}

The authors cordially thank P. S. Bhupal Dev for his collaboration in the earlier stage of the project.
AD would like to thank Valentin Hirschi for useful discussions on {\tt MadGraph}.
This work is supported in part by the United States Department of Energy Grant, No. DE-SC 0013680.


\end{document}